\documentclass[twocolumn,aps,prx,footinbib,10pt]{revtex4-2}

\usepackage{amsmath}
\usepackage{amssymb}
\usepackage{wasysym}
\usepackage{graphicx}
\usepackage{hyperref}
\usepackage{dsfont}
\usepackage{newtxtext}
\usepackage[varvw]{newtxmath}
\usepackage{bbm}
\usepackage[percent]{overpic}
\usepackage{tabularx}
\usepackage{mathtools}

\hypersetup{
    colorlinks,
    linkcolor={blue},
    citecolor={blue},
    urlcolor={blue}
}

\graphicspath{{./}{./figs/}}

\newcommand{\rme}{\mathrm{e}}
\newcommand{\rmi}{\mathrm{i}}

\newcommand{\eg}{{e_\mathrm{g}}}
\newcommand{\tg}{{t_\mathrm{2g}}}

\usepackage{xcolor}
\usepackage[normalem]{ulem}

\newcommand\rowincludegraphics[2][]{\raisebox{-0.5\height}{\includegraphics[#1]{#2}}}

\begin{document}

\title{Exact nematic and mixed magnetic phases driven by competing orders on the pyrochlore lattice}

\author{Niccol\`o Francini}
\author{Lukas Schmidt}
\author{Lukas Janssen}
\author{Daniel Lozano-G\'omez}

\affiliation{Institut f\"ur Theoretische Physik and W\"urzburg-Dresden Cluster of Excellence ctd.qmat, Technische Universit\"at Dresden, 01062 Dresden, Germany}
\begin{abstract}
Pyrochlore magnets are a paradigmatic example of three-dimensional frustrated systems and provide an excellent platform for studying a variety of exotic many-body phenomena, including spin liquids, nematic phases, fragmentation, and order by disorder. In recent years, increasing attention has been devoted to bilinear spin models on this lattice, where multiple magnetic phases can be degenerate in energy, often stabilizing unconventional magnetic states. In this work, we focus on one such model, parametrized by the interaction coupling $J_{z\pm}$, which defines a line in parameter space corresponding to the phase boundary between three distinct magnetic phases. Using a combination of analytical and numerical methods, we show that this model exhibits an order-by-disorder mechanism at low temperatures, giving rise to a \emph{mixed} magnetic phase. 
This realizes a $\mathbf q=0$ long-range-ordered phase in a pyrochlore magnet characterized by two distinct order parameters, which we denote as the $A_2 \oplus \psi_2$ phase. Furthermore, at $J_{z\pm} = 1/\sqrt{2}$, the model acquires a subextensive number of discrete symmetries, which preclude the stabilization of conventional long-range order and instead lead to the emergence of a novel nematic phase. We characterize this nematic phase, describe how its ground-state configurations are constructed, and analyze its stability at higher temperatures and under small deviations from $J_{z\pm} = 1/\sqrt{2}$.
\end{abstract}

\date{\today}

\maketitle

\tableofcontents

\section{Introduction}
Competing interactions, broadly referred to as frustration, are key drivers in the emergence of novel many-body phenomena. In magnetic systems, this competition often results in a complex energy landscape, where a  system cannot simultaneously minimize all energetic terms. Frustration is generally categorized into two types: geometric and exchange frustration. 
Geometric frustration arises when the lattice geometry prevents all spin interactions from being simultaneously satisfied~\cite{Springer_frust,Balents2010,savary2012}. This setting is characteristic of lattices containing triangular patterns, such as the two-dimensional triangular, Kagome, and square Kagome lattices~\cite{kagome_PhysRevB.71.024401,Yogendra_PRB}, and the three-dimensional pyrochlore lattice, a periodic network consisting of corner-sharing tetrahedra~\cite{Rau2018FrustratedQR}.
On the other hand, exchange frustration takes place when the spin interactions in the system do not share a common spin configuration which minimizes all interactions simultaneously independently of the lattice geometry.
A notable example is provided by anisotropic exchange interactions, which can result from strong spin–orbit coupling~\cite{krempa_14}, as in Kitaev materials~\cite{trebst22}.

In this scenario, magnetic pyrochlore oxides of the type $R_2^{3+} M_2^{4+} \mathrm{O}_7^{-2}$, where $R$ is a rare earth ion and $M$ is (typically) a nonmagnetic transition metal, represent the paradigmatic platform to study three-dimensional frustrated magnets, where geometric and exchange frustration coexist, paving the way to exotic many-body phenomena~\cite{Rau2018FrustratedQR}.
Indeed, in recent years, different works on pyrochlore magnets have further investigated how distinct many-body phenomena arise in this lattice. Examples of these are the realization of spin liquids~\cite{Castelnovo2008,Yan2020_rank2_u1,Moessner1998_low_temp,Taillefumier2017_xxz,Benton2016_Pinch_line,Benton2012_seeing,EvansPhysRevX.12.021015,lozano-gomez2023,chung_gingras_2023arxiv,lozano_2024arxiv}, thermal and quantum order by disorder~\cite{villain1980,belorizky1980,zhitomirsky2012,zhitomirsky2013,savary2012,chern10,Noculak_HDM_2023,Hickey_2024arxiv}, magnetic fragmentation~\cite{Brooks-Bartlett2014,Petit2016,Benton2016_quantum_origins,lozanogomez2025fragmentedSL}, topological magnons~\cite{Onose2010}, and nematic phases~\cite{Taillefumier2017_xxz,francini25}.
For the characterization of these phenomena, the corner-sharing architecture of the pyrochlore lattice can be exploited to track down the classical ground-state phase diagram for the most general classical bilinear nearest-neighbor Hamiltonian~\cite{Rau2018FrustratedQR,Yan2017_theory_multiphase}. This analysis identifies five $\mathbf{q}=0$ long-range ordered phases whose spin configurations are defined in a single tetrahedron~\cite{Yan2017_theory_multiphase,Wong2013}. In these phase diagrams, numerous works have focused on regions in the interaction-parameter space where the competition between multiple of these $\mathbf{q}=0$ phases is most extreme, possibly resulting in the realization of exotic spin phases~\cite{atlas_2024,Taillefumier2017_xxz,lozano-gomez2023,Benton2016_Pinch_line,chung2025quadrupolespyrochlore,Gresista_QPLSL}.

In the classical model, these regions of amplified competition take place at phase boundaries where multiple magnetic orders possess the same ground-state energy. While such boundaries have a central role in the realization of classical spin liquids~\cite{Taillefumier2017_xxz,Benton2016_Pinch_line,lozano-gomez2023,atlas_2024}, they do \emph{not} guarantee the extensive ground-state degeneracy typical of classical spin-liquid phases, and may instead displayed $\mathbf{q}=0$ ordering due to order by disorder~\cite{Yan2017_theory_multiphase,Hickey_2024arxiv}, as well as nematic phases~\cite{francini25}.
In all of these cases, however, a recurring feature resulting from the energy degeneracy of multiple magnetic order is the emergence of accidental symmetries, an ubiquitous aspect the realization of all the aforementioned phenomena in pyrochlore magnets~\cite{Yan2017_theory_multiphase,Taillefumier2017_xxz,Benton2014_thesis,NoculakPRB2024,francini25,Gresista_QPLSL}.
Accidental symmetries typically emerge at the mean-field level, reflecting an enlarged ground-state manifold where the spin configurations of the competing magnetic orders deform continuously into each other with no additional cost. These enlarged ground-state manifolds are usually parametrized by the additional accidental symmetry transformations, which are not part of the symmetry group of the microscopic model, thereby reflecting the enlargement of the mean-field ground-state manifold. Since these manifolds lack the protection that an exact symmetry possesses, spin configurations associated by these accidental symmetries result in different spin fluctuations and therefore nonequivalent corrections to their ground-state energy in the quantum state (quantum order by disorder)~\cite{khatua23,hickey2025arXiv_ObD,khatua2025finitesizespectralsignaturesorder,savary2012}, and to additional entropy contributions to the free-energy when $T\neq 0$ (thermal order by disorder)~\cite{Hickey_2024arxiv,schick_FCC_2020}, which ultimately lead to the selection of particular states in the accidentally degenerate manifold.

On the pyrochlore lattice, one of the most studied case of emergent symmetries takes place whenever the magnetic order associated with the $E$ irreducible representation (irrep) of the cubic symmetry group is one of the low-energy states of the system. The spin configurations of the $E$ irrep are defined in the so-called $\Gamma_5$ manifold, an accidentally degenerate U(1) manifold relevant for $XY$ pyrochlore magnets such as $\mathrm{Er_2Ti_2O_7}$ and $\mathrm{Yb_2Ge_2O_7}$~\cite{Rau2018FrustratedQR,Wong2013,zhitomirsky2012,savary2012,Sarkis2020YbGeO}. In the general case, accidental degeneracies typically result in additional accidental continuous compact manifolds. 
%
In the present work, we investigate the $A_2\oplus E \oplus T_{1-}$ triple line, where three energetically-degenerate long-range magnetic orders meet. This region is characterized by the couplings of the bilinear nearest-neighbor Hamiltonian, and can be effectively parametrized by a line in coupling space with a single exchange parameter, namely $J_{z\pm}$. At the $A_2 \oplus E \oplus T_{1-}$ line, a spin liquid is not expected to be realized since the energy landscape of this model does not reflect the typical extensive degeneracy~\footnote{This extensive degeneracy is typically seen in low-energy flat bands in the interaction matrix, which reflects an extensive ground-state degeneracy, hinting at a spin-liquid state.} associated with a spin liquid~\cite{KTC_2024_phase,atlas_2024}.
The lack of a macroscopic (system-size-scaling) ground-state degeneracy, characteristic of such spin liquids, does not preclude the emergence of exotic many-body phenomena~\cite{francini25}.

\begin{figure*}
    \centering
    \begin{overpic}[width=\textwidth]{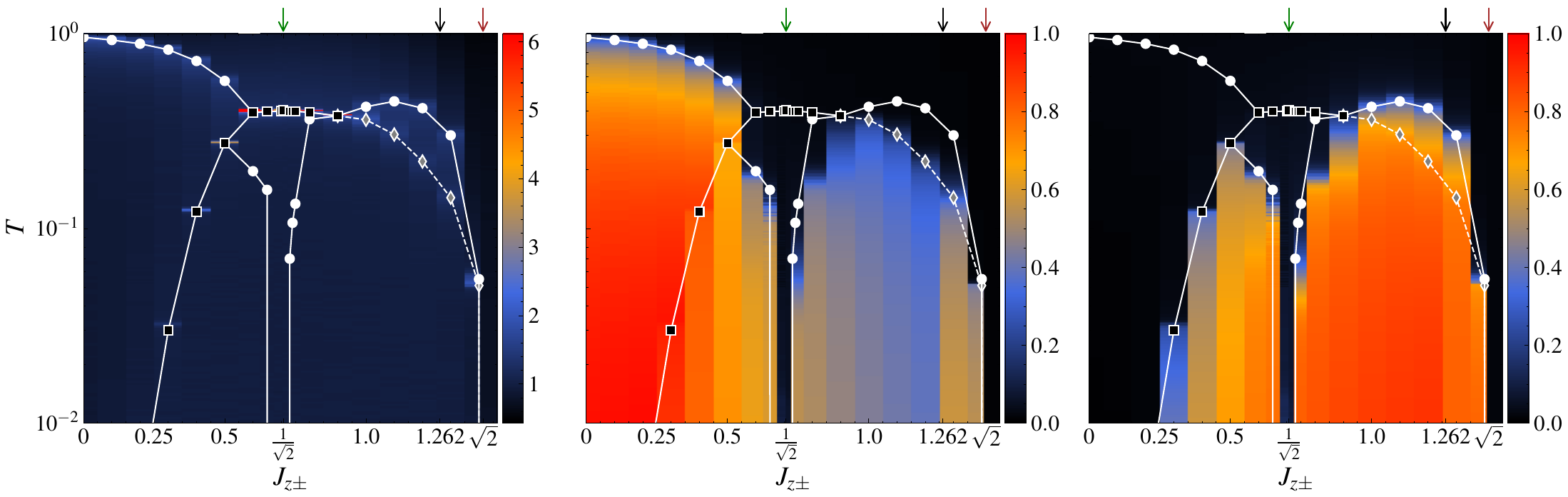}
    \put(5,31){(a)}
    \put(32,30){$C$}
    \put(37,31){(b)}
    \put(64,31){$|m_{A_2}|$}
    \put(70,31){(c)}
    \put(96,31){$|m_E|$}
    \put(7,20){\color{white} $A_2$}
    \put(39,20){\color{black} $A_2$}
    \put(71,20){\color{white} $A_2$}
    \put(10.5,6){\color{white} $A_2\oplus\psi_2$}
    \put(42.5,6){$A_2\oplus\psi_2$}
    \put(74.5,6){$A_2\oplus\psi_2$}
    \put(21,6){\color{white} $A_2\oplus\psi_2$}
    \put(54,6){$A_2\oplus\psi_2$}
    \put(86,6){$A_2\oplus\psi_2$}
    \put(16.5,22){\color{white}SN}
    \put(48.5,22){\color{white}SN}
    \put(80.5,22){\color{white}SN}
    \put(28,27){\color{white}PM}
    \put(60,27){\color{white}PM}
    \put(92,27){\color{white}PM}
    \end{overpic}
\caption{Phase diagram as a function of temperature $T$ and coupling $J_{z\pm}$ along the $A_2\oplus E\oplus T_{1-}$ line from classical Monte Carlo simulations, with the color scale indicating (a) specific heat, (b) $A_2$ order parameter $|m_{A_2}|$, and (c) $E$ order parameter $|\mathbf{m}_E|$. The green arrow marks the position of the exact nematic point at $J_{z\pm}=1/\sqrt{2}$. The black arrow highlights the position of the jump of the $A_2\oplus E$ mixing angle. Lastly, the red arrow signals the Heisenberg antiferromagnetic point at $J_{z\pm}=\sqrt{2}$ where a spin-liquid phase is stabilized down to zero temperature. The specific heat peaks and the order parameters define several regions in the $(J_{z\pm},T)$ phase diagram. The high-temperature paramagnetic phase freezes into different phases depending on $J_{z\pm}$. For $J_{z\pm}\lesssim 1/\sqrt{2}$, there is an intermediate $\mathbf{q}=0$ $A_2$ phase which is driven into a $A_2\oplus \psi_2$ long-range order at lower temperatures, while for $J_{z\pm}\gtrsim1/\sqrt{2}$, this mixed phase is realized directly from the paramagnetic phase, with an intermediate region characterized by $|m_E| \gg |m_{A_2}| > 0$, separated from the low-temperature region by a crossover. 
From the exact spin-nematic (SN) point at $(J_{z\pm},T)=(1/\sqrt{2},0)$ stems a finite-temperature nematic phase fan.
Lines are guides to the eye, with solid lines indicating true phase transitions and dashed lines indicating crossovers.
White (black) markers refer to continuous (first-order) phase transition, while gray diamonds mark crossover temperatures.
}
\label{fig:phase-diagram}
\end{figure*}
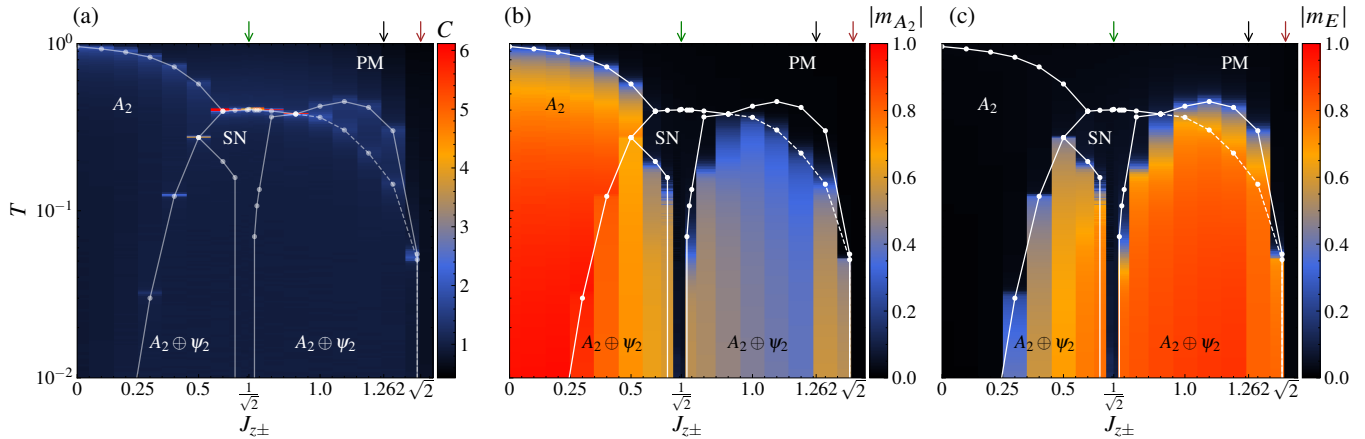

In this work, we aim to fill this gap. Employing a variety of analytical and numerical methods, we expose a thermal order-by-disorder selection into a low-temperature $\mathbf{q}=0$ long-range ordered \emph{mixed} phase for $J_{z\pm}\neq\pm 1/\sqrt{2}$. 
This phase extends the conventional single-irrep description of $\mathbf{q}=0$ order by exhibiting coexisting $A_2$ and $E$ order parameters.
The novel $A_2\oplus E$ phase is characterized analytically, and it is exposed in the classical Monte Carlo phase diagram. Furthermore, we show that at the points $J_{z\pm}=\pm 1/\sqrt{2}$, the system acquires an additional $\mathbb{Z}_2$ subsystem symmetry involving spin-component flips along the chains forming the pyrochlore lattice~\cite{Benton2014_thesis}. The presence of these new symmetries drastically changes the nature of the low-temperature phase; the ground state acquires a subextensive degeneracy, and the low-temperature $\mathbf{q}=0$ long-range order is suppressed, instead realizing nematic phase, which breaks rotational symmetry but preserves time reversal symmetry. We characterize the nematic phase through exact analytical results at zero temperature supported by finite-temperature classical Monte Carlo data. Additionally,  duality relations reveal the presence of three additional nematic points on a different degenerate phase boundary, namely the $T_2 \oplus T_{1-}$ plane. The main results of this work are summarized in the phase diagram, constructed via Monte Carlo simulations, in Fig.~\ref{fig:phase-diagram},
which displays the evolution of the specific heat and order parameters as functions of temperature and $J_{z\pm}$, tuning the system along the $A_2 \oplus E \oplus T_{1-}$ line. Table~\ref{tab:resume-table} summarizes the finite-temperature phases, their symmetries, structure-factor signatures, and representative single-tetrahedron configurations.
\begin{table}[t]
\caption{Characterization of the finite-temperature phases identified in this work. Each phase breaks either time-reversal symmetry ($\tau$), $2\pi/3$ rotational symmetry about the [111] axis ($C_3$), or both. While the $A_2$ all-in-all-out phase and the mixed $A_2 \oplus \psi_2$ phase exhibit isolated Bragg peaks, the nematic phase is characterized by line features in the equal-time spin structure factor (SSF). The last column shows representative spin configurations on a single tetrahedron for each phase.}
    \centering
\begin{tabular*}{\linewidth}{@{\extracolsep{\fill} }>{\raggedright\arraybackslash}p{6em} c c >{\raggedright\arraybackslash}p{6.5em} c}
    \hline 
    \hline
        \multicolumn{1}{l}{Phase} & $\tau$ & $C_3$ &  SSF & Spin configuration\\ \hline
        $A_2$\newline (all-in-all-out) & $ \times$ & \checkmark & Bragg peak at $\mathbf{q}=0$ & \rowincludegraphics[width=0.4\columnwidth]{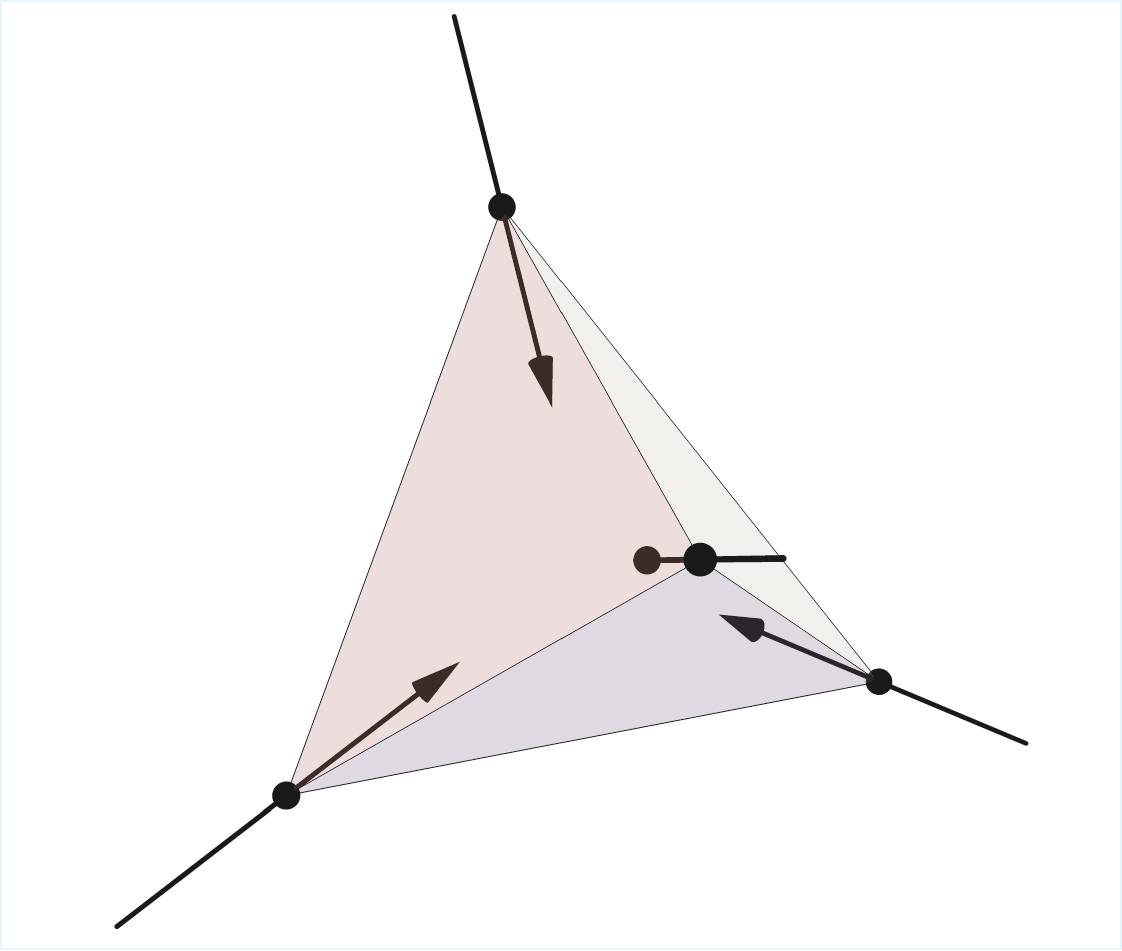} \\  \hline
        $A_2\oplus\psi_2$\newline (mixed) & $\times$ & $\times$ & Bragg peak at $\mathbf{q}=0$ & \rowincludegraphics[width=0.3\columnwidth]{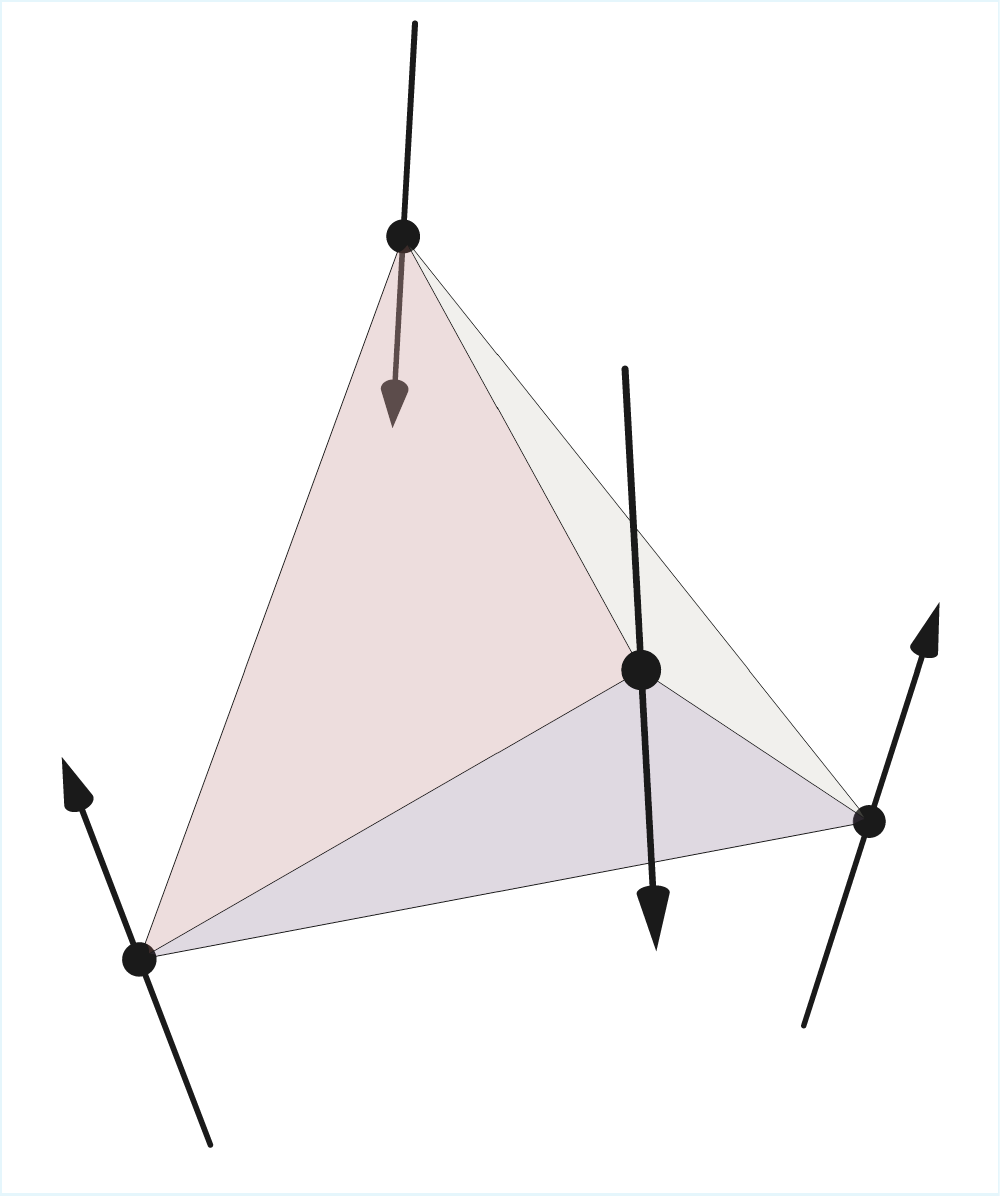}  \\  \hline
        Spin nematic & \checkmark & $\times$ &High-intensity lines along high-symmetry directions  & 
        \rowincludegraphics[width=0.4\columnwidth]{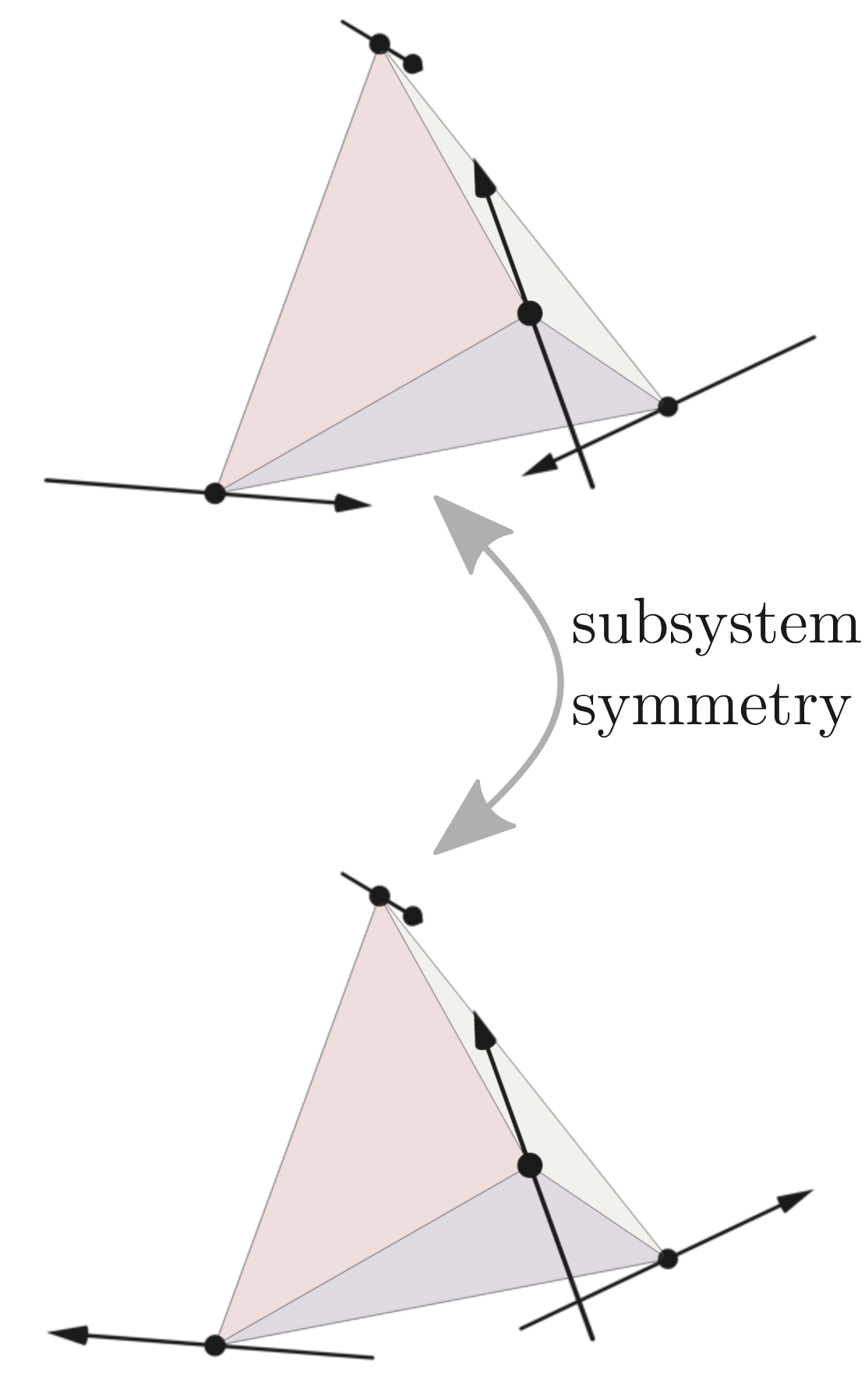} \\
        \hline
        \hline
\end{tabular*}
    \label{tab:resume-table}
\end{table}
The paper is organized as follows. In Sec.~\ref{sec:irreps-analysis}, we parametrize the $A_2 \oplus E \oplus T_{1-}$ line and characterize the accidental symmetries emerging from irrep analysis. 
The entropic effects on the accidentally degenerate manifolds are analyzed in Sec.~\ref{sec:CLTE}, revealing an order-by-disorder mechanism driving the system into the low-temperature mixed $A_2\oplus E$ phase.
The mixed phase is phenomenologically captured within a Landau theory framework in Sec.~\ref{sec:Landau-theory}.
In Sec.~\ref{sec:cMC}, we provide classical Monte Carlo results for the $\mathbf{q}=0$ phases, showing that the low-temperature order is reached through one or two phase transitions, depending on $J_{z\pm}$ value.
The nematic phase is studied in detail in Sec.~\ref{sec:SN-phase} with analytical results and Monte Carlo simulations.
Our conclusions are given in Sec.~\ref{sec:discussion-conclusion}.
%

\section{Irreducible-representation analysis}
\label{sec:irreps-analysis}
Single-tetrahedron irrep analysis has proven to be a powerful tool to determine the classical ground-state phase diagram for pyrochlore magnets with a nearest-neighbor bilinear spin Hamiltonian~\cite{Yan2017_theory_multiphase,atlas_2024,KTC_2024_phase}. Consider the most generic nearest-neighbor bilinear Hamiltonian compatible with the tetrahedron symmetry $T_d$~\cite{Rau2018FrustratedQR},
\begin{eqnarray}
\mathcal{H}&=&\sum_{\langle ij \rangle} \Big\{ J_{zz} S_i^z S_j^z -J_{\pm}( S_i^+ S_j^- + S_i^- S_j^+) + J_{\pm\pm } (\gamma_{ij}S_i^+ S_j^+ +\rm{h.c.})\nonumber\\
&& \qquad +\ J_{z\pm}(\zeta_{ij}S_i^z( S_j^++S_j^-) +\rm{h.c.})\Big\},
\label{eq:general_bilinear_hamiltonian} 
\end{eqnarray}
with $S^z_i$ and $S_i^\pm = S_i^x \pm \rmi S_i^y$ being the components of the (pseudo-) spin-$1/2$ on site $i$ in a \emph{local} coordinate frame as defined in Appendix~\ref{appendix:irreps-basis}. In Eq.~\eqref{eq:general_bilinear_hamiltonian}, the bond-dependent phase factors $\gamma_{ij}=-\zeta_{ij}^\ast$ in the four-site unit cell are defined as
\begin{eqnarray}
    \gamma=\begin{pmatrix}
        0 & 1 & c & c^\ast\\
        1 & 0 & c^\ast & c\\
        c & c^\ast & 0 & 1\\
         c^\ast & c & 1 & 0
    \end{pmatrix}
\end{eqnarray}
where $c=e^{2\pi i/3}$ as reported in Ref.~\cite{atlas_2024,Ross2011quantum_excitations}.
Exploiting the corner-sharing structure of the pyrochlore lattice, the Hamiltonian in Eq.~\eqref{eq:general_bilinear_hamiltonian} can be reformulated as
\begin{equation}
    \label{eq:hamiltonian_sum_tetrahedron}
    \mathcal{H}=\sum_{\boxtimes}\mathcal{H}_{\boxtimes}\,,
\end{equation}
where $\mathcal{H}_{\boxtimes}$ represents the single-tetrahedron Hamiltonian, and the summation extends over all tetrahedra labeled by $\boxtimes$. As a consequence, the minimization of the original Hamiltonian reduces to the minimization of the single-tetrahedron Hamiltonian $\mathcal{H}_{\boxtimes}$. 
In the classical limit, the spin components commute with each other, allowing the single-tetrahedron Hamiltonian to be decomposed in terms of the five irreps of the single-tetrahedron symmetry group $T_d$~\cite{KTC_2024_phase,Wong2013,Yan2017_theory_multiphase}:
\begin{eqnarray}
    \mathcal{H}_\boxtimes   &=& \frac{1}{2}\left[ a_{A_2} \mathrm{m}_{A_2}^2 + a_{E} \mathbf{m}_{E}^2 + a_{T_2} \mathbf{m}_{T_2}^2 \right. \nonumber\\
   && \left. +\  a_{T_{1-}} \mathbf{m}_{T_{1 -}}^2
  +  a_{T_{1+}} \mathbf{m}_{ T_{1 +}}^2 \right]\,.
  \label{eq:irrep_decomp}
\end{eqnarray}
Here, $A_2$, $E$, $T_2$, $T_{1\pm}$ denote the five different irreps entering the single-tetrahedron Hamiltonian, each one of them representing a different spin mode. Proceding in order, the one-dimensional $A_2$ irrep describes the all-in-all-out (AIAO) mode~\cite{sadeghiSpinHamiltonianOrder2015,Petit2016}, the two-dimensional $E$ irrep describes a U(1) manifold of states also known as the $\Gamma_5$ manifold~\cite{McClarty_2009,zhitomirsky2012,Sarkis2020YbGeO}, the three dimensional $T_2$ irrep designates the Palmer-Chalker state~\cite{PC2000PRB,yahne2021_reentrance}, and, roughly speaking, the three dimensional $T_{1\pm}$ irreps are related to two types of splayed ferromagnetic~\footnote{For a specific combination of the interaction parameters $\{J_{zz},J_\pm, J_{\pm\pm},J_{z\pm}\}$ one of the $T_1$ irreps is identified with a collinear ferromagnet while the remaining one is a coplanar antiferromagnet~\cite{Noculak_HDM_2023}. Aside from these cases, the two $T_1$ irreps are always ferromagnetic.}
orders~\cite{Thompson2017,Ross2011quantum_excitations}.
In Eq.~\eqref{eq:irrep_decomp}, $\{\mathbf{m}_I\}$ are the spin modes corresponding to the $I$-th irrep while $\{a_I\}$ are functions of the couplings $\{J_{zz},J_\pm,J_{\pm\pm},J_{z\pm}\}$ and represent the energy cost associated to the modes $\mathbf{m}_I$. Details on the formulations of the irreps in terms of spins, their energy cost in terms of the interaction parameters, and coordinate frame conventions are given in Appendix~\ref{appendix:irreps-basis}.

The decomposition in Eq.~\eqref{eq:irrep_decomp} has a key role in determining the classical ground-state phase diagram~\cite{Yan2017_theory_multiphase,Wong2013,KTC_2024_phase}. In particular, it serves to identify the irrep with the lowest associated energy cost $a_{I}$, and their corresponding spin mode, effectively characterizing the classical ground-state manifold. In case the ground-state manifold is characterized by multiple irreps and a spin-liquid phase is stabilized, recent work has shown that these representations serve as a starting point in the identification of effective gauge theories describing the low-energy theory of these phases~\cite{Yan_2024_PRB_classification,atlas_2024}. 
In the present work, we focus on a region where the minimal energy irreps are the $A_2$, $E$, and $T_{1-}$ irreps, while $T_2$ and $T_{1+}$ are gapped at higher energy. We refer to the set of interaction parameters defining this model as the $A_2\oplus E\oplus T_{1-}$ line. The set of interaction parameters characterizing this region is obtained from the condition $a_{A_2}=a_{E}=a_{T_{1-}}=\min\{a_{A_2},a_{E},a_{T_2},a_{T_{1\pm}}\}$ yielding the constraints
\begin{align}
\label{eq:A2ET1minus_line}
        J_{zz}&<0\,, &
        |J_{z\pm}|&<\sqrt{2}|J_{zz}|\,, \\
        J_\pm &= -\frac{J_{zz}}{2}\,, &
        J_{\pm\pm}&= \frac{J_{zz}^2-J_{z\pm}^2}{J_{zz}}.       
\end{align}
Fixing the overall energy scale $J_{zz}=-1$, the three remaining interaction parameters are constrained by two equations, so the effective region in the coupling space is a line parametrized by the $J_{z\pm}$ coupling, namely the $A_2\oplus E\oplus T_{1-}$ line. 
The boundaries $J_{z\pm}=\pm\sqrt{2}|J_{zz}|$ correspond to the antiferromagnetic Heisenberg point ($J_{z\pm}>0$) and its dual ($J_{z\pm}<0$)~\cite{atlas_2024}. At these points, rank-1 U(1) spin liquids persist as stable phases in the classical limit, extending down to the lowest temperatures~\cite{Moessner1998_low_temp,atlas_2024}. Although $J_{z\pm}$ covers positive and negative values, we may restrain the study only to $J_{z\pm} \geq 0$: the model in Eq.~\eqref{eq:general_bilinear_hamiltonian} has a duality transformation mapping the Hamiltonian into itself via a change of $J_{z\pm}$ sign followed by a local $C_2$ rotation of the pseudo-spins around the local $\hat{\mathbf{z}}_i$ axis~\cite{Rau2018FrustratedQR}, defined in Appendix~\ref{appendix:irreps-basis}. Consequently, the thermodynamics observed for the negative $J_{z\pm}$ region are equivalent to those for the positive $J_{z\pm}$, which in turn implies that the temperature-vs-$J_{z\pm}$ phase diagram is symmetric about the  $J_{z\pm}=0$ line up to local $C_2$ rotations~\cite{francini25}.
Note that the $J_{z\pm}$ parametrization can be classified in two cases: the $J_{z\pm}=0$ case (non-Kramers) and the $J_{z\pm} \neq 0$ case (Kramers).
The nomenclature is due to the different transformation laws of the pseudospin components under time reversal, allowing or forbidding a finite $J_{z\pm}$~\cite{Rau2018FrustratedQR}.

In the classical limit and at zero temperature, the spins satisfy the hard spin-length constraint
\begin{align} \label{eq:spin-length}
\mathbf{S}_i \cdot \mathbf{S}_i=S^2 
\end{align}
at each site $i$, where $S$ is the spin length.

Along the $A_2 \oplus E \oplus T_{1-}$ line, the spin-length constraint imposes the relation
\begin{equation}
    \label{eq:irreps-constraint}
    \mathbf{m}_{\boxtimes}^2=m_{A_2}^2+\mathbf{m}_{E}^2+\mathbf{m}_{T_{1-}}^2 = 4S^2,
\end{equation}
at zero temperature, implying that the spin configuration in every single tetrahedron results from a linear combination of the low-energy degenerate irreps~\cite{atlas_2024}. 
Consequently, the weights of the irreps in a single-tetrahedron configuration in the ground-state manifold can be parametrized with two angles as
\begin{equation}
    \label{eq:irreps mixing}
    \mathbf{m}_{\boxtimes}=2S(\sin{\theta}\cos{\phi}\,\mathbf{e}_{A_2} + \sin{\theta}\sin{\phi}\,\mathbf{e}_{E} + \cos{\theta}\,\mathbf{e}_{T_{1-}})\,,
\end{equation}
where the angle $\phi$ measures the possible mixing between $A_2$ and $ E$ irreps, the angle $\theta$ quantifies the $T_{1-}$ irrep, and $\mathbf{e}_{A_2},\mathbf{e}_{E},\mathbf{e}_{T_{1-}}$ are fictious directions quantifying the $A_2$, $E$ and $T_{1-}$ irreps' magnitude, respectively.
Using this angular parametrization, the magnitude of the order parameters is obtained from the expressions
\begin{equation}
\label{eq:A2ET1_mixing_angles}
    \frac{|m_{A_2}|}{2S}=\sin{\theta}\cos{\phi}, \quad \frac{|\mathbf{m}_E|}{2S}=\sin{\theta}\sin{\phi}, \quad \frac{|\mathbf{m}_{T_{1-}}|}{2S}=\cos\theta\,,
\end{equation}
with ${\theta,\phi}\in[0,\pi/2]$, parametrizing a sphere of unit radius in the first octant in the coordinate system spanned by $\{\mathbf{e}_{A_2},\mathbf{e}_{E},\mathbf{e}_{T_{1-}}\}$.
%

%

\subsection{Accidental symmetries}
\label{subsec:mixing-angles}
The energy degeneracy of the irreps of the Hamiltonian describing the $A_2\oplus E \oplus T_{1-}$ line results in the identification of accidental symmetries defining continuous manifolds where the spin configuration can fluctuate with no associated energy cost. Indeed, the simplest low-energy manifold in the ground-state of the $A_2\oplus E \oplus T_{1-}$ line is associated with the $E$ irrep; in the Hamiltonian in Eq.~\eqref{eq:irrep_decomp}, the two-dimensional $E$ irrep takes the form
\begin{equation}
    a_E\mathbf{m}_E^2=a_E\left[(m_{E}^x)^2+ (m_{E}^y)^2\right]\,,
\end{equation}
where $m_{E}^x$ and $m_{E}^y$ correspond to the two different components of the irreps, and are often refer as the $\psi_2$ and $\psi_3$ states~\cite{Wong2013,Yan2017_theory_multiphase}. Previous works have shown that the states in the $E$ manifold are characterized by an accidental U(1) manifold
spanned by the $\psi_2$ and $\psi_3$ states, resulting in the so-called
$\Gamma_5$ manifold~\cite{McClarty_2009,zhitomirsky2012,Sarkis2020YbGeO}.
We note that the magnetic phase associated with a dominant $E$ irrep can be labeled either as an $E$ or a $\Gamma_5$ phase. The former label is more commonly used when the magnetic phase is classified according to the point group $T_d$, whereas the latter is used when the space group $Fd\bar{3}m$.

Apart from the $\Gamma_5$ manifold, which only considers the spin configurations spanned by the $E$ irrep, the degeneracy of the $E$ irrep with the $A_2$ and the $T_{1-}$ irrep results in the emergence of additional accidental symmetries. To identify these manifolds, we consider a generic spin configuration in the ground-state manifold of the Hamiltonian along the $A_2\oplus E \oplus T_{1-}$ line  parametrized by the six components comprising the three low-energy irreps~\footnote{The three low-energy degenerate irreps $A_2$, $E$ and $T_{1-}$ have different dimensions, being one-, two-, and three-dimensional, respectively, yielding a total of six degrees of freedom.} constrained by Eq.~\eqref{eq:irreps-constraint}. 
The constraint in Eq.~\eqref{eq:irreps-constraint} defines an $S^5$ hypersphere of radius $2S$. Consequently, the spin configurations in the ground-state manifold described by this hypersphere are parametrized by five different angles $(\theta,\phi,\eta,\beta,\gamma)$. In terms of these angles, the low-energy irrep components take the form
\begin{equation}
    \label{eq:angle-representation}
    \begin{split}
    &m_{A_2}=2S \sin{\theta}\cos{\phi}, \\
    &m_{E}^x=2S\sin{\theta}\sin{\phi}\cos{\eta},\\
    &m_{E}^y=2S\sin{\theta}\sin{\phi}\sin{\eta},\\
    &m_{T_{1-}}^x=2S\cos{\theta}\sin{\beta}\cos{\gamma},\\
    &m_{T_{1-}}^y=2S\cos{\theta}\sin{\beta}\sin{\gamma},\\
    &m_{T_{1-}}^z=2S\cos{\theta}\cos{\beta},
\end{split}
\end{equation}
where $\theta$ measures the mixing of $T_{1-}$ irrep with $A_2$ and $E$ irreps, while the $\phi$ angle quantifies the mixing between $A_2$ and $E$ irreps, as introduced in Eq.~\eqref{eq:A2ET1_mixing_angles}. The angle $\eta$ specifies the $\psi_2$ and $\psi_3$ components of the $E$ irrep,  while the $T_{1-}$ internal angles $\beta$ and $\gamma$ characterize the $T_{1-}$ irrep orientation. 
The ground-state spin configurations defined by Eq.~\eqref{eq:angle-representation} are characterized by a tuple of five angles $\boldsymbol{\pi}=(\theta,\phi,\eta,\beta,\gamma)$. However, these angles are strongly constrained by the local spin-length constraint in Eq.~\eqref{eq:spin-length}. For example, the spin-length constraint may be fulfilled by configurations with vanishing $T_{1-}$ contribution, where the angle $\theta$ in Eq.~\eqref{eq:angle-representation} is fixed to $\theta=\pi/2$, and the angles $\phi$ and $\eta$ are unconstrained. This particular choice of $\theta$ defines a 2-sphere, which we refer to as the $A_2\oplus E$ sphere. The unconstrained character of the $\phi$ and $\eta$ angles identifies two directions in which the spins can fluctuate with no energy cost, i.e., zero modes. These two zero modes are present in all the spin configurations in the $A_2\oplus E$ sphere. 
Generic spin configurations in the ground-state manifold of the $A_2\oplus E \oplus T_{1-}$ line are expected to have a nonvanishing $T_{1-}$ contribution, i.e., without constraining the angle $\theta$. These can be constructed starting from a configuration in the $A_2\oplus E$ sphere by applying a local rotation to the spins, which introduces a $T_{1-}$ contribution without violating the spin-length constraint. This implies that there are sets of angles $\phi$ and $\eta$ from which an introduction of a nonvanishing $T_{1-}$ irrep fulfills the spin-length constraint. At these particular points, additional zero-modes \emph{away} from the $A_2\oplus E$ sphere into a manifold of mixed states are associated with the spin configurations.
To identify the set of points for which fluctuations into multiple degenerate manifolds are possible, we employ a gradient expansion of Eq.~\eqref{eq:spin-length}, yielding the four equations
\begin{equation}
\label{eq:fluctuation-equation}
    \delta\boldsymbol{\pi}\cdot \left.\nabla(\mathbf{S}_\mu\cdot \mathbf{S}_\mu)\right|_{\boldsymbol{\pi}_0}=0,
\end{equation}
to leading order in $\delta\boldsymbol{\pi} = \boldsymbol{\pi} - \boldsymbol{\pi}_0$, where $\boldsymbol{\pi}_0$ is a ground-state spin configuration that satisfies the spin-length constraint, $\mu$ is the sublattice index, and $\nabla=(\partial_\theta,\partial_\phi,\partial_\eta,\partial_\beta,\partial_\gamma)$.

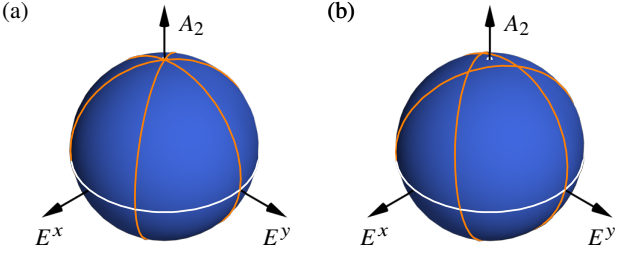
\begin{figure}
    \centering
    \begin{overpic}[width=\columnwidth]{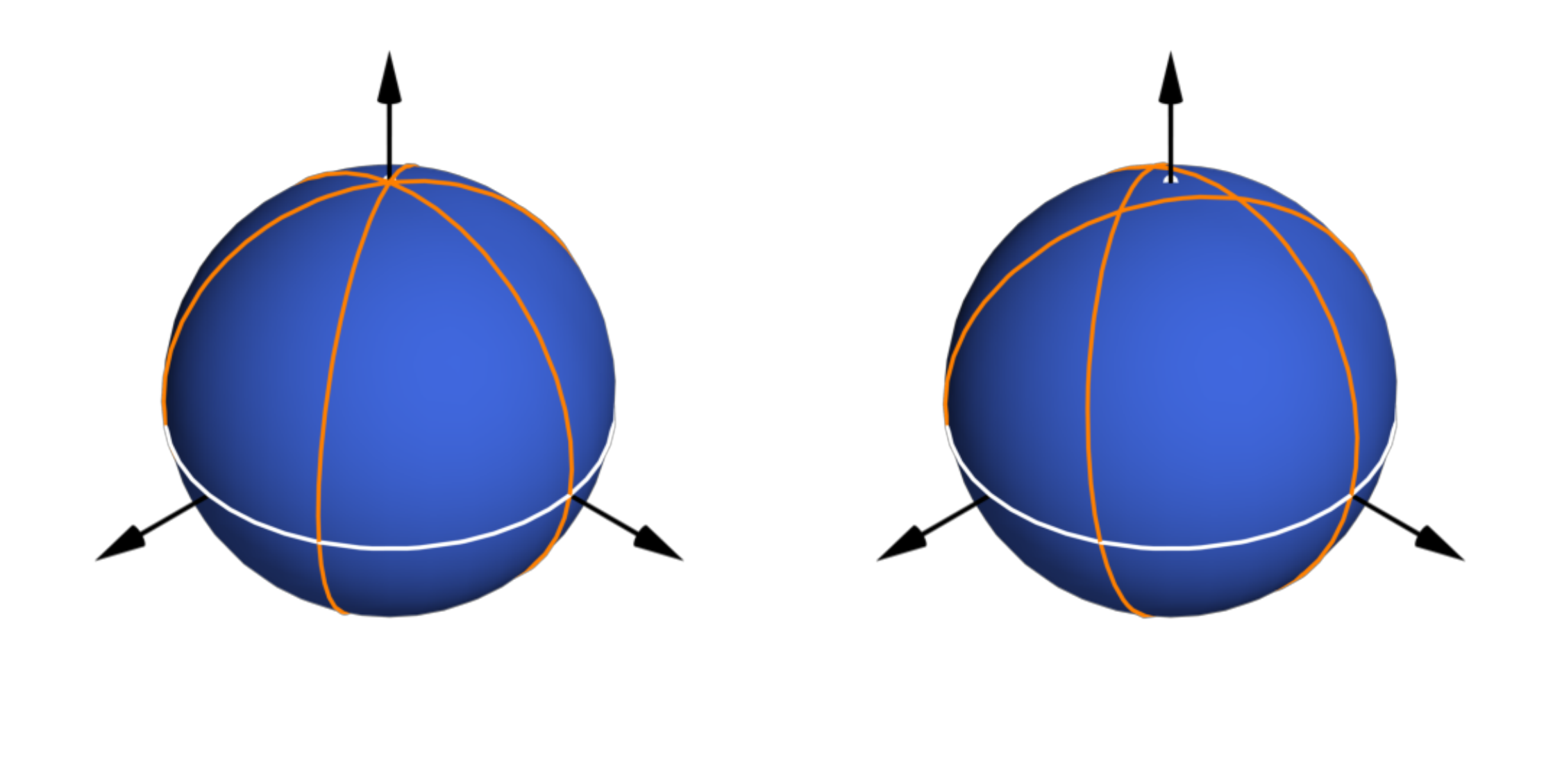}
    \put(0,45){(a)}
    \put(50,45){(b)}
    \put(5,10){$E^x$}
    \put(40,10){$E^y$}
    \put(19,4){$J_{z\pm}=0$}
    \put(27,43){$A_2$}
    \put(55,10){$E^x$}
    \put(90,10){$E^y$}
    \put(77,43){$A_2$}
    \put(67,4){$J_{z\pm}\approx0.14$}
    \put(50,45){(b)}
    \end{overpic}
    \caption{%
$A_2\oplus E$ sphere, defined by $\theta = \pi/2$ in Eq.~\eqref{eq:angle-representation}, for (a) $J_{z\pm} =0$ and (b) $J_{z\pm} \approx0.14$.
The white dot at the north pole corresponds to a spin configuration that is uniquely described by the $A_2$ irrep. The white circle along the equator marks the $\Gamma_5$ manifold, corresponding to spin configurations that are composed of the $E$ irrep mode only.
The orange circles correspond to spin configurations in the U(1) manifolds parametrized by Eqs.~\eqref{eq:A2E-U1-manifolds}, which can fluctuate away from the $A_2\oplus E$ sphere with no associated energy cost.
}            
    \label{fig:U1-manifolds}
\end{figure}

For more details on the derivation of the above equations, we refer the reader to Appendix~\ref{appendix:accidental_symm}. For configurations in the $A_2\oplus E$ sphere, Eq.~\eqref{eq:fluctuation-equation} identifies three $J_{z\pm}$-dependent U(1) manifolds labeled as circles in Fig.~\ref{fig:U1-manifolds}. These three U(1) manifolds are parametrized by the equations
\begin{equation}
\begin{split}
     \tan{\phi}\tan{\alpha_{\mathrm{mix}}}&=\frac{2}{\cos{\eta}+\sqrt{3}\sin{\eta}},\\
        \tan{\phi}\tan{\alpha_{\mathrm{mix}}}&=\frac{2}{\cos{\eta}-\sqrt{3}\sin{\eta}},\\
        \tan{\phi}\tan{\alpha_{\mathrm{mix}}}&=-\frac{1}{\cos{\eta}},    
\end{split}
\label{eq:A2E-U1-manifolds}      
\end{equation}
where $\alpha_{\mathrm{mix}}$ is the splay angle defining the $T_{1-}$ irrep. We refer the reader to Appendix~\ref{appendix:irreps-basis} for the definition of this quantity. For $J_{z\pm}=0$, the three circles intersect at the poles of the $A_2\oplus E$ sphere. At these points, the spin configuration is described by a single irrep, namely the $A_2$ irrep~\footnote{This should not be surprising as for $J_{z\pm}=0$ both the $E$ and the $T_{1-}$ irreps modes have all spins lying in the local $xy$. As such, the $A_2$ mode, where the spins are aligned with the local $z$ direction, can smoothly fluctuate into all other states.}.
As $J_{z\pm}$ is varied, the three circles move away from the poles and closer to the equator where $m_{A_2}=0$.

The presence of these additional zero-modes for the spin configurations defined by the U(1) manifolds shown in Fig.~\ref{fig:U1-manifolds} plays a pivotal role in the entropy associated with these spin configurations. Spin configurations with these additional zero-modes are typically associated with higher entropy compared to other energetically degenerate configurations. In the next sections, we study the impact of these additional zero-modes on the low-temperature thermal selection of a spin configuration in the $A_2\oplus E \oplus T_{1-}$ line.

\section{Classical Low-Temperature Expansion}
\label{sec:CLTE}
Classical low-temperature expansion is a method that analyzes the thermal spin fluctuation about a classical ground-state configuration~\cite{Noculak_HDM_2023}. At low temperatures, the spins are mostly polarized along a (local) ordering direction with small perpendicular fluctuations $\delta n_i^{\bar{\alpha}}$. In this low-temperature limit, the spin configurations may be approximated by
\begin{equation}
    \label{eq:CLTE-spins}
    \mathbf{S}_i \simeq \left( \delta n_i^{\tilde{x}}, \delta n_i^{\tilde{y}}, S\left(1-\frac{(\delta n_i^{\tilde{x}})^2}{2S^2}-\frac{(\delta n_i^{\tilde{y}})^2}{2S^2} \right)  \right)\,,
\end{equation}
where the coordinate system is chosen such that the $\tilde{z}$ axis aligns with the $i$-site ground-state spin, and  $\delta n^{\tilde{\alpha}} \ll 1$. For a particular ground-state configuration $\boldsymbol{\pi}_0$, applying the ansatz in Eq.~\eqref{eq:CLTE-spins} to the Hamiltonian in Eq.~\eqref{eq:general_bilinear_hamiltonian} allows us to obtain a low-temperature effective Hamiltonian governing spin fluctuations $\delta n^{\tilde{\alpha}}$ about the spin state $\boldsymbol{\pi}_0$. 
This approximation results in a single-tetrahedron effective Hamiltonian where the spins interact via the $8\times 8$ Hessian matrix $\mathcal{H}_2$. In terms of $\mathcal{H}_2$, the low-temperature free energy of a spin configuration in the ground-state manifold yields
\begin{align}
    \label{eq:CLTE-free-energy}
    \mathcal{F} &=\mathcal{H}_0-T\mathcal{S}
    \nonumber \\
    & = \mathcal{H}_0 - T \left( N\ln{T} - \frac{1}{2}\sum_{\mathbf{q}}\ln{\operatorname{det}\mathcal{H}_2(\mathbf{q})} \right),   
\end{align}
where $\mathcal{H}_0$ is the classical ground-state energy, $\mathcal{S}$ corresponds to the entropy resulting from thermal spin fluctuations, and $\mathcal{H}_2(\mathbf{q})$ represents the Hessian matrix in reciprocal space. 
Equation~\eqref{eq:CLTE-free-energy} is used to compute the entropy associated with different spin configurations with the same classical ground-state energy $\mathcal{H}_0$. In this approximation, the state with the highest associated entropy is expected to be thermally favored at low temperatures. 
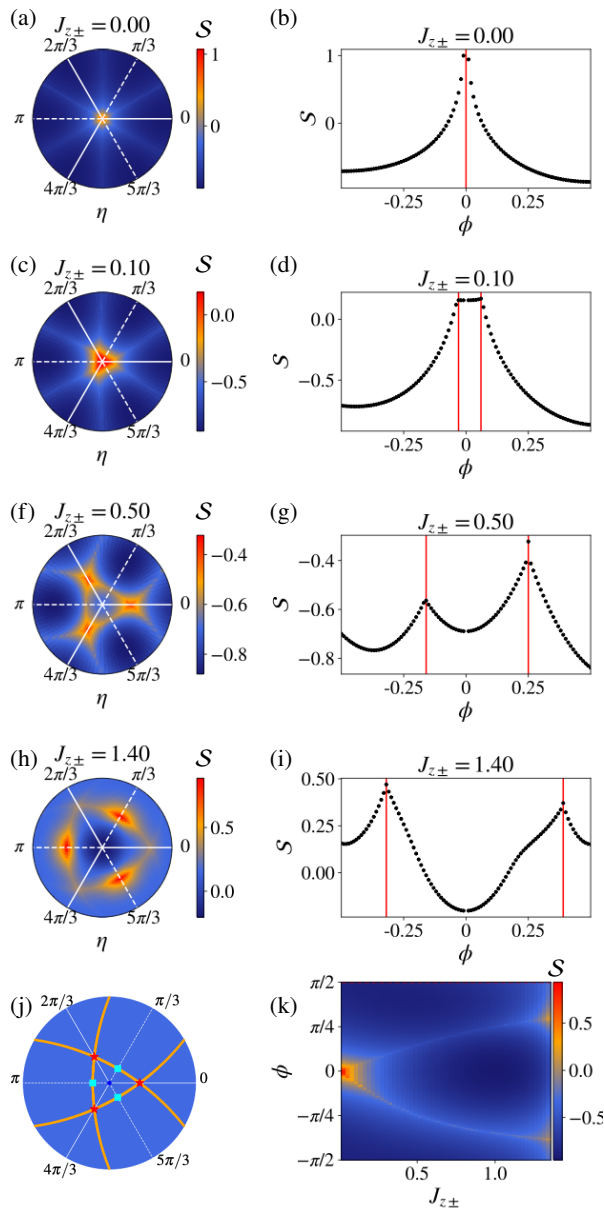
\begin{figure}
    \centering
    \begin{overpic}[width=0.95\columnwidth]{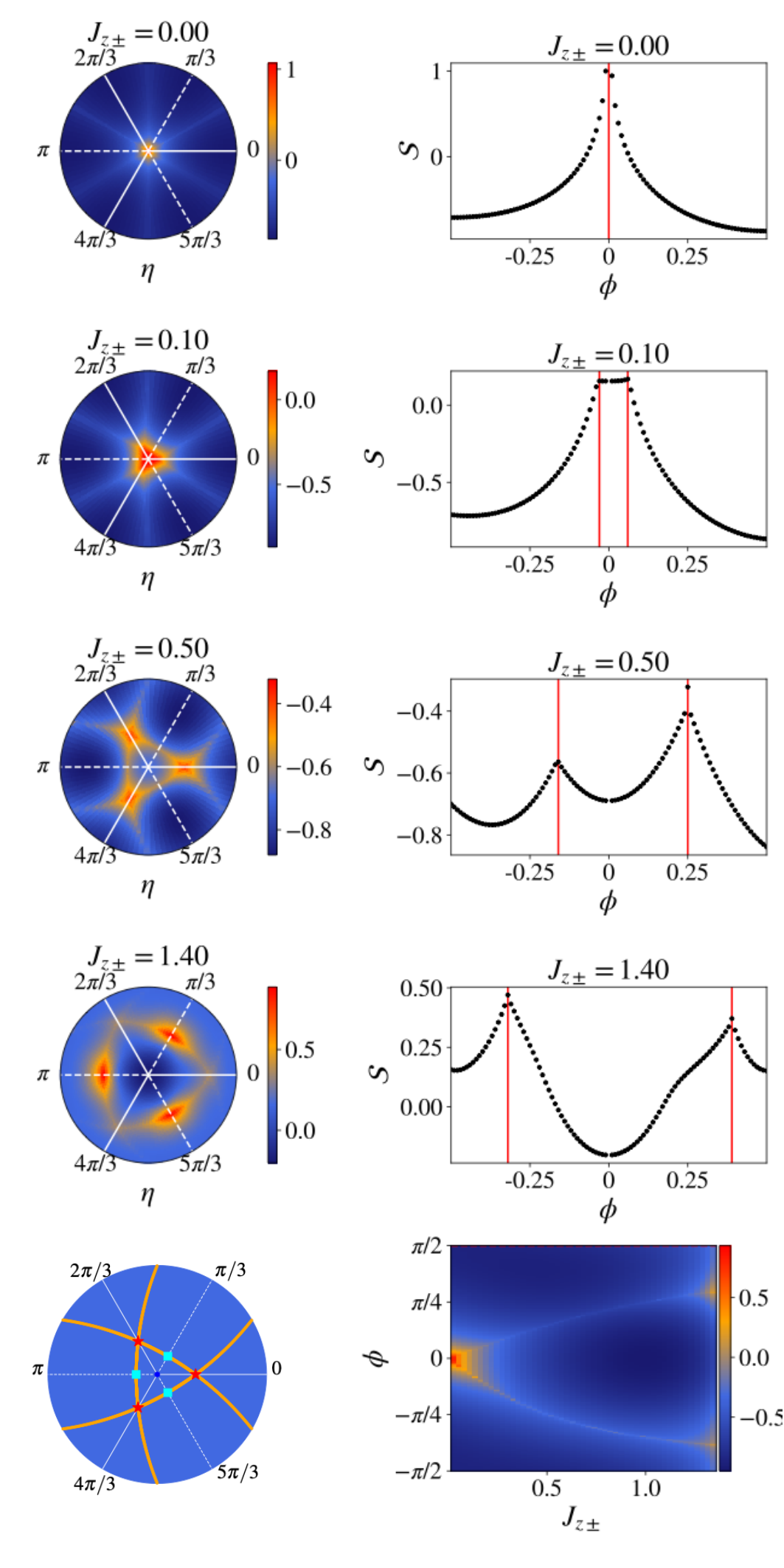}
    \put(2,98){(a)}
    \put(17,97){$\mathcal{S}$}
    \put(23,98){(b)}
    \put(2,78){(c)}
    \put(17,78){$\mathcal{S}$}
    \put(23,78){(d)}
    \put(2,58){(e)}
    \put(17,58){$\mathcal{S}$}
    \put(23,58){(f)}
    \put(2,38){(g)}
    \put(17,38){$\mathcal{S}$}
    \put(23,38){(h)}
    \put(2,18){(i)}
    \put(23,18){(j)}
    \put(45.5,21){$\mathcal{S}$}
    \end{overpic}
    \caption{Entropy $\mathcal S$ (modulo a $T$ and $J_{z\pm}$ dependent constant) on the $A_2\oplus E$ sphere for different values of $J_{z\pm}$ from classical low-temperature expansion. Panels (a), (c), (e), and (g) show the stereographic projection of the entropy, where we identify the north pole as the point where $A_2$ is maximal [cf.~Fig.~\ref{fig:U1-manifolds}]. The azimuthal angle $\eta$ parametrizes the $\Gamma_5$ manifold, and the polar angle $\phi$ parametrizes the mixing with the $A_2$ configuration. The white full (dashed) lines denote spin configurations that are linear combinations of the $A_2$ and $\psi_2$ ($-\psi_2$) states. Panels (b), (d), (f), and (h) illustrate the evolution of the entropy $\mathcal{S}$ for $\eta = 0$. Panel (i) illustrates the stereographic projection of the high-entropy paths defined by Eq.~\eqref{eq:A2E-U1-manifolds} and shown in Fig.~\ref{fig:U1-manifolds} in orange, taking $J_{z\pm}=0.4$ as a representative example.  The red dots (light blue squares) in panel (i) indicate the locations in the $A_2\oplus E$ manifold where the entropy is maximal for $J_{z\pm} <1.26$ ($J_{z\pm} > 1.26$). Panel (j) illustrates the evolution of the entropy along the $A_2\oplus \psi_2$ line, defined by $\eta=0$, as a function of the coupling $J_{z\pm}$. In panels (b), (d), (f), and (h), the vertical red lines mark the location of the two high-entropy branches.
}
\label{fig:CLTE_map}
\end{figure} 

\subsection{Entropy in the \texorpdfstring{$A_2\oplus E$}{A2+E} sphere as a function of \texorpdfstring{$J_{z\pm}$}{Jzpm}}
We study the entropy associated with the states in the $A_2\oplus E$ sphere via a classical low-temperature expansion, and study its dependence as a function of $J_{z\pm}$~\footnote{States with nonvanishing contribution of $T_{1-}$ irrep have fewer soft modes than any states on the $A_2\oplus E$ sphere. Thus, to quadratic order, the inclusion of $T_{1-}$ states does not introduce any additional fluctuation mechanism leading to its selection.}. Figure~\ref{fig:CLTE_map}(a)-(i) illustrates a stereographic projection of the entropy (modulo a $J_{z\pm}$ and $T$ dependent constant) for the states in the upper half of the $A_2\oplus E$ sphere and 
along a one-dimensional cut of the $A_2\oplus E$ sphere for $\eta=0$. Details on the stereographic projection are given in Appendix \ref{appendix:stereo-proj}.
Both the stereographic projections and the cuts are illustrated for four distinct values of $J_{z\pm}$. In this figure, the entropy obtained via the classical low-temperature expansion exposes three circular trajectories, where the entropy is higher, corresponding to the U(1) manifolds shown in Fig.~\ref{fig:U1-manifolds}. For $J_{z\pm}=0$, the spin configuration with the highest entropy is obtained for the $A_2$ configuration, corresponding to the point where the three U(1) manifolds intersect, see Fig.~\ref{fig:CLTE_map}(a) and (b).
Furthermore, we note that, at $J_{z\pm}=0$, the $\mathcal{H}_2(\mathbf{q})$ matrix in Eq.~\eqref{eq:CLTE-free-energy} presents two zero-energy flat bands.
This suggests one of two possible scenarios: the first being that there are two spin fluctuation modes about the $A_2$ configuration whose first $\mathbf{q}$-dependent contribution appears at higher-orders in $\mathbf{q}$. The second is that the two flat bands represent true zero modes which have no energy cost at all orders.
As $J_{z\pm}$ is varied, the U(1) manifolds drift away from the $A_2$ configuration. It is worth noting that, for $J_{z\pm}\neq 0$, the entropy reaches its maximum value at states where $\eta=2\pi n/3$ with $n\in\{0,1,2\}$, i.e., states for which the contribution of the $E$ irrep to the ground-state configuration is solely described by a $\psi_2$ configuration. The one-dimensional cuts along $\eta=0$ for distinct values of $J_{z\pm}$ reveal how the maximal entropy, and therefore minimal free energy, takes place at two possible values, which we label $\phi_1$ and $\phi_2$. The value of $\phi_1$ is obtained at points where the U(1) manifolds defined by Eq.~\eqref{eq:A2E-U1-manifolds} intersect. These intersections occur at six points in the $A_2\oplus E $ sphere defined by the constraints
\begin{equation}
    \label{eq:A2E-two-soft-modes}
    \begin{split}
        \tan\phi\tan\alpha_{\mathrm{mix}}=2\quad\mathrm{at}\quad \eta=0,\frac{2\pi}{3},\frac{4\pi}{3}, \\
    \tan\phi\tan\alpha_{\mathrm{mix}}=-2\quad\mathrm{at}\quad \eta=\frac{\pi}{3},\pi,\frac{5\pi}{3},
    \end{split}
\end{equation}
indicated by the red stars in Fig.~\ref{fig:CLTE_map}(i). On the other hand, for $J_{z\pm}$ slightly below the boundary $J_{z\pm}=\sqrt{2}$ of the $A_2\oplus E \oplus T_{1-}$ line, the six states with the highest entropy are now defined by the angle $\phi_2$ which is defined by the constraints
\begin{equation}
    \label{eq:A2E-second states}
    \begin{split}
        \tan\phi\tan\alpha_{\mathrm{mix}}=-1\quad\mathrm{at}\quad \eta=0,\frac{2\pi}{3},\frac{4\pi}{3} \\
    \tan\phi\tan\alpha_{\mathrm{mix}}=1\quad\mathrm{at}\quad \eta=\frac{\pi}{3},\pi,\frac{5\pi}{3} 
    \end{split}
\end{equation}
that also live on the U(1) manifolds defined by Eq.~\eqref{eq:A2E-U1-manifolds}, as marked by the light blue squares in Fig.~\ref{fig:CLTE_map}(i). A detailed study of the entropy as a function of $J_{z\pm}$ along the $\eta=0$ line identifies two high-entropy states which smoothly evolve from the $A_2$ state at $J_{z\pm}=0$ to two mixed states at nonzero $J_{z\pm}$, see the high-intensity lines in Fig.~\ref{fig:CLTE_map}(j). These states correspond to the  $J_{z\pm}$-dependent angles $\phi_1$ and $\phi_2$. 

According to our classical low-temperature expansion study of the entropy, the maximal-entropy configuration swaps from $(\eta,\phi)=(0,\phi_1)$ to $(\eta,\phi)=(0,\phi_2)$ at the value $J_{z\pm}^\star\simeq 1.26$. For values of $J_{z\pm}$ in the close proximity to $J_{z\pm}^\star$, the states defined by the combined triangle-shaped segments of the U(1) manifolds around the north pole feature an approximately equal entropy. We refer to Appendix~\ref{appendix:CLTE} for further discussion regarding the classical low-temperature expansion and the definition of $J_{z\pm}^\star$. 

Altogether, the entropy analysis of the distinct ground-state configurations in the $A_2\oplus E $ sphere identifies the spin configurations with high-entropy contribution as those which have more zero-fluctuation directions in the ground-state space, resulting in its selection at low temperatures. We note that a similar effect has been discussed in previous works, stemming from both classical~\cite{ObS_classical_PRB2024,NoculakPRB2024} and quantum~\cite{Canals_HDM_2008,Noculak_HDM_2023,ObS_1,ObS_2} phenomena. For the classical phenomenon, it was found that states lying at ``self-intersecting'' points of a ground-state manifold dominate the low-temperature physics and are therefore expected to be selected. This observation is consistent with our CLTE calculations and is further supported by our numerical simulations presented in Sec.~\ref{sec:cMC}. On the other hand, in the quantum case, two alternative routes for the selection were identified. In Ref.~\cite{Canals_HDM_2008,Noculak_HDM_2023}, a distinct pyrochlore system is considered, where the ground-state configuration space is parametrized by concentric intersecting circles on a sphere. In these works, the spin configurations at the intersection of these circles is said to be selected via a quantum order-by-disorder phenomenon in the low-temperature limit. On the other hand, Ref.~\cite{ObS_1,ObS_2} characterize a phenomenon dubbed order by singularity, where spin configurations lying at highly-connected points in the ground-state space (referred to as ``singularities'') are selected. In both scenarios, the selection either by quantum order by disorder or order by singularity is associated with the connectivity of the ground-state manifold, in analogy with the results presented for the classical system in this section. Given the similarities regarding the origin of the selection mechanism of the $A_2\oplus\psi_2$ state~\cite{ObS_classical_PRB2024} and those discussed in the quantum order-by-singularity phenomenon~\cite{ObS_1,ObS_2}, the selection phenomenon described in this section may be understood as a classical analogue of order by singularity.

\section{Landau theory}
\label{sec:Landau-theory}
In this section, we construct a minimal Landau theory and provide further insights on the possible $\mathbf{q}=0$ phases, the transitions between them, and the evolution of the order parameters $m_I$. Given that our Monte Carlo simulations identify contributions from only the $A_2$ and $E$ irreps to the $\mathbf{q}=0$ phases (see Sec.~\ref{sec:cMC}), we construct a Landau theory based on these two order parameters.
This approximation is further motivated by the fact that the only terms in the Landau free-energy theory which mix the $\mathbf{m}_{T_{1-}}$ order parameter with the remaining order parameters are fully invariant under the tetrahedral symmetry group $T_d$, and thus do not introduce additional symmetry-breaking effects.
Following Ref.~\cite{Javanparast2015}, we consider the local magnetic order parameters defined as
\begin{align}
m_{E}^x & =\frac{1}{4N}\sum_{i}S^x_i,  &
m_{E}^y & =\frac{1}{4N}\sum_{i}S^y_i, & 
m_{A_2} & =\frac{1}{4N}\sum_{i}S^z_i, 
\end{align}
where $m_E^x$, $m_E^y$, and $m_{A_2}$ measure the local $x$, $y$, and $z$ magnetic moment, respectively, and correspond to global order parameters associated with the components of the $E$ and $A_2$ irreps. 
It is useful to introduce the complex order parameter
\begin{align}
m_{xy} =|\mathbf{m}_E|e^{i\eta},
\end{align}
which measures the local $xy$ magnetic moment, where the angle $\eta$ indicates the direction of the magnetic moment in the local $xy$ planes, whereas the magnitude of the complex parameter agrees with that of the $E$ order parameter, i.e. $|m_{xy}|=|\mathbf{m}_E|$.
We construct the Landau theory by considering the action of the group $T_d$ over the local magnetic moments. It is worth noting that the $C_3$ symmetry around local $\langle111\rangle$ direction constrains the terms of the form $m_{xy}^k+(m_{xy}^\ast)^k=2|\mathbf{m}_E|^k\cos k\eta$ to those for which $k\in    3\mathbb{Z}$. For an extensive discussion of all  symmetry-allowed terms up to the sixth order, we refer the reader to Appendix~\ref{appendix:landau}.
Here, we consider the minimal Landau theory of the form
\begin{equation} \label{eq:landau}
    \begin{split}
        \mathcal{F}_\mathrm{minimal} &= \mathcal{F}_0 + r_0 |\mathbf{m}_E|^2 + r_1 m_{A_2}
        ^2 + r_2 |\mathbf{m}_E|^4 + r_{xyz}m_{A_2}^2|\mathbf{m}_E|^2\\
        &+ r_3 m_{A_2}^4 + r_4 |\mathbf{m}_E|^6  + \omega m_{A_2} |\mathbf{m}_E|^3 \cos{3\eta} \\
        &+ f_6|\mathbf{m}_E|^6\cos{6\eta}+\mathcal{F}[\mathbf{m}_{T_{1-}}] + \mathcal{O}[m_I^8]
    \end{split}
\end{equation}
which resembles the Landau theory provided by Ref.~\cite{Javanparast2015}, with the additional terms parameterized by $r_{xyz}$ and $r_3$. In the above, $\mathcal{F}[\mathbf{m}_{T_{1-}}]$ corresponds to terms in the Landau theory which possess a $T_{1-}$ dependence. In the following analysis, we set these to zero as this particular magnetic order is not observed in our Monte Carlo results discussed in Sec.~\ref{sec:cMC}~\footnote{We note that this simplification implies that this Landau theory can also be used to characterize models describing the phase boundary between the $A_2$ and $E$ irreps.}.
For a given set of parameters $\{r_0,r_1,r_2,r_{xyz},r_3,\omega,f_6\}$, we find the solution with the lowest free energy  numerically. We note that this theory explicitly forbids a coupling between the $\psi_3$ and the $A_2$ orders, as expected by symmetry and previously noted in Ref.~\cite{Javanparast2015}.

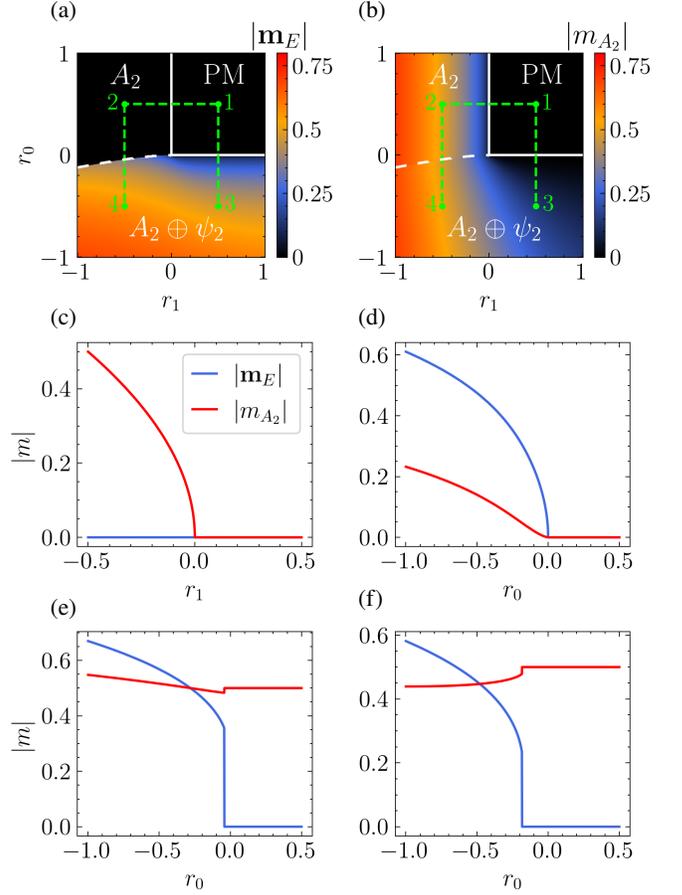
\begin{figure}
    \centering
    \begin{overpic}[width=\columnwidth]{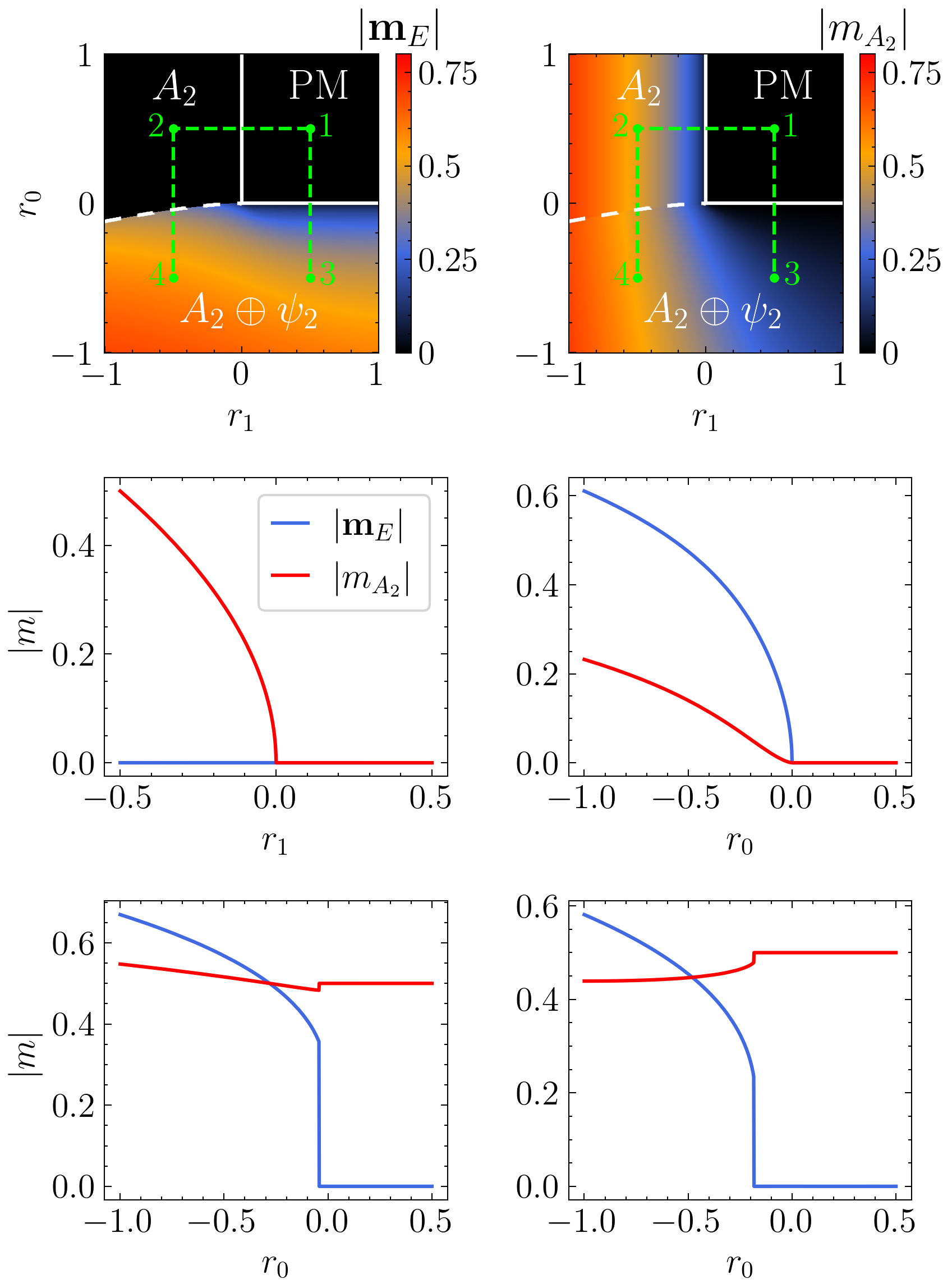}
    \put(5,100){(a)}
    \put(40,100){(b)}
    \put(5,65){(c)}
    \put(40,65){(d)}
    \put(5,32){(e)}
    \put(40,33){(f)}
    \end{overpic}
\caption{%
Mean-field phase diagram as a function of $r_1$ and $r_0$ for fixed $(r_2,r_{xyz},r_3,r_4,\omega,f_6)=(1,1,1,1,2,0.5)$ from Landau theory, with the color scale indicating
(a)~the $E$ order parameter $|\mathbf{m}_E|$ and (b)~the $A_2$ order parameter $|m_{A_2}|$.
Panels (c)-(e) illustrate the evolution of the order parameters $|\mathbf{m}_E|$ (blue curve) and $|m_{A_2}|$ (red curve) along the different one-dimensional cuts marked in green in panels (a) and (b): Panel (c) shows the order parameters along the path $1\rightarrow2$, panel (d) along the path $1\rightarrow 3$, and panel (e) along the path $2\rightarrow 4$.
Panel (f) shows the order parameters also along the path $2 \rightarrow 4$, but for a smaller value $\omega = 1$ of the coupling between $m_{A_2}$ and $m_E$.}
\label{fig:LG}
\end{figure}

Figure~\ref{fig:LG} shows the mean-field phase diagram and the evolution of the magnetic order parameters as a function of the tuning parameters $r_0$ and $r_1$, for fixed values of the remaining parameters.
From the evolution of the order parameters, we identify three distinct phases; a paramagnetic phase, where $|m_{A_2}|=0$ and $|\mathbf{m}_E|=0$, an $A_2$ phase, where  $|m_{A_2}|\neq0$ and $|\mathbf{m}_E|=0$, and a mixed $A_2\oplus \psi_2$ phase, where $|m_{A_2}|\neq0$ and $|\mathbf{m}_E|\neq0$. 
Importantly, we find that $\eta=\frac{n\pi}{3}$ with $n\in\{0,1,2,3,4,5\}$ in the mixed phase, corresponding to an $A_2 \oplus \psi_2$ state. 

The color plots shown in Figs.~\ref{fig:LG}(a) and (b) suggest continuous phase transitions from the paramagnetic phase to the ordered phases $A_2$ and $A_2\oplus\psi_2$, and first-order transition between the $A_2$ and the $A_2\oplus\psi_2$ phases.
To more clearly expose the nature of the transitions obtained within Landau theory, we illustrate in Fig.~\ref{fig:LG}(c)-(f) line cuts illustrating the evolution of the order parameters as a phase boundary is crossed. These line cuts are marked in Figs.~\ref{fig:LG}(a) and (b) by dashed green paths. Figure~\ref{fig:LG}(c) illustrates the evolution of the order parameters in the line cut between points marked 1 and 2 in Figure~\ref{fig:LG}(a) and (b) in the paramagnetic and $A_2$ phases, respectively. Along this line, the $m_{A_2}$ order parameter shows a square-root-like behavior, while $|\mathbf{m}_{E}|=0$. 
On the other hand, for the line cut between points 1 and 3, corresponding to points in the paramagnetic and $A_2\oplus\psi_2$ phase, respectively, both order parameters become nonzero as soon as the phase boundary is crossed. It is worth noting that the $A_2$ order parameter remains small compared to the $E$ order parameter in the vicinity of the phase boundary and for larger values of $r_1$, see Fig.~\ref{fig:LG}(b). The continuous evolution of the parameter across these two phase boundaries indicates a continuous phase transition between the paramagnetic phase and the ordered phases. 
Lastly, Figs.~\ref{fig:LG}(e) and (f) illustrate the evolution of the order parameters for the line cut between points 2 and 4 corresponding to points in the $A_2$ phase and $A_2\oplus\psi_2$ phase for two distinct sets of $(r_2,r_{xyz},r_3,r_4,\omega,f_6)$ parameters. In both cases, we observe how the $E$ order parameter presents a large finite jump when crossing into the $A_2\oplus\psi_2$ phase, while the $A_2$ order parameter presents a small jump. The size of the small discontinuous jump in the $A_2$ parameter and its evolution in the $A_2\oplus\psi_2$ phase are parameter dependent, as seen when comparing Figs.~\ref{fig:LG}(e) and (f). The discontinuous nature of the order parameters, which is more drastically evidenced in the $E$ order parameter, indicates that these two phases are separated by a first-order phase transition.

We note that the resulting Landau theory discussed in this section also describes models whose interaction parameters fall in the $A_2\oplus E$ phase boundary. We return to this point in Section~\ref{sec:beyond_triple}.
%

\section{Classical Monte Carlo simulations}
\label{sec:cMC}
In this section, we present numerical Monte Carlo simulations used to obtain the finite-temperature phase diagram shown in Fig.~\ref{fig:phase-diagram}. 
A wide range of temperatures is considered for several values of $J_{z\pm}$ on the $A_2\oplus E \oplus T_{1-}$ line. 
Our simulations expose a rich phase diagram with a variety of symmetry-breaking phases being realized. In particular, we find two $\mathbf{q}=0$ long-range ordered phases, namely the $A_2$ and the mixed $A_2 \oplus \psi_2$ phases, which are defined at the level of a single tetrahedron. We reiterate that $\psi_2$ corresponds to the $x$ component of the $E$ irrep. 
Furthermore, we find a spin nematic phase that is stabilized at low and intermediate temperatures around the point $J_{z\pm}=\pm 1/\sqrt{2}$, where the Hamiltonian in Eq.~\eqref{eq:general_bilinear_hamiltonian} acquires additional subsystem symmetries. At $J_{z\pm}=\pm 1/\sqrt{2}$, these additional symmetries prevent any type of conventional magnetic order, resulting in the stabilization of the spin nematic phase down to the lowest temperatures. The spin nematic phase is characterized by broken cubic spin rotational symmetry, but unbroken time reversal symmetry. 

The section is organized as follows. First, we introduce algorithmic details and the measured observables in Secs.~\ref{subsec:algorithmic-detail} and \ref{subsec:observables}, respectively. Our results for the $\mathbf{q}=0$ phases are shown in Sec.~\ref{subsec:q=0-phases}. We also compare the Monte Carlo results with the predictions from classical low-temperature expansion and Landau theory. The spin nematic phase stemming from the zero temperature $J_{z\pm}=1/\sqrt{2}$ point deserves its own discussion as it presents several additional features. We postpone its discussion to Sec.~\ref{sec:SN-phase}.

\subsection{Algorithmic details}
\label{subsec:algorithmic-detail}
We study the Hamiltonian in Eq.~\eqref{eq:general_bilinear_hamiltonian} using large-scale classical Monte Carlo simulations. We employ Metropolis \cite{metropolis1953} and over-relaxation algorithms \cite{creutz1987} combined  with parallel tempering exchange \cite{marinari92,hukushima96,hukushima99}. The spins are treated as classical vectors of unit length $\mathbf{S}_i^2=1$.
A single Metropolis step is performed locally, with a trial spin sampled randomly from a cone around the original spin direction, using a solid angle that depends on the temperature. The solid angle is initially chosen from a feedback loop, to ensure an acceptance rate between 0.2 and 0.5 throughout the temperature range.
Each lattice sweep of a Metropolis update is combined with 20 over-relaxation steps followed by a replica exchange trial of the configurations at adjacent temperatures. To minimize autocorrelations, the procedure is repeated 20 times between successive measurements.
Note that the initial condition affects thermalization. We have tested two different starting configurations: a random configuration, in which each spin is sampled independently, and a biased configuration corresponding to all-in-all-out $A_2$ long-range order. The random initial condition exhibits a longer thermalization time and, in some cases, fails to thermalize within the finite number of Monte Carlo samples. In all cases where thermalization is achieved, the results are consistent with those obtained from the biased initialization. We therefore use the more stable biased starting condition for the results presented in this work.
Errors on the observables are computed with the jackknife method combined with blocking techniques~\cite{Landau_Binder_2014}.
We consider pyrochlore lattices with $L^3$ tetrahedra and $N=4L^3$ sites, using $L=6, 8, 10, 12$ with periodic boundary conditions.
The accumulated statistics are of the order $\mathcal{O}(10^5)$ for each point in parameter space. The simulations are carried out on the high-performance computer Barnard at the NHR Center of TU Dresden~\cite{nhr-alliance}.
%

\subsection{Observables}
\label{subsec:observables}
We measure the following thermodynamic observables in the simulations:

\paragraph{Specific heat.} From energy fluctuations, we measure the specific heat as
\begin{equation}
    \label{eq:specific_heat-MC}
    C \coloneqq \frac{1}{NT^2}({\langle \mathcal{H}^2 \rangle - \langle \mathcal{H}\rangle^2}),
\end{equation}
with $T$ the absolute temperature and $N$ the total number of sites. 

\paragraph{Magnetic order parameters.} Magnetic order parameters are constructed on the single-tetrahedron irreps $\mathbf{m}_{I}$, $I = A_2, E, T_2, T_{1\pm}$ as
\begin{equation}
    \label{eq:irreps_measures}
   \langle m_I\rangle = \left\langle \left| \frac{1}{2SL^3} \sum_{\boxtimes} \mathbf{m}_I^\boxtimes \right| \right\rangle,
\end{equation}
where $\sum_{\boxtimes}$ denotes the sum over all tetrahedra on the lattice and $\mathbf{m}_I^\boxtimes$ corresponds to the $I$-th irrep mode in the $\boxtimes$ tetrahedron. The single-tetrahedron irreps modes are expressed in terms of the spin components, as reported in the Table~\ref{tab:irreps-def} of Appendix~\ref{appendix:irreps-basis}. The observables in Eq.~\eqref{eq:irreps_measures} are normalized such that $\sum_I \langle m_I \rangle^2 = 1$ in the low-temperature limit. As a consequence, if an ordered state associated with a single irrep $I$ is realized at low temperatures, then we obtain $\langle m_I \rangle =1$ and $\langle m_J \rangle =0$ for $J\neq I$. However, in a mixed-irreps phase, the respective order parameters acquire a value between 0 and 1, such that the normalization condition $\sum_I \langle m_I \rangle^2 = 1$ is respected.
\paragraph{Sublattice magnetic order parameter.}
The sublattice magnetic order parameter provides a measure of whether any form of magnetic long-range order develops on the individual sublattices and is defined as
\begin{equation}
    \label{eq:sublattice-OP}
    \langle m^2_{\mathrm{sub}} \rangle =\left\langle \frac{1}{4}\sum_\mu \left|\frac{1}{2SL^3}\sum_{\boxtimes} \mathbf{S}_{\boxtimes,\mu}\right|^2 \right\rangle\,,
\end{equation}
where $\mu$ denotes the sublattice index in the tetrahedron $\boxtimes$, and $\mathbf{S}_{\boxtimes,\mu}$ is the corresponding spin vector. 
\paragraph{Magnetic susceptibilities.}
We also measure the susceptibilities of the magnetic order parameters,
\begin{align}
\chi_I = \frac{N}{T} (\langle m_I^2 \rangle - \langle m_I \rangle^2), \qquad I = A_2, E, T_2, T_{1\pm}.
\end{align}

\paragraph{Binder cumulants.}
To extract the location of the transition points from the finite-size simulations and to analyze the natures of the transition, it is also useful to measure renormalization-group-invariant Binder cumulants defined as
\begin{align}
U_I = \frac{\langle m_I^4 \rangle}{\langle m_I^2 \rangle^2}, \qquad I = A_2, E, T_2, T_{1\pm}.
\end{align}

\paragraph{$E$ irrep order parameter.} 
In order to distinguish between the $\psi_2$ and $\psi_3$ components of the $E$ irrep in the case of $E$ long-range order, we also measure the so-called $\langle m_{E6} \rangle$ parameter, defined as
\begin{equation}
\label{eq:e6}
    \langle m_{E6} \rangle =\left\langle m_E \cos6\eta  \right\rangle,
\end{equation}
where $\eta$ quantifies the angle of the spin configurations in the $E$ manifold~\cite{Andrade_xy_glass_PhysRevLett.120.097204,Noculak_HDM_2023,Zhitomirsky_2014}, see Eq.~\eqref{eq:angle-representation}. It reduces to $\eta=n\pi/3$ if the $\psi_2$ component is realized, while it assumes the values $\eta=(2n+1)\pi/6$ when the $\psi_3$ component is achieved, with $n\in \mathbb{Z}$ in both cases. Consequently, the $\langle m_{E6} \rangle$ parameter acquires a positive (negative) value when the spin configurations in the low-temperature phase are described by a $\psi_2$ ($\psi_3$) configuration.

\paragraph{Spin-nematic order parameter.} The single-site spin-nematic order parameter is a five-component vector \cite{Nematic_order_Shanon2006PRL,Nematic_order_Shanon2015PRB,janssen15,Taillefumier2017_xxz,francini24,hecker_24} obtained by combining the five symmetric traceless $3\times3$ matrices $\Lambda^{(\alpha)}$,
\begin{equation}
\begin{split}
    \Lambda^{(\alpha)}= \Biggl\{ \frac{1}{\sqrt{3}}
    &\begin{pmatrix}
        -1 & 0 & 0 \\
        0 & -1 & 0 \\
        0 & 0 & 2
    \end{pmatrix},
    \begin{pmatrix}
        -1 & 0 & 0 \\
        0 & 1 & 0 \\
        0 & 0 & 0
    \end{pmatrix},
    \begin{pmatrix}
        0 & 0 & 0 \\
        0 & 0 & 1 \\
        0 & 1 & 0
    \end{pmatrix}, \\
    &\begin{pmatrix}
        0 & 0 & 1 \\
        0 & 0 & 0 \\
        1 & 0 & 0
    \end{pmatrix},
    \begin{pmatrix}
        0 & 1 & 0 \\
        1 & 0 & 0 \\
        0 & 0 & 0
    \end{pmatrix}
    \Biggl\},
\end{split}
    \label{eq:symmetric_traceless_matrices}
\end{equation}
with the traceless symmetric tensor $\mathsf{Q}^{\mu\nu}_i=S^\mu_i S^\nu_i - \frac{1}{3}\delta^{\mu\nu}$, as
\begin{equation}
    \label{eq:single_site_spin_nematic_vector}
    d^{\alpha}_i = \frac{\sqrt{3}}{2}\operatorname{Tr}[\mathsf{Q}_i\Lambda^{(\alpha)}]\,,
\end{equation}
where the normalization factor is fixed by demanding a unitary absolute value of $\mathbf{d}_i$ if any a nematic state is realized.
Explicitly, the five components~\footnote{In the global basis, the nematic-order-parameter components decompose into a doublet and a triplet under the symmetry group $T_d$. However, our use of a local definition of the nematic parameter leads to more involved transformation rules; see Appendix~\ref{appendix:nem-OP-transformation} for details.} of $\mathbf{d}_i$ read
\begin{equation}
    \begin{split}
        d^1_i&=\frac{1}{2}\left[2(S_i^z)^2-(S_i^x)^2-(S_i^y)^2\right]\,,\\
        d^2_i&=-\frac{\sqrt{3}}{2}\left[(S_i^x)^2-(S_i^y)^2\right]\,,\\
        d_i^3&=\sqrt{3}S_i^xS_i^y\,,\quad
        d_i^4=\sqrt{3}S_i^xS_i^z\,, \quad
        d_i^5=\sqrt{3}S_i^yS_i^z\,.
    \end{split}
\end{equation}
Further details on the single-site nematic order parameter in the global and local bases are given in Appendix~\ref{appendix:nem-OP-transformation}.
In the Monte Carlo simulations, we measure the long-range spin-nematic order parameter as 
\begin{equation}
\label{eq:long_range_spin_nematic_OP}
     \langle m_\mathrm{nem}\rangle =\left< \left| \frac{1}{4S^2L^3} \sum_{i} \mathbf{d}_i \right| \right>\,.
\end{equation}
By construction, this order parameter is also nonzero in the $\mathbf{q}=0$ ordered phases we observe~\footnote{For example, for an $A_2$ state, we obtain $\mathbf{d}=(1,0,0,0,0)$.}.
However, in the regions where the ordinary long-range irreps order parameters in Eq.~\eqref{eq:irreps_measures} vanish, the quantity in Eq.~\eqref{eq:long_range_spin_nematic_OP} can still be finite, signaling a spin-nematic phase with unbroken time-reversal symmetry and broken spin rotational symmetry.

\paragraph{Equal-time spin structure factor.} The equal-time spin structure factor is defined as
\begin{equation}
    \label{eq:spin_sructure_factor}
    S({\mathbf{q}}) = \frac{1}{N} \sum_{i,j} \langle \mathbf{S}_i \cdot \mathbf{S}_j \rangle \rme^{- \rmi\mathbf{q}\cdot(\mathbf{R}_i - \mathbf{R}_j)},
\end{equation}
where $\mathbf{R}_i$ is the position vector of the site $i$ on the lattice. 
%

\subsection{Magnetic q = 0 phases}
\label{subsec:q=0-phases}
In the following, we present the numerical results from Monte Carlo simulations. First, the non-Kramers case is covered, then the finite $J_{z\pm}$ Kramers results are reported. Details on the finite-size-scaling procedure used to analyze continuous phase transitions are given in Appendix~\ref{appendix:FSS-continuous-transition}. 

\subsubsection{Non-Kramers case \texorpdfstring{$J_{z\pm}=0$}{J=0}}
\begin{figure}
    \centering
    \begin{overpic}[width=\columnwidth]{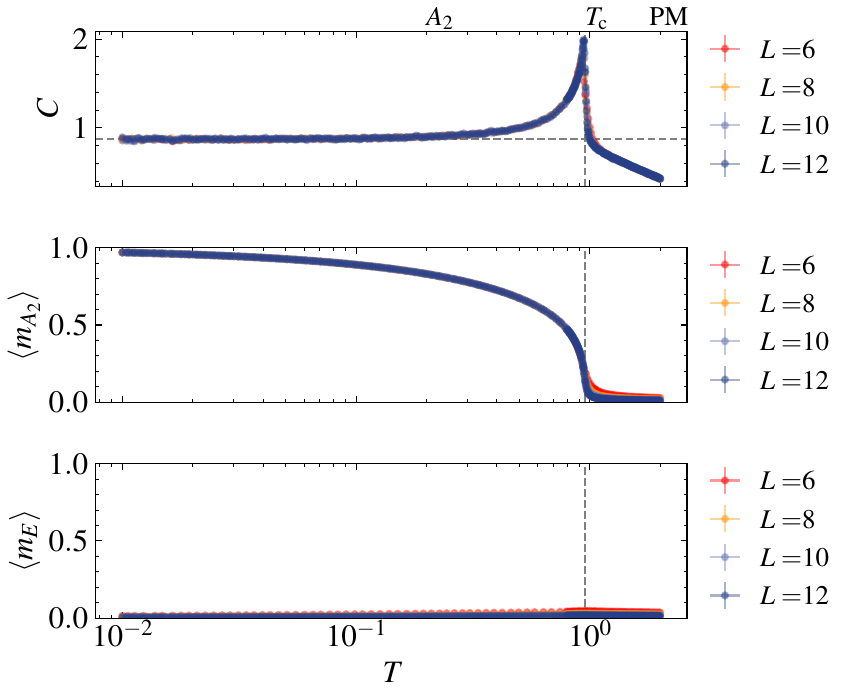}
    \put(0,79){(a)}
    \put(0,53){(b)}
    \put(0,27){(c)}
    \end{overpic}
    \caption{Monte Carlo results in the non-Kramers case $J_{z\pm} = 0$. (a)~Specific heat, (b)~${A_2}$ order parameter, and (c)~$E$ order parameter as functions of the temperature $T$ for different lattice sizes $L$. The continuous transition at $T_{\mathrm{c}}$, indicated by the vertical dashed line, separates the paramagnetic phase at high temperatures from the $A_2$ long-range-ordered phase at low temperatures. In (a), the horizontal dashed line at $C = 7/8$ highlights the low-temperature limit of the specific heat, as expected from the equipartition theorem with $n_2=6$ quadratic modes and $n_4=2$ quartic modes.}
    \label{fig:jzpm-0-cMC}
\end{figure}

At $J_{z\pm}=0$, the Hamiltonian in Eq.~\eqref{eq:general_bilinear_hamiltonian} describes a non-Kramers spin system, where the local $z$ components of pseudospins transform as magnetic dipoles and the local $xy$ components transform as magnetic quadrupoles~\cite{Rau2018FrustratedQR,Petit2016PRBPrZr}. In this case, the $T_{1-}$ irrep mode corresponds to a splayed ferromagnet where the spins lie on the local $xy$ planes of each sublattice, see Appendix~\ref{appendix:irreps-basis} for details. Consequently, both the $E$ and the $T_{1-}$ irreps are regarded as local $xy$ irreps, with the spin configurations for these irreps lying on the local $xy$ planes, while the $A_2$ irrep is a local $z$ irrep. 
For $J_{z\pm} = 0$, the Monte Carlo data indicate a phase transition from a paramagnetic phase into an $A_2$ long-range-ordered phase at $T_\mathrm{c}=0.957(1)$. The $A_2$ long-range order extends down to the lowest temperatures we considered and agrees with the predictions from the classical low-temperature expansion discussed in Sec.~\ref{sec:CLTE}, see Fig.~\ref{fig:jzpm-0-cMC}. 
The $A_2$ irreps transform as a $\mathbb{Z}_2$ object, so the critical behavior is expected to fall in the three-dimensional Ising universality class, at least asymptotically, if the transition is continuous. Finite-size scaling allows us to extract the exponent $\nu=0.635(8)$ and $\gamma=1.205(15)$, consistent with the three-dimensional Ising universality class~\cite{pelissetto02,chang2024}. Details on the scaling analysis are provided in Appendix~\ref{appendix:FSS-continuous-transition}.

As the temperature approaches zero, the specific heat does not reach a unitary value, as expected from the equipartition theorem for purely quadratic fluctuations, implying that the low-energy fluctuations about the ground-state configuration are of higher order~\cite{ChalkerHoldsworthKagomePhysRevLett.68.855,lozano_2024arxiv,francini25}. Indeed, as we already briefly mentioned in Sec.~\ref{sec:CLTE}, the classical low-temperature expansion finds that the matrix $\mathcal{H}_2(\mathbf{q})$ has two zero-energy flat bands in momentum space, meaning that two out of the eight fluctuation modes are not captured by the quadratic approximation, therefore identifying two of these fluctuation modes as quartic or higher modes. In the case the fluctuations modes are only quadratic and quartic in nature, the specific heat in the low-temperature limit is expected to asymptotically approach the value
\begin{equation}
\label{eq:spec-heat-equipartition}
\frac{C}{k_{\rm B}}=\frac{1}{4}\left(\frac{n_2}{2}+\frac{n_4}{4}\right)
\end{equation}
where $n_2$ is the number of quadratic modes and $n_4$ is the number of quartic modes, and $n_2+n_4=8$ is the total number of fluctuation modes. Under the assumption that the two zero modes obtained in the second-order classical low-temperature expansion evolve into quartic modes when higher-order fluctuations are considered, corresponding to $n_2 = 6$ and $n_4 = 2$, the predicted low-temperature specific heat is $\frac{C}{k_\mathrm{B}} = 7/8$.
As illustrated in Fig.~\ref{fig:jzpm-0-cMC}(a), this value is asymptotically approached by the specific heat in the low-temperature regime, validating the quartic-modes hypothesis.

\subsubsection{Kramers case \texorpdfstring{$J_{z\pm}\neq0$}{J!=0}}
In contrast to the non-Kramers system, all the spin components in a Kramers system transform as magnetic dipoles, allowing for a nonvanishing $J_{z\pm}$ coupling. In this case, the classical low-temperature expansion identifies mixed $A_2 \oplus \psi_2$ states as the maximal entropy spin configurations, see Fig.~\ref{fig:CLTE_map}(j). 
Our Monte Carlo simulations confirm that the low-temperature state along the $A_2\oplus E\oplus T_{1-}$ line realizes a mixed $A_2\oplus\psi_2$ phase, where the order parameters associated to these two magnetic orders acquire finite values in the low-temperature regime, with the exception of the immediate vicinity of the point $J_{z\pm}=1/\sqrt{2}$. 
Our low-temperature Monte Carlo simulations in fact show a remarkable quantitative agreement with the classical low-temperature expansion predictions at low temperatures (see Sec.~\ref{sec:CLTE}). The qualitative features associated with the transitions and order parameter evolutions are well captured by the Landau theory (see Sec.~\ref{sec:Landau-theory}). A detailed comparison is provided at the end of this section. 
For the low-temperature regime, it is crucial to highlight that the predictions of the classical low-temperature expansion are only valid in this low-temperature region of the phase diagram. Indeed, and as elucidated in recent works~\cite{schick_FCC_2020,NoculakPRB2024,Andrade_xy_glass_PhysRevLett.120.097204}, the state selected at high temperatures can differ from those selected at intermediate and low temperatures. Indeed, our Monte Carlo simulations reveal a more intricate phase diagram, characterized by a cascade of phase transitions between the $\mathbf{q}=0$ phases and the nematic phase, which are not exposed by the classical low-temperature expansion. Ultimately, the $A_2 \oplus \psi_2$ phase is approached in different ways as temperature is lowered, depending on the value of $J_{z\pm}$.
It is worth noting that, in our Monte Carlo simulations, the long-range $A_2\oplus E$ order is well captured by the long-range order parameters in Eq.~\eqref{eq:irreps_measures}. These parameters, however, are not able to differentiate between $\psi_2$ and $\psi_3$ components of the $E$ irreps. 
For this purpose, we have also studied the evolution of the so-called $\langle m_{E6} \rangle$ parameter, see Eq.~\eqref{eq:e6}.
In what follows, we discuss the results obtained for the $\mathbf{q}=0$ phases in two parameter regimes, one for a small coupling  $J_{z\pm}\lesssim 1/\sqrt{2}$ and one for a large coupling  $J_{z\pm}\gtrsim 1/\sqrt{2}$.

\paragraph{Small coupling $J_{z\pm} \lesssim 0.5$.}
For sufficiently small $J_{z\pm}$, the low-temperature phase is reached through two phase transitions located at $T_{\mathrm{c}1}$ and $T_{\mathrm{c}2}$, with $T_{\mathrm{c}1}<T_{\mathrm{c}2}$, see Fig.~\ref{fig:phase-diagram}. 
The higher-temperature transition at $T_{\mathrm{c}2}$ is continuous and takes the system from the paramagnetic phase to an $A_2$ ordered phase.
The lower-temperature transition is discontinuous and takes the system from the $A_2$ order phase to the mixed $A_2 \oplus \psi_2$ phase.

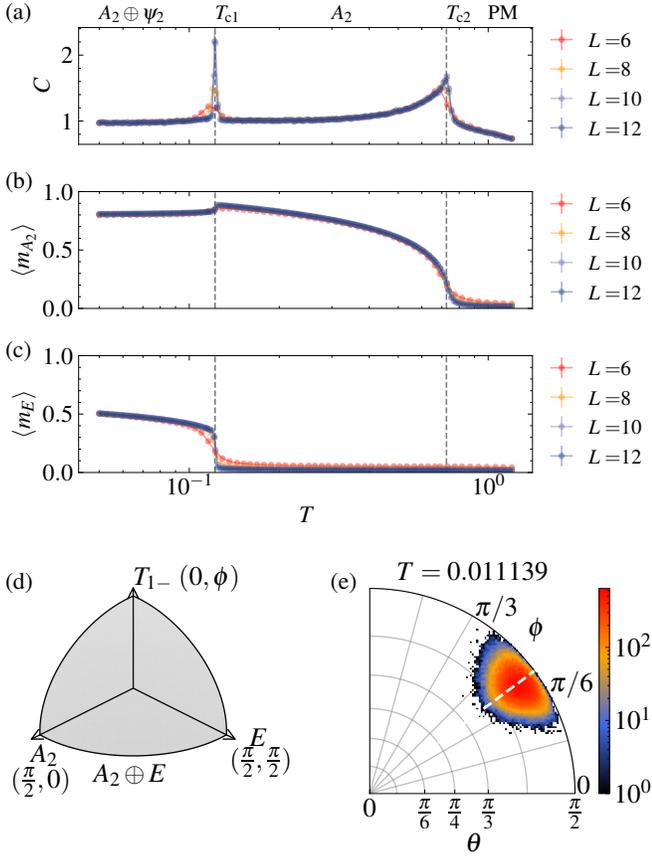
\begin{figure}
    \centering
    \begin{overpic}[width=\columnwidth]{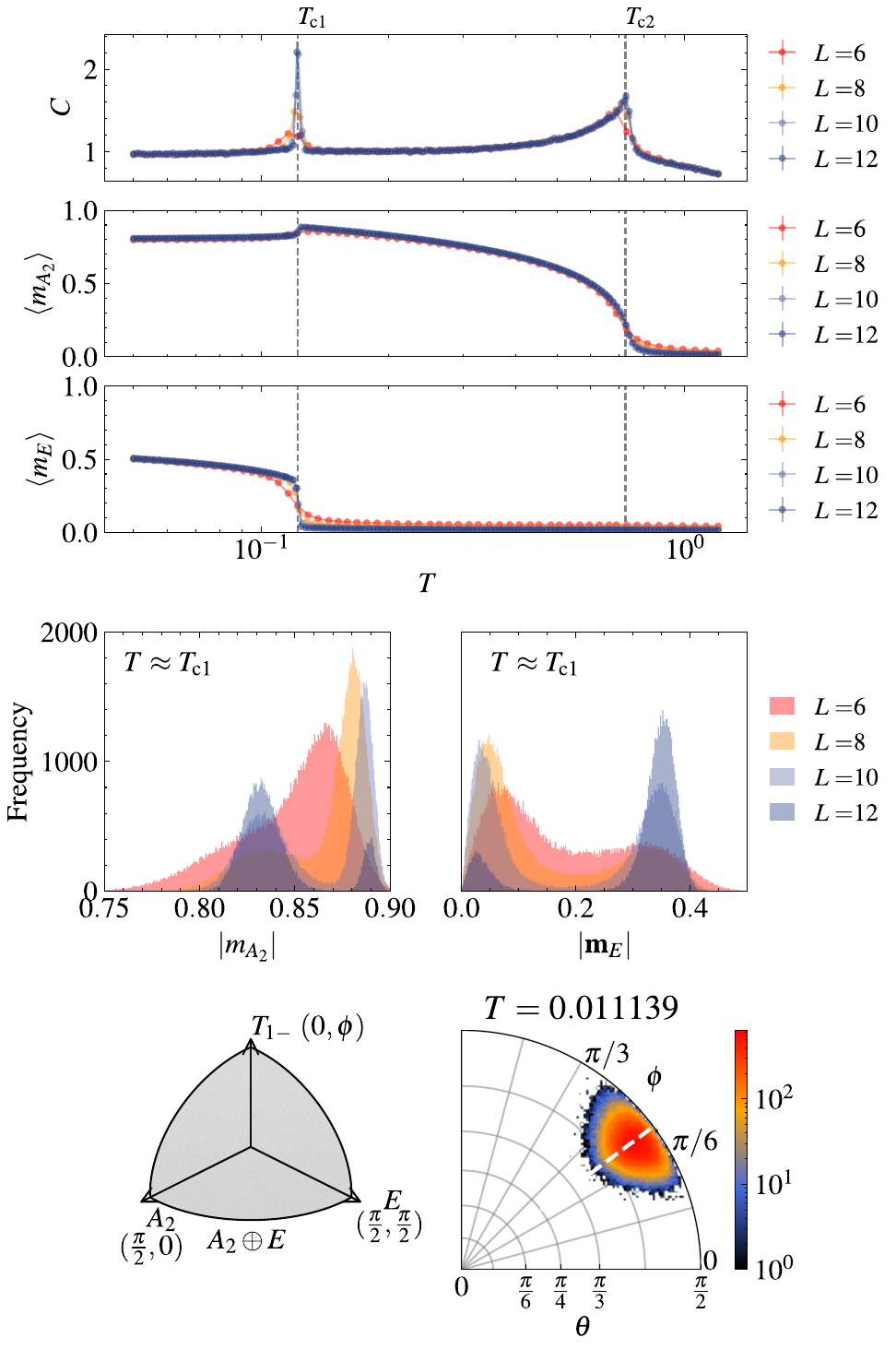}
    \put(-2,96){(a)}
    \put(-2,83){(b)}
    \put(-2,70){(c)}
    \put(-2,52){(d)}
    \put(31,52){(e)}
    \put(-2,24){(f)}
    \put(31,24){(g)}
    \put(10,98.5){$A_2\oplus\psi_2$}
    \put(33,98.5){$A_2$}
    \put(50,98.5){PM}
    \end{overpic}
\caption{ Monte Carlo results in the Kramers case for $J_{z\pm} = 0.4$. (a)~Specific heat, (b)~$A_2$ order parameter, and (c)~$E$ order parameter as functions of the temperature $T$ for different lattice sizes $L$. The two dashed lines indicate the positions of the phase transitions. The continuous transition at $T_{\mathrm{c}2}\approx0.752$ separates the paramagnetic phase at high temperatures from the $A_2$ long-range order phase at intermediate temperatures. The first-order transition at $T_{\mathrm{c}1}\approx0.122$ separates the $A_2$ phase at intermediate temperatures from the mixed $A_2\oplus \psi_2$ phase at low temperatures.
(d)~Histogram histogram of the $A_2$-order-parameter magnitude at the critical temperature $T_{\mathrm{c}1}\approx0.122$.
(e)~Same as (d), but for the $E$-order-parameter magnitude.
(f)~Sketch of the $A_2\oplus E\oplus T_{1-}$ first octant, generated by the mixing angles $\theta$ and $\phi$ in Eq.~\eqref{eq:irreps mixing}. 
(g)~Irrep-magnitude distribution on the $A_2\oplus E\oplus T_{1-}$ first octant, stereographically projected, measured on many spin configurations obtained from Monte Carlo simulations at a low temperature. The value of the mixing angle $\phi$ at the center of the distribution agrees well with the prediction from classical low-temperature expansion (white dashed line).}
    \label{fig:jzpm-0.4-cMC}
\end{figure}

Figure~\ref{fig:jzpm-0.4-cMC} illustrates a representative example of the specific heat, order parameters, and irrep-magnitude distribution for $J_{z\pm}=0.4$. 
The specific heat depicted in Fig.~\ref{fig:jzpm-0.4-cMC}(a) develops two sharp features at $T_{\mathrm{c1}}=0.122(2)$ and $T_{\mathrm{c2}}=0.752(2)$, signaling the phase transitions. We note that the peak at $T_{\mathrm{c2}}$ slowly increases with system size, indicating a continuous transition with a small $\alpha/\nu$ critical-exponent ratio. 
In this temperature regime, the $m_{A_2}$ order parameter varies smoothly, see Fig.~\ref{fig:jzpm-0.4-cMC}(b) with a diverging slope at the transition, similar to the behavior seen in the Landau theory.
Finite-size scaling of the respective Binder parameter around $T_{\mathrm{c}2}$ yields a value of the correlation-length exponent $\nu=0.62(3)$, in well agreement with the theoretical value $\nu = 0.62997$ for the three-dimensional $\mathbb{Z}_2$ universality class. However, the finite-size scaling on the susceptibility $\chi_{A_2}$ yields a ratio $\gamma/\nu=1.66(4)$ which is quite far from the expected value $\gamma/\nu = 1.9637$ for the three-dimensional $\mathbb{Z}_2$ universality class~\cite{pelissetto02,chang2024}. Further details on the scaling analysis are given in Appendix~\ref{appendix:FSS-continuous-transition}.
This disagreement may be caused by a slow renormalization group flow in our microscopic model on the $A_2 \oplus E \oplus T_{1-}$ line, leading to sizable finite-size corrections to the asymptotically expected critical behavior~\cite{chern10,pelissetto02}.
Although possibly slightly more pronounced in our case, similar deviations from the respective expected universality were also observed in the context of Heisenberg-Dzyaloshinskii-Moriya pyrochlores~\cite{chern10}.

As the system cools down, our Monte Carlo simulations detect a rapid change at $T =  T_{\mathrm{c}1}$ in the ${A_2}$ and $E$ order parameters, shown in Figs.~\ref{fig:jzpm-0.4-cMC}(b) and \ref{fig:jzpm-0.4-cMC}(c). This jump, in addition to the sharp peak of the specific heat, shown in Fig.~\ref{fig:jzpm-0.4-cMC}(a), and the double-peak structure in the histograms of the order-parameter-magnitude distributions at $T\approx T_{\mathrm{c}1}$, displayed in Figs.~\ref{fig:jzpm-0.4-cMC}(d) and \ref{fig:jzpm-0.4-cMC}(e), allows us to identify this transition as a first-order transition. 
In the low-temperature phase, the $E$ order parameter $\langle m_E \rangle$ jumps to a finite value, effectively reducing the magnitude of $\langle m_{A_2}\rangle$.

The phase transitions and their nature can be interpreted in the following way. The relevant symmetries are represented by time reversal symmetry $\Theta$ and three-fold rotation symmetry $C_3$, which are isomorphic to a $\mathbb{Z}_2 \times \mathbb{Z}_3$ symmetry group structure. Time reversal ($\mathbb{Z}_2$) symmetry breaking is spontaneously broken at $T_{\mathrm{c}2}$ with the possibility of a continuous transition in the three-dimensional Ising universality class, which seems to be numerically realized. The residual rotational $C_3$ symmetry ($\mathbb{Z}_3$) is broken at $T_{\mathrm{c}1}$, realizing a first-order phase transition as expected from three-state clock models or three-state Potts model~\cite{Wu82}. 

To further characterize the low-temperature mixed phase observed below $T_\mathrm{c1}$, we employ Eq.~\eqref{eq:A2ET1_mixing_angles} to extract the $A_2 \oplus \psi_2$ mixing angle $\phi$. The mixing angle between these irreps is obtained by measuring the distribution of the magnitude of the \emph{local} single-tetrahedron irreps, i.e., the distribution obtained from the magnitude of the irreps in the $L^3$ tetrahedra in a single full-system configuration~\cite{lozano-gomez2023,francini25}
[see Fig.~\ref{fig:jzpm-0.4-cMC}(f) for the coordinate frame used for these distributions]. Since the $T_{1-}$ irrep is strongly thermally depopulated at temperatures below $T_\mathrm{c1}$, the distribution of the single-tetrahedron irreps lies close to the equator line of the $A_2 \oplus E $ mixed phase, with $\theta\approx \pi/2$. This is illustrated in the histogram of the single-tetrahedron irreps shown in Fig.~\ref{fig:jzpm-0.4-cMC}(g). Below $T_\mathrm{c1}$, the distribution is concentrated around a finite value of $\phi$, which effectively quantifies the mixing between the $A_2$ and $E$ irreps, see the white dashed line in Fig.~\ref{fig:jzpm-0.4-cMC}(g). The precise location at which the distribution concentrates at low temperatures depends on $J_{z\pm}$, implying that the mixing angle $\phi$ defining the low-temperature phase drifts as a function of $J_{z\pm}$, in agreement with the results of the classical low-temperature expansion.

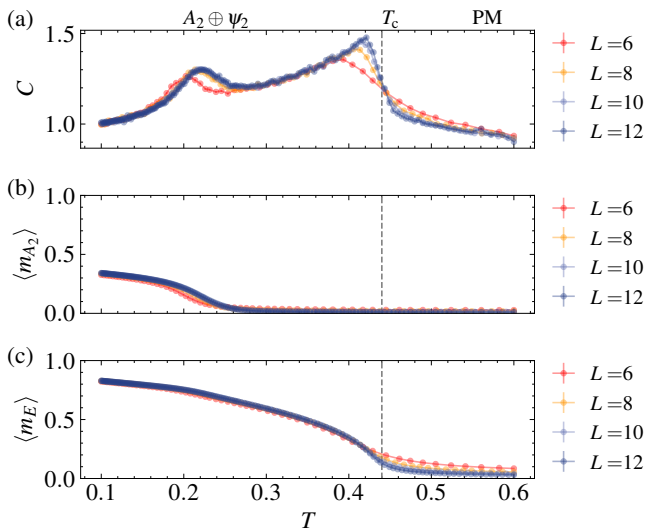
\begin{figure}
    \centering
    \begin{overpic}[width=\columnwidth]{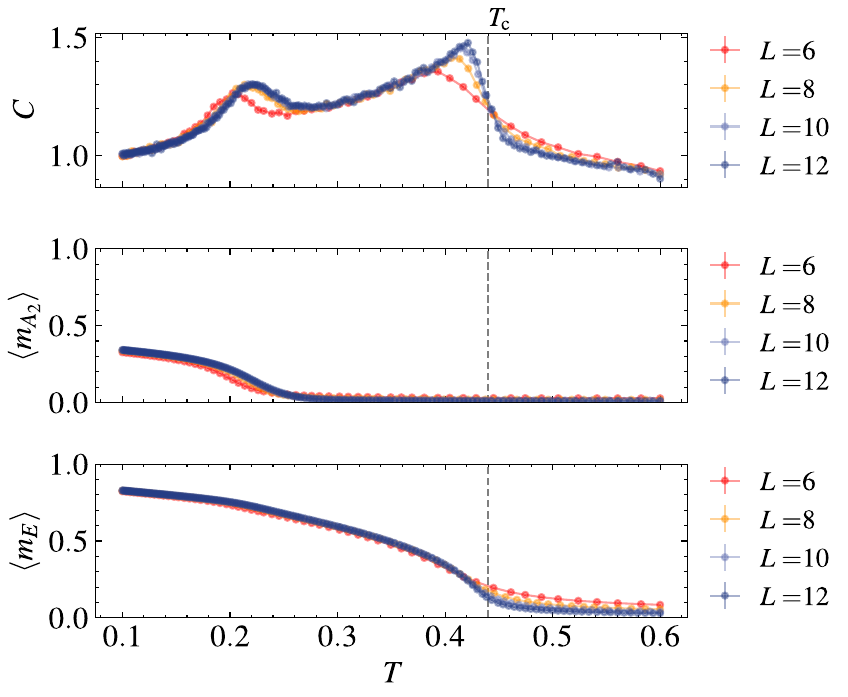}
    \put(0,79){(a)}
    \put(0,53){(b)}
    \put(0,27){(c)}
    \put(30,79.5){$A_2\oplus\psi_2$}
    \put(66,79.5){PM}
    \end{overpic}
    \caption{Monte Carlo results in the Kramers case for $J_{z\pm} = 1.2$. (a)~Specific heat, (b)~$A_2$ order parameter, and (c)~$E$ order parameter as functions of the temperature $T$ for different lattice sizes $L$. The dashed line at $T_{\mathrm{c}}\approx0.440$ indicates the position of the continuous transition that separates the paramagnetic phase at high temperatures from the $A_2\oplus \psi_2$ long-range ordered phase at intermediate and low temperatures. The states below and above the broad hump observed in the specific heat (as long as $T < T_\mathrm{c}$) are characterized by the same symmetries, such that the hump corresponds to a crossover rather than a genuine phase transition, as also evidenced by the behavior of the $A_2$ Binder cumulant and susceptibility shown in Appendix~\ref{appendix:E6-op-and-details}.}
    \label{fig:jzpm-1.2-cMC}
\end{figure}

\paragraph{Large coupling $J_{z\pm}\gtrsim 0.9$.}
For sufficiently large $J_{z\pm}$, approximately $J_{z\pm} \gtrsim 0.9$, the low-temperature $A_2 \oplus \psi_2$ phase is reached through a single continuous phase transition at critical $T_\mathrm{c}$, see Fig.~\ref{fig:phase-diagram}. Figure~\ref{fig:jzpm-1.2-cMC} illustrates a representative example of the specific heat and irrep order parameters for $J_{z\pm}=1.2$. In this large-coupling regime, the specific heat peaks at $T_\mathrm{c}$, followed by a broad hump at lower temperatures, see Fig.~\ref{fig:jzpm-1.2-cMC}(a). Reducing the temperature below $T_\mathrm{c}$, $\langle m_E \rangle$ continuously increases to a moderate value, while $\langle {m}_{A_2}\rangle$ initially increases only very slowly, see Figs.~\ref{fig:jzpm-1.2-cMC}(b) and (c). As the system is cooled below temperatures where the hump in the specific heat is observed, $\langle {m}_{A_2}\rangle$ starts to also increase significantly. Concurrently, the $\langle m_E \rangle$ keeps on increasing as the system is further cooled, with only minor changes in the slope of this order parameter.
Given that the $\psi_2$ and $A_2 \oplus \psi_2$ spin configurations break the same symmetries, the low-temperature hump in the specific heat should be interpreted as a crossover from a region with dominant $\psi_2$ order to one with moderate order in both the $A_2$ and $\psi$ irreps, rather than as a genuine phase transition~\footnote{Here we have excluded the possibility of a first-order phase transition, as the finite-size scaling of the hump in the specific heat does not indicate a linear dependence of this feature with $L$.}.
This interpretation is further supported by the behavior of the $A_2$ Binder cumulant and susceptibility. The Binder cumulant shows no crossing point, as would be expected for a genuine phase transition, and the susceptibility exhibits no divergence with increasing system size (see Appendix~\ref{appendix:E6-op-and-details}). Given that the internal energies of the $\psi_2$ and $A_2 \oplus \psi_2$ spin configurations are degenerate by construction, the low-temperature hump in the specific heat reflects the entropy release associated with thermal fluctuations of the $A_2$ irrep, while all symmetry breaking occurs at $T_\mathrm{c}$.

The precise location of $T_\mathrm{c}=0.440(2)$ is determined from the $E$ Binder cumulant, showing a common crossing point at different sizes at the critical temperature. We also attempted to characterize the critical behavior by extracting critical indices from the finite-size scaling of the Binder cumulant and the $E$ susceptibility, obtaining $\nu=0.65(2)$ and $\gamma/\nu=1.60(2)$, see Appendix~\ref{appendix:FSS-continuous-transition} for details.
The transition and its nature can be interpreted within the following scenario. The symmetry group $\mathbb{Z}_2 \times \mathbb{Z}_3 \simeq \mathbb{Z}_6$ is spontaneously broken, both $\mathbb{Z}_2$ and $\mathbb{Z}_3$, at $T_\mathrm{c}$. At criticality, the $\mathbb{Z}_6$ anisotropies are irrelevant and the system experiences a symmetry enlargement to the continuous U(1) group~\cite{Hove2003}, so the critical behavior could fall, at least asymptotically, into the three-dimensional U(1) universality class, characterized by $\nu = 0.6717$ and $\gamma/\nu = 1.9618$~\cite{pelissetto02, hasenbusch19, chester20}.
Comparing the finite-size scaling critical exponents with those of the three-dimensional U(1) universality class, we find that $\nu$ agrees with the theoretical value within error bars, whereas $\gamma/\nu$ deviates significantly. A possible explanation is a slow renormalization-group flow in our microscopic model along the $A_2 \oplus E \oplus T_{1-}$ line, similar to the case for small $J_{z\pm}$ discussed in the previous paragraph (see also Ref.~\cite{chern10}).
%

\begin{figure}
    \centering
    \begin{overpic}[width=\columnwidth]{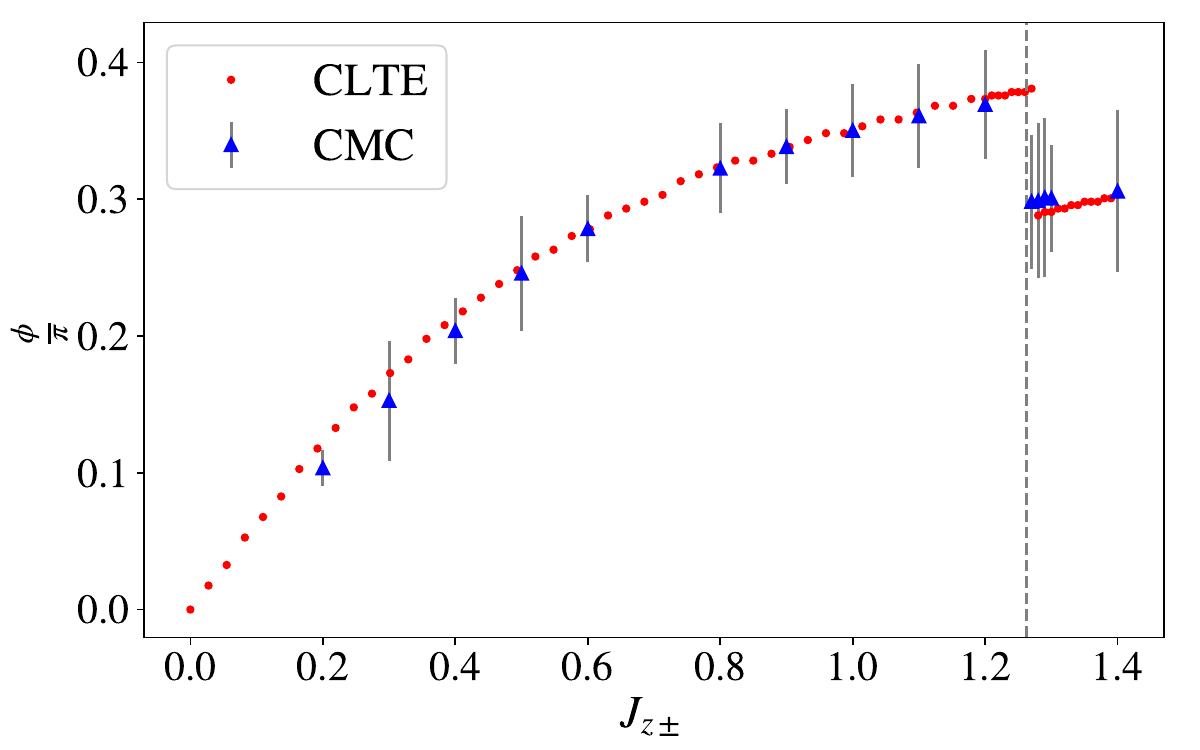}
    \end{overpic}
    \caption{$A_2 \oplus E$ mixing angle $\phi$ in the low-temperature state as a function of $J_{z\pm}$ from classical low-temperature expansion (CLTE) and classical Monte Carlo (CMC) simulations. The vertical dashed line at $J_{z\pm}^\star\simeq 1.262$ indicates the value of the coupling at which the mixing angle abruptly changes.
}
\label{fig:CLTE_v_CMC}
\end{figure}

\paragraph{Comparison with predictions from classical low-temperature expansion and Landau theory.}
\label{paragraph:CLTE}
Except in the immediate vicinity of the point $J_{z\pm} = 1/\sqrt{2}$, our Monte Carlo simulations reveal a mixed $A_2 \oplus \psi_2$ low-temperature ground state throughout the phase diagram, in agreement with the expectation from the classical low-temperature expansion (see Sec.~\ref{sec:CLTE}). In fact, the agreement with the classical low-temperature expansion expectations can be assessed on a quantitative level by studying the single-tetrahedron-irrep-magnitude distribution on the full lattice configurations obtained from Monte Carlo simulations, and measuring the resulting mixing angles in the low-temperature regime, see, e.g., the irrep-magnitude distribution on the $A_2\oplus E \oplus T_{1-}$ manifold shown in Fig.~\ref{fig:jzpm-0.4-cMC}(g). From the distributions, we extract the average $A_2\oplus E$ mixing angle $\phi$, which may be compared with the corresponding expectation from the classical low-temperature expansion, see, e.g., Fig.~\ref{fig:CLTE_map}.
Since we measured the magnitude of the irreps in the classical Monte Carlo simulations, the corresponding mixing angle is restricted to $\phi\in[0,\pi/2]$. This means that mixing angles $\phi > \pi/2$ obtained in the classical low-temperature expansion need to be reflected as $\phi\rightarrow \pi-\phi$ to facilitate comparison with the corresponding angles measured in the Monte Carlo simulations.
Figure~\ref{fig:CLTE_v_CMC} shows, within error bars, excellent agreement between classical Monte Carlo simulations and the classical low-temperature expansion across the entire phase diagram, thereby linking the low-temperature order-by-disorder phenomenon discussed in Sec.~\ref{sec:CLTE} to the corresponding regions of the Monte Carlo phase diagram.
In particular, the angle extracted from the Monte Carlo data also presents a jump around the position $J_{z\pm}^\star\approx1.26$ predicted from the classical low-temperature expansion. It is caused by the change in the highest-entropy configuration described by the two $A_2\oplus \psi_2$ branches, see Sec.~\ref{sec:CLTE}, and can be interpreted as an unnecessary first-order transition occurring within a single thermodynamic phase. This transition may be understood as a magnetic analogue of the liquid-gas transition~\cite{jimenez21} and is here driven by tuning $J_{z\pm}$ at fixed temperature, separating two $A_2\oplus\psi_2$ configurations characterized by different mixing angles $\phi_1$ and $\phi_2$.
Up to our maximal sizes, we cannot resolve this transition in the specific heat, while it is well represented in the order parameters.

Our Monte Carlo results can also be compared with the Landau theory analysis presented in Sec.~\ref{sec:Landau-theory}. As discussed above, both approaches yield continuous transitions between the paramagnetic phase and the magnetically ordered phases, while the transition between the two distinct magnetic phases ($A_2$ and $A_2 \oplus \psi_2$) observed for $J_{z\pm} < 1/\sqrt{2}$ is identified as first order.
The evolution of the order parameters shown in Figs.~\ref{fig:jzpm-0.4-cMC}(b) and (c) near the continuous transition at $T_\mathrm{c2}$ closely resembles the corresponding Landau transition in Fig.~\ref{fig:LG}(c). However, the critical exponent $\beta$ in the power law $\langle m_{A_2} \rangle \propto (T_\mathrm{c2} - T)^\beta$ differs from the value predicted by Landau theory, as expected below the upper critical dimension.
The behavior near the low-temperature transition at $T_\mathrm{c1}$, between the $A_2$ and $A_2 \oplus \psi_2$ phases [Figs.~\ref{fig:jzpm-0.4-cMC}(b) and (c)], can be qualitatively described by the Landau transition shown in Fig.~\ref{fig:LG}(f).
For $J_{z\pm} > 1/\sqrt{2}$, the evolution of the order parameters [Figs.~\ref{fig:jzpm-1.2-cMC}(b) and (c)] is similar to that captured by the Landau transition in Fig.~\ref{fig:LG}(d), where the $E$ order parameter dominates near the phase transition, while the $A_2$ order parameter becomes comparable only deeper in the ordered phase.
%
%
%

\section{Spin nematic phase}
\label{sec:SN-phase}
In this section, we present a detailed discussion of the nematic phase found in the vicinity of the point $J_{z\pm}=1/\sqrt{2}$. We discuss the emergence of additional symmetries at this point and identify dual models that consequently realize analogous ground states at low temperatures. We illustrate how the ground-state configurations of the nematic state can be explicitly constructed, and demonstrate that these can be classified according to their total magnetization. We then present a detailed Monte Carlo analysis illustrating how the nematic phase is reached from higher temperatures, and discuss the correlation functions and irrep-magnitude distributions measured in this phase.
%

\subsection{Additional symmetries at \texorpdfstring{$J_{z\pm}=\pm1/\sqrt{2}$}{Jzpm=1/sqrt(2)}}
\label{subsec:exact-symmetry-nematic}

The point $J_{z\pm}=\pm1/\sqrt{2}$ plays a special role along the $A_2\oplus E \oplus T_{1-}$ line, with no conventional magnetic order detected down to the lowest temperatures~\cite{Benton2014_thesis}. At this point, the Hamiltonian in Eq.~\eqref{eq:general_bilinear_hamiltonian} has a subextensive number of additional exact symmetries that prevent dipole ordering throughout the whole system. To discuss the symmetries and construct the ground-state manifold, it is convenient to first discuss the case $J_{z\pm} = -1/\sqrt{2}$, which is dual to $J_{z\pm} = 1/\sqrt{2}$~\cite{Rau2018FrustratedQR, francini25}.
At this point, the Hamiltonian takes the simplified compass-model-like~\cite{nussinov-15} form
\begin{equation}
    \label{eq:nematic-hamiltonian-dual-global}
    \mathcal{H}_{\rm nem}^{<}=J_2\sum_{\boxtimes}(s_0^z s_1^z + s_2^z s_3^z + s_0^y s_2^y+ s_1^y s_3^y+ s_0^x s_3^x + s_1^x s_2^x)
\end{equation}
in the \emph{global} basis defined in Appendix~\ref{appendix:irreps-basis}. Here, $J_2=3J_{zz}<0$, and the global spin components  $s_\mu^\alpha$ are obtained using Eq.~\eqref{eq:global-coordinate-components}.
We refer to the Hamiltonian in Eq.~\eqref{eq:nematic-hamiltonian-dual-global} as the ``nematic'' Hamiltonian $\mathcal{H}^<_{\rm nem}$, where $<$ signifies the negative sign of the coupling $J_{z\pm}$.
To expose the additional symmetry realized by the nematic Hamiltonian, we first note that the pyrochlore lattice can be constructed as a set of nonoverlapping chains in three possible ways; (i) the first is obtained by considering chains running along the (110) and (1$\bar{1}$0) directions, (ii) the second way is by considering chains running along the (011) and (01$\bar{1}$) directions, (iii) and the last possible way is by considering chains running along the (101) and (10$\bar{1}$) directions~\footnote{We note that these three ways are related by a cubic $C_3$ symmetry.}.
In each of these cases, a chain is composed of only two specific sublattice sites, e.g., the chains running along the (110) direction are composed of sublattice sites $\mu=0$ and $\mu=1$.

\begin{figure}
    \centering
    \begin{overpic}[width=0.7\columnwidth]{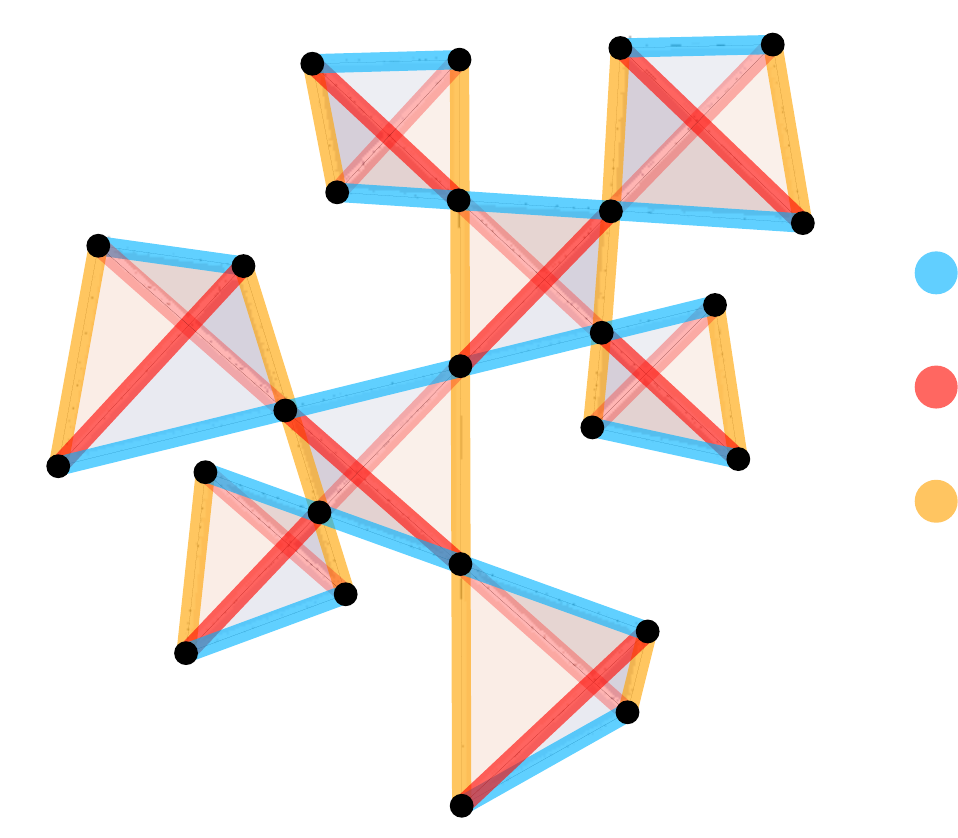}
        \put(102,58){$z$-chain}
        \put(102,46){$y$-chain}
        \put(102,34){$x$-chain}
        \put(43,0){0}
        \put(65,7){1}
        \put(70,20){2}
        \put(50,30){3}
    \end{overpic}
\caption{Illustration of the three sets of spin chains along which the three different terms occurring in the nematic Hamiltonian $\mathcal H_\text{nem}^<$ are defined. $z$ terms defined on the blue chains correspond to interactions of the form $s_0^zs_1^z+s_2^zs_3^z$, $y$ terms (red chains) correspond to $s_0^ys_2^y+s_1^ys_3^y$, and $x$ terms (yellow chains) correspond to $s_0^xs_3^x+s_1^xs_2^x$.}
    \label{fig:chain-symmetry-lattice}
\end{figure}

The nematic Hamiltonian consists of three different terms with Ising-type interactions along the three different cubic axes in the global basis, namely an $x$ term $J_2\sum_{\boxtimes} s_0^xs_3^x+s_1^xs_2^x$, a $y$ term $J_2\sum_{\boxtimes} s_0^ys_2^y+s_1^ys_3^y$, and a $z$ term $J_2\sum_{\boxtimes} s_0^zs_1^z+s_2^zs_3^z$. 
The three sets of chains corresponding to the three different terms in the nematic Hamiltonian are illustrated in Fig.~\ref{fig:chain-symmetry-lattice}.
On the level of a single tetrahedron, each of these terms describes a pair-wise ferromagnetic (recall that $J_2<0$) interaction between the global Cartesian components of the spins on two perpendicular bonds of a single tetrahedron. In the full lattice, each of these sets of interactions defines a system of nonoverlapping infinite spin chains, 
corresponding to the three ways in which the pyrochlore can be constructed as mentioned above. The decomposition of the nematic Hamiltonian into the $x$, $y$, and $z$ terms results in an additional $\mathbb{Z}_2$ symmetry on \emph{every} spin chain over a fixed cartesian component of the spins. Focusing on the $z$ term, for example, the Hamiltonian is invariant under the transformation $(s_0^z,s_1^z) \to -(s_0^z,s_1^z)$, with the spins in sublattices $0$ and $1$ lying on a chain along the (110) direction. 
As these additional $\mathbb{Z}_2$ symmetry transformations only involve a subset of the spins in the lattice, we identify these as subsystem symmetries~\cite{vijay16, nandkishore19}. 
\begin{figure}
    \centering
    \begin{overpic}[width=0.8\columnwidth]{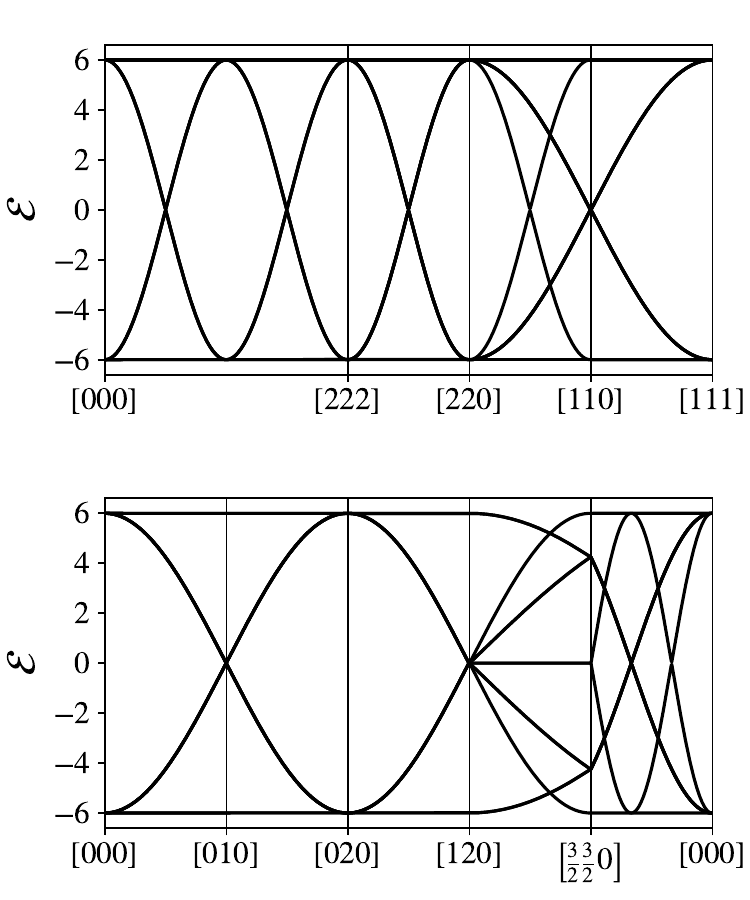}
    \put(0,95){(a)}
    \put(0,45){(b)}
    \end{overpic}
    \caption{Interaction-matrix bands in reciprocal space of the Hamiltonian in Eq.~\eqref{eq:nematic-hamiltonian-dual-global} along a high-symmetry path in the (a)~$[hh\ell]$ and (b)~$[hk0]$ planes.}
    \label{fig:nem-bands}
\end{figure} 

The subsystem symmetries also result in an enlarged ground-state manifold. Indeed, since each spin chain is associated with a $\mathbb{Z}_2$ symmetry, there are of the order of $6L^2$ additional subsystem symmetries on the system. These imply that the ground-state manifold of the system must present a similar number of ground-state configurations, resulting in a subextensive ground-state manifold.
This subextensive degeneracy can be exposed by studying the spectrum of the interaction matrix $J_{ij}^{\alpha\beta}$ associated to the nematic Hamiltonian in reciprocal space~\footnote{For a general bilinear spin models, a Hamiltonian can be rewritten in the form $\mathcal{H}=\frac{1}{2}\sum_{i,j} s_i^\alpha J_{ij}^{\alpha\beta}s_j^\beta$.}.
Figure~\ref{fig:nem-bands} illustrates the energy bands of the interaction matrix along two high-symmetry paths. We note that, in the $[hh\ell]$ plane [Fig.~\ref{fig:nem-bands}(a)], the lowest-energy band is completely flat, whereas it becomes dispersive in the $[hk0]$ plane [Fig.~\ref{fig:nem-bands}(b)]. 
From a self-consistent Gaussian approximation (a soft-spin approximation where the spin-length constraint is fulfilled on average in the system)~\cite{Chung_PRL,SCGA_Canals_kagome,SCGA_Canals_pyrochlore,francini25}, this degeneracy in the lowest-energy bands implies that there are $\mathcal{O}(L^2)$ spin modes in the ground-state manifold, agreeing with the $6L^2$ additional subsystem symmetries $\mathbb{Z}_2$ present in the model with $J_{z\pm}=-1/\sqrt{2}$. 
Furthermore, due to the duality of the Hamiltonian in Eq.~\eqref{eq:general_bilinear_hamiltonian} under the sign change of the $J_{z\pm}$ coupling~\cite{Rau2018FrustratedQR}, all the properties we have discussed for the Hamiltonian $\mathcal{H}^<_{\rm nem}$, where $J_{z\pm}=-1/\sqrt{2}$, are also applicable to its dual, namely  $\mathcal{H}^>_{\rm nem}$, where $J_{z\pm}=1/\sqrt{2}$.
In particular, the spectrum of the interaction matrix possesses the same form.
\begin{table*}
\caption{Exact nematic point with coordinates in local and global basis (cf.~Appendix~\ref{appendix:irreps-basis}).}
    \centering
    \setlength{\extrarowheight}{2pt}
    \begin{tabularx}{\linewidth}{>{\hsize=.2\hsize\centering\arraybackslash}X >{\hsize=.4\hsize\centering\arraybackslash}X >{\hsize=.4\hsize\centering\arraybackslash}X}
    \hline 
    \hline
        Degenerate irreps & Local basis $(J_{zz},J_{\pm},J_{\pm\pm},J_{z\pm})$ &  Global basis $(J_1,J_2,J_3,J_4)$ \\[1pt]
        \hline
        $A_{2}\oplus E \oplus {T_{1-}}$ & $\Bigl(J_{zz}<0$, $-\frac{J_{zz}}{2}$, $\frac{J_{zz}}{2}$, $\frac{J_{zz}}{\sqrt{2}}\Bigl)$ &  $(0,3,0,0)J_{zz}$ \\ \\
         $A_{2}\oplus E \oplus {T_{1-}}$ & $\Bigl(J_{zz}<0$, $-\frac{J_{zz}}{2}$, $\frac{J_{zz}}{2}$, $-\frac{J_{zz}}{\sqrt{2}}\Bigl)$ & $(-\frac{4}{3},\frac{1}{3},-\frac{4}{3},\frac{2}{3})J_{zz}$ \\ \\
        $T_{2} \oplus {T_{1-}}$ & $\Bigl(J_{zz}>0$, $-\frac{J_{zz}}{2}$, $\frac{J_{zz}}{2}$, $\frac{J_{zz}}{\sqrt{2}}\Bigl)$ &  $(0,3,0,0)J_{zz}$ \\ \\
         $T_{2} \oplus {T_{1-}}$ & $\Bigl(J_{zz}>0$, $-\frac{J_{zz}}{2}$, $\frac{J_{zz}}{2}$, $-\frac{J_{zz}}{\sqrt{2}}\Bigl)$ & $(-\frac{4}{3},\frac{1}{3},-\frac{4}{3},\frac{2}{3})J_{zz}$   \\[8pt]
        \hline
        \hline
    \end{tabularx}
    \label{tab:nem-points}
\end{table*}

The spectrum of the interaction matrix shown in Fig.~\ref{fig:nem-bands} exposes an additional symmetry of the system: the bands of the interaction matrix are symmetric about the zero-energy line. This symmetry implies that the subsystem symmetries and ground-state degeneracy realized in the nematic Hamiltonians $\mathcal{H}^<_{\rm nem}$ and $\mathcal{H}^>_{\rm nem}$ must also hold for their inverse additives $-\mathcal{H}^<_{\rm nem}$ and $-\mathcal{H}^>_{\rm nem}$.
The additional symmetry of the spectrum is also reflected in the irreps' energies. Indeed, we identify two sets of degenerate irreps with opposite energies, each one containing six out of the twelve degrees of freedom encoded in the irreps; one  with $a_{A2}=a_E=a_{T_{1-}}=3J_{zz}$ and  the other $a_{T_2}=a_{T_{1+}}=-3J_{zz}$. Hence, we can identify the six negative-energy bands in Fig.~\ref{fig:nem-bands} with the six negative-energy modes, while the positive-energy bands link to the six positive-energy irreps. Depending on the sign of $J_{zz}$, the negative-energy modes are identified with $A_2\oplus E \oplus T_{1-}$ for $J_{zz}<0$ and with $T_{1-}\oplus T_2$ for $J_{zz}>0$. 
Below, we demonstrate that the point $J_{z\pm} = 1/\sqrt{2}$ on the $A_2 \oplus E \oplus T_{1-}$ line realizes an exact nematic point. The duality involving the $J_{zz}$ sign, combined with the duality relating positive and negative $J_{z\pm}$ regions, allows us then to expose four exact nematic points, two on the $A_2\oplus E\oplus T_{1-}$ degenerate line and two on the $T_{1-}\oplus T_2$ degenerate plane, as summarized in Tab.~\ref{tab:nem-points}. 
%

\subsection{Characterization of ground-state manifold}
\label{subsec:GS-nematic-characterization}
\begin{figure}
    \centering
    \begin{overpic}[width=\columnwidth]{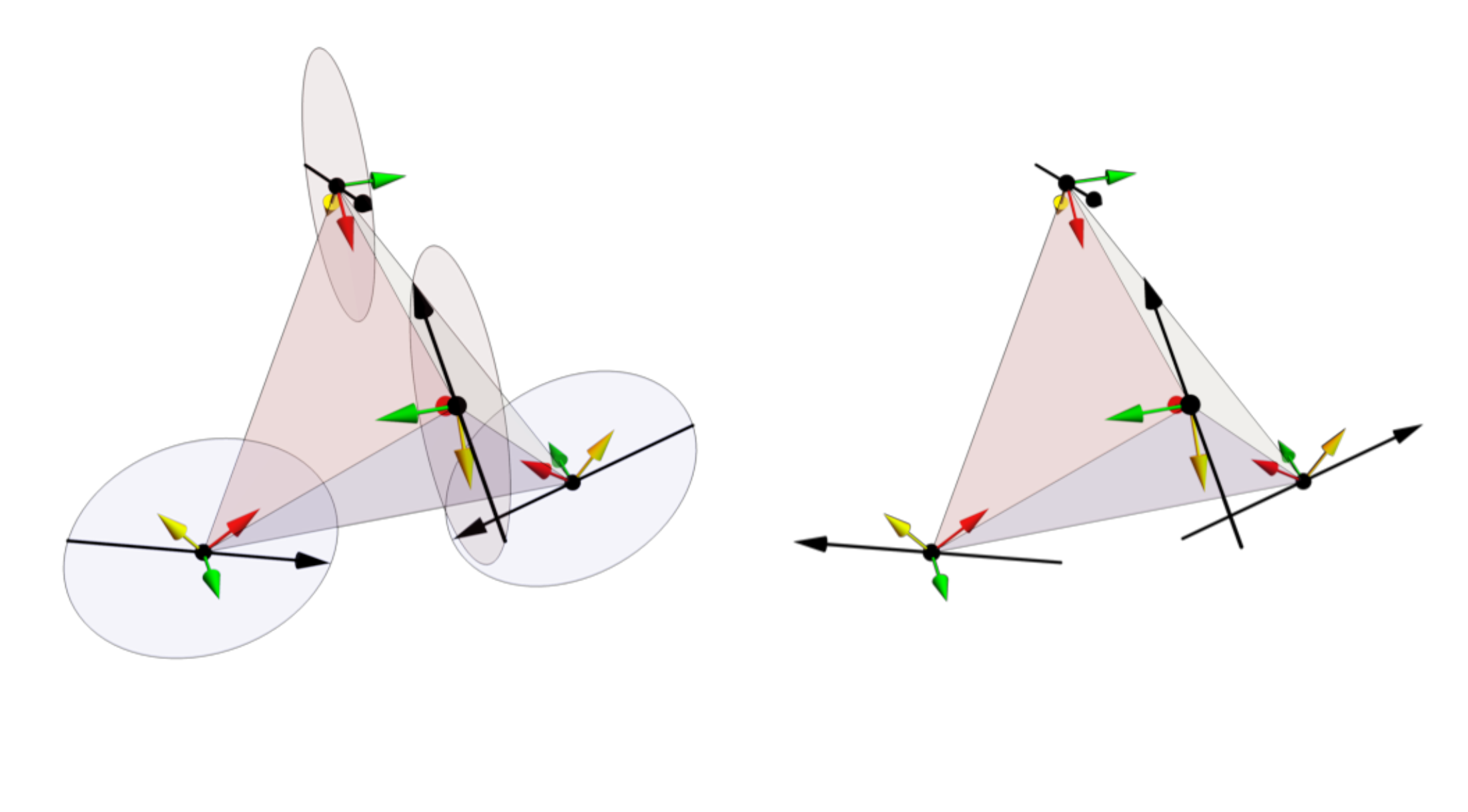}
    \put(10,50){(a)}
    \put(60,50){(b)}
    \put(23,10){$A_2\oplus \psi_2$}
    \put(73,10) {$T_{1-}$}
    \put(10,15){\tiny{0}}
    \put(40,21){\tiny{1}}
    \put(28,29){\tiny{2}}
    \put(19,40){\tiny{3}}
    \put(60,15){\tiny{0}}
    \put(90,21){\tiny{1}}
    \put(78,29){\tiny{2}}
    \put(69,40){\tiny{3}}
    \end{overpic}
\caption{Representative ground-state configuration on a single tetrahedron obtained from (a) $A_2\oplus \psi_2$ and (b) $T_{1-}$ irreps at $J_{z\pm}=1/\sqrt{2}$. The additional $\mathbb{Z}_2$ symmetries allow for a flipping of the spins on sublattices 0 and 1, yielding a mapping between the $A_2 \oplus \psi_2$ mode and the $T_{1-}$ mode. There are a total of six states for both the $A_2\oplus \psi_2$ and $T_{1-}$ spin configurations, which are mapped one-to-one into each other, i.e., $A_2\oplus \psi_2 \leftrightarrow T_{1-}$, by the additional symmetry transformation.
Colored arrows indicate local bases vectors $\mathbf x_\mu$, $\mathbf y_\mu$, $\mathbf z_\mu$, while black arrows indicate spin directions.
The opaque disks in (a) represent the $A_2 \oplus \psi_2$ manifold.
}
\label{fig:jzpm-nematic-irreps}
\end{figure}

%
The presence of the additional $\mathbb{Z}_2$ subsytem symmetries has important consequences for the single-tetrahedron irreps. At the level of a single tetrahedron, these symmetries provide a transformation relating the $A_2 \oplus \psi_2$ and $T_{1-}$ spin configurations.
At the nematic point $J_{z\pm}=1/\sqrt{2}$, the $A_2\oplus \psi_2$ configurations with the highest-entropy contribution consist of six antiferromagnetic configurations, for example
\begin{align}
\label{eq:nematic-A2-plus-E-gs-1}
        \mathbf{S}_0&= \frac{2\mathbf{e}_x+2\mathbf{e}_y-\mathbf{e}_z}{3}S\,, &
        \mathbf{S}_1&= \frac{-2\mathbf{e}_x-2\mathbf{e}_y-\mathbf{e}_z}{3}S\,, \\
\label{eq:nematic-A2-plus-E-gs-2}
        \mathbf{S}_2&=\frac{-2\mathbf{e}_x+2\mathbf{e}_y+\mathbf{e}_z}{3}S\,,&
        \mathbf{S}_3&= \frac{2\mathbf{e}_x-2\mathbf{e}_y+\mathbf{e}_z}{3}S\,.
\end{align}
on a single tetrahedron, as sketched in Fig.~\ref{fig:jzpm-nematic-irreps}(a).
On the other hand, the $T_{1-}$ subspace consists of six splayed ferromagnetic states, such as 
\begin{align}
 \label{eq:nematic-Tminus-gs-1}
        \mathbf{S}_0&= \frac{-2\mathbf{e}_x-2\mathbf{e}_y+\mathbf{e}_z}{3}S\,, &
        \mathbf{S}_1&= \frac{2\mathbf{e}_x+2\mathbf{e}_y+\mathbf{e}_z}{3}S\,, \\
 \label{eq:nematic-Tminus-gs-2}    
        \mathbf{S}_2&=\frac{-2\mathbf{e}_x+2\mathbf{e}_y+\mathbf{e}_z}{3}S\,, &
        \mathbf{S}_3&= \frac{2\mathbf{e}_x-2\mathbf{e}_y+\mathbf{e}_z}{3}S\,,
\end{align}
shown in Fig.~\ref{fig:jzpm-nematic-irreps}(b).
We note that in these two configurations the spins $\mathbf{S}_0$ and $\mathbf{S}_1$ are inverted, whereas the remaining spins $\mathbf{S}_2$ and $\mathbf{S}_3$ possess the same orientation. The agreement between the spin configurations in sublattices $\mu=2$ and $\mu=3$ is associated with the additional subsystem $\mathbb{Z}_2$ symmetry.
In other words, starting from a single-tetrahedron $A_2\oplus \psi_2$ spin configuration, the application of a $\mathbb{Z}_2$ subsystem symmetry 
on the spin chain containing sites $\mu=0$ and $\mu=1$ yields a single-tetrahedron $T_{1-}$ state, illustrated in Fig.~\ref{fig:jzpm-nematic-irreps}(b). In more general terms, each of the six states of the $A_2\oplus \psi_2$ subspace are mapped by the $\mathbb{Z}_2$ subsystem symmetry to a unique $T_{1-}$ state.
The fact that the relation between these spin configurations arises from an exact symmetry at $J_{z\pm}=1/\sqrt{2}$ implies that the $T_{1-}$ irrep presents the same entropy and therefore \emph{must} be present at higher temperatures, as also confirmed by our Monte Carlo results discussed in Sec.~\ref{subsec:cMC-at-nematic}~\footnote{The additional $\mathbb{Z}_2$ subsystem symmetry at the point $J_{z\pm}=1/\sqrt{2}$ yields a low-temperature phase featuring a reappearance of the $T_{1-}$ irreps, which contrasts with the depopulation of the $T_{1-}$ mode throughout the rest of the phase diagram.}.

Beyond the single-tetrahedron level, the fact that the $A_2\oplus \psi_2$ and $T_{1-}$ modes feature the same $\mathbf{S}_2$ and $\mathbf{S}_3$, but different $\mathbf{S}_0$ and $\mathbf{S}_1$, allows us to tile single tetrahedra with different spin configurations, effectively constructing full spin configurations in the entire lattice. 
Following this construction results in spin configurations of ordered chains with each tetrahedron realizing either a $A_2\oplus \psi_2$ or $T_{1-}$ state. This leads to a multitude of full-lattice spin configurations which, in general, do not preserve translational symmetry. 
Due to the subsystem symmetries, each chain of spins is independent from the others, suppressing any long-range dipolar order. As a consequence, time reversal symmetry is preserved, but the preferred directions of the spins break the rotational symmetry of the microscopic Hamiltonian, resulting in a spin nematic state at low temperatures.

Let us now use the $\mathbb{Z}_2$ subsystem symmetries to identify and characterize the spin configuration in the ground-state manifold. As a starting point, consider the 
Hamiltonian $\mathcal{H}_{\mathrm{nem}}^<$ in Eq.~\eqref{eq:nematic-hamiltonian-dual-global}. In this case, the $T_{1-}$ irrep describes a collinear ferromagnetic spin configuration. Taking one of the three cubic domains, say the $z$ direction, we can explicitly construct the ground-state manifold. 
Consider the fully aligned state with all chains pointing along the $+z$ direction with magnetization $4L^3S$, corresponding to all tetrahedra being in the $+z$ component of the $T_{1-}$ mode.
The application of the $\mathbb{Z}_2$ subsystem symmetries along the chains results in full-lattice configurations belonging to distinct magnetization sectors. Each magnetization sector has a total magnetization of $M_n=(4L^3-4Ln)S$, obtained by flipping the $z$ components of the spins in $n$ randomly selected spin chains of the fully aligned configuration. Following this procedure until all spin chains have been flipped, results in the fully aligned state along the $-z$ direction, which features all the tetrahedra in the $-z$ component of the $T_{1-}$ mode. 
Each magnetization sector can be characterized by the number of flipped spin chains $n$, which varies from 0 in the fully aligned state along the $+z$ direction to $2L^2$ for the fully aligned state along the $-z$ direction.
\begin{figure}[tb!]
    \begin{overpic}[width=\columnwidth]{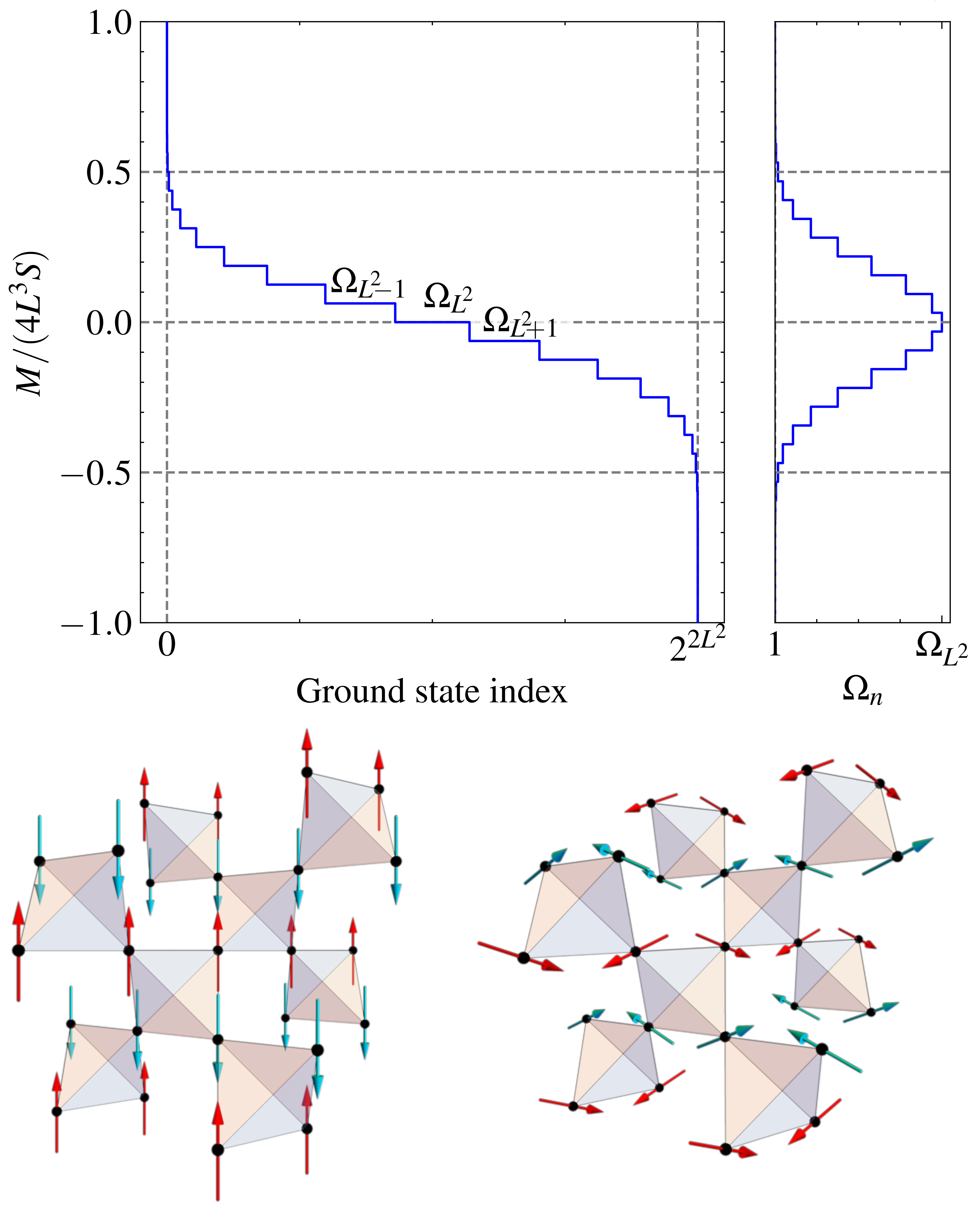}
    \put(0,92){(a)}
    \put(65,92){(b)}
    \put(10,-1){$J_{z\pm}=-1/\sqrt{2}$}
    \put(50,-1){$J_{z\pm}=1/\sqrt{2}$}
    \put(0,37){(c)}
    \put(40,37){(d)}
    \end{overpic}
\caption{
(a)~Normalized magnetization $M / (4L^3 S)$ of the $2^{2L^2}$ distinct states in the ground-state manifold at the exact nematic point $J_{z\pm} = -1/\sqrt{2}$. States are indexed by decreasing magnetization, starting from the fully aligned configuration with all chains along $+z$. The $\mathbb{Z}_2$ subsystem symmetries generate distinct sectors, visible as magnetization plateaus. Each plateau $n$ has width $\Omega_n$, corresponding to the number of microstates with magnetization $M_n$.
(b)~Distribution of the normalized magnetization $M / (4L^3 S)$ across magnetization sectors. The distribution is peaked near zero magnetization and approaches a delta-like form around $M = 0$ in the thermodynamic limit.
(c)~Representative ground-state configuration for $J_{z\pm}=- 1/\sqrt{2}$, with blue (red) arrows corresponding to spin chains that have been flipped (have not been flipped) with respect to the fully aligned configuration.
(d)~Representative ground-state configuration for $J_{z\pm}=1/\sqrt{2}$, obtained from the configuration in (b) by a duality transformation involving a $C_2$ spin rotation about the local $\mathbf{z}_\mu$ axes.
}
\label{fig:nem-GS-and-degeneracy}
\end{figure} 
This identification allows us to count the number of microstates per magnetization sector
\begin{equation}
    \label{eq:microstate-magnetization sector}
    \Omega_n=\begin{pmatrix}
        2L^2\\ n
    \end{pmatrix},
\end{equation}
contributing to the degeneracy of the ground state, see Fig.~\ref{fig:nem-GS-and-degeneracy}(a), and leading to a subextensive residual entropy of
\begin{equation}
    \label{eq:nem-residual-entropy}
    \mathcal S=\ln\left(3\sum_{n=0}^{2L^2}\Omega_n\right)=2L^2\ln2 + \ln{3}\,,
\end{equation}
where the multiplicative factor $3$ accounts for the three cubic directions. 

The magnetization sector with the largest number of microstates is reached when half of the spin chains are flipped, i.e., for $n=L^2$, and the total magnetization of the system vanishes. All other sectors have a nonvanishing total magnetization. We note that time reversal symmetry links positive and negative magnetization sectors of the ground-state manifold, enforcing the magnetization to be an odd function with center $n=L^2$, as shown in Fig.~\ref{fig:nem-GS-and-degeneracy}(a), and proved by the relation
$M_{2L^2-n}=-4L(L^2-n)S=-M_n$. 
The microstate degeneracy is an even function around the point $n=L^2$, with $\Omega_{2L^2-n}=\Omega_n$. 
The magnetization distribution shown in Fig.~\ref{fig:nem-GS-and-degeneracy}(b) is peaked around $M = 0$. As the thermodynamic limit is approached, the width of the distribution decreases proportionally to $1/L$, while the height of the peak at $M = 0$ increases as $2^{2L^2}/L$. In this limit, the distribution converges to a delta-like form centered at zero magnetization, indicating that time-reversal symmetry remains unbroken in the ground state at the nematic point $J_{z\pm} = 1/\sqrt{2}$.

A representative ground-state configuration of the nematic Hamiltonian $\mathcal{H}^{<}_{\rm nem}$ with zero magnetization ($n=L^2$) is displayed in  Fig.~\ref{fig:nem-GS-and-degeneracy}(c), where all tetrahedra are in a collinear antiferromagnetic configuration. In Fig.~\ref{fig:nem-GS-and-degeneracy}(d), we provide the equivalent dual configuration, obtained for the dual nematic Hamiltonian $\mathcal{H}^{>}_{\rm nem}$ for $J_{z\pm}=1/\sqrt{2}$. The duality preserves the degeneracy counting and the symmetry, which remains valid also in the $\mathcal{H}^{>}_{\rm nem}$ case.
The duality transformation connecting the positive and negative $J_{z\pm}$ cases acts as a $C_2$ rotation about the local $\mathbf{z}_\mu$ axes. An example is shown in Fig.~\ref{fig:nem-GS-and-degeneracy}(c)-(d), with spin transforming
\begin{equation}
    \label{eq:nematic-duality-example}
    \begin{split}
        \mathbf{S}_0&=S\mathbf{e}_z \mapsto \frac{2\mathbf{e}_x+2\mathbf{e}_y-\mathbf{e}_z}{3}S\,,\\
        \mathbf{S}_1&=S\mathbf{e}_z\mapsto \frac{-2\mathbf{e}_x-2\mathbf{e}_y-\mathbf{e}_z}{3}S\,,\\
        \mathbf{S}_2&=-S\mathbf{e}_z \mapsto  \frac{-2\mathbf{e}_x+2\mathbf{e}_y+
        \mathbf{e}_z}{3}S\,,\\
        \mathbf{S}_3&=-S\mathbf{e}_z\mapsto \frac{2\mathbf{e}_x-2\mathbf{e}_y+\mathbf{e}_z}{3}S\,,
    \end{split}
\end{equation}
from the $J_{z\pm}=-1/\sqrt{2}$ spin configuration on the left-hand side to $J_{z\pm}=1/\sqrt{2}$ spin configuration on the right-hand side, where the spin components are expressed in the global basis defined in Appendix~\ref{appendix:irreps-basis}. 

\begin{figure}[tb!]
    \begin{overpic}[width=\columnwidth]{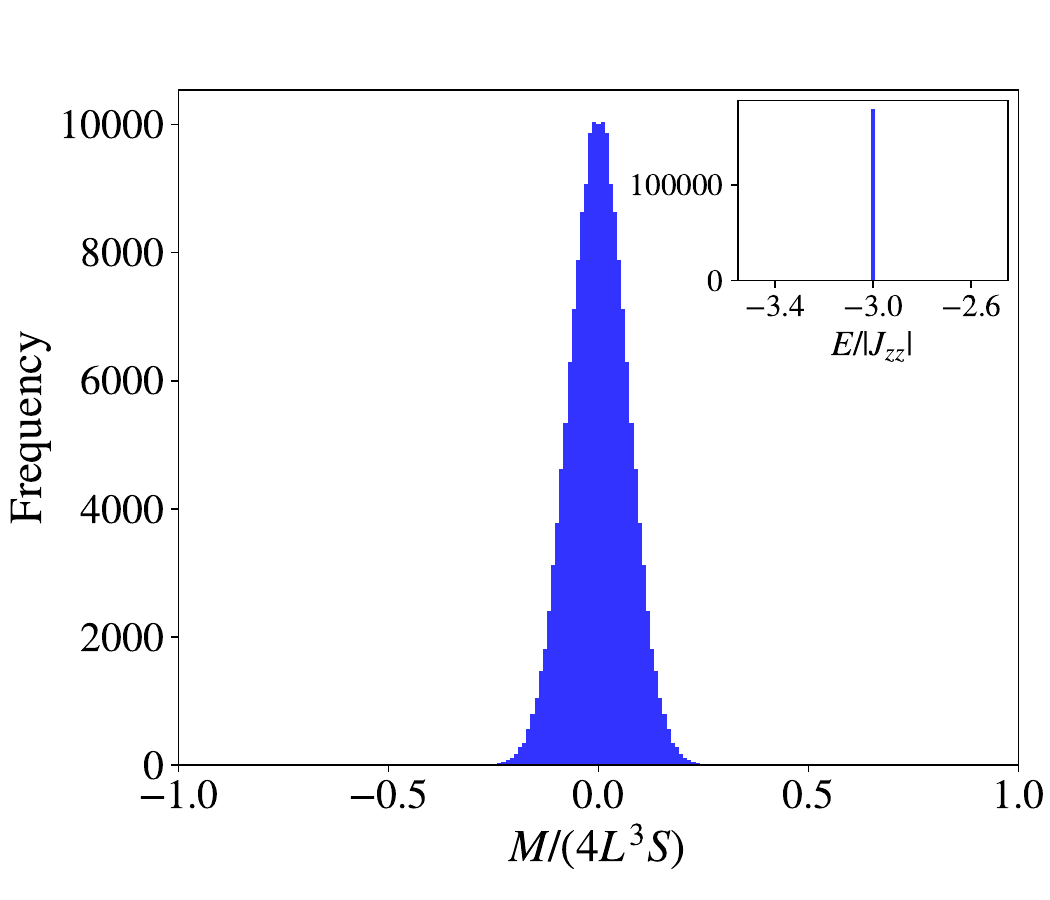}
    \end{overpic}
\caption{Histogram of the total magnetization in one of the three $C_3$ sectors of the nematic phase. Inset shows the histogram of the energy for the same set of states, confirming the degeneracy of the whole set of states considered.}
\label{fig:nematic_distribution}
\end{figure}

The preceding arguments deserve an additional remark. As discussed in the previous subsection, the interaction-matrix band analysis in Sec.~\ref{subsec:exact-symmetry-nematic} reveals flat planes rather than flat bands among the low-energy modes, as shown in Fig.~\ref{fig:nem-bands}. This structure suggests a scaling of the ground-state degeneracy as $\mathcal{O}(L^2)$~\footnote{This conclusion is based on a soft-spin system where the spin-length constraint would only be enforced on average~\cite{SCGA_Canals_pyrochlore,SCGA_Canals_checker,SCGA_Canals_kagome}.},
consistent with the subextensive residual entropy in Eq.~\eqref{eq:nem-residual-entropy}. However, the states considered in the above reasoning do \emph{not} exhaust the full set of energetically degenerate states. In particular, the pure $A_2$ all-in-all-out and the $E$ long-range-ordered states also belong to the degenerate manifold, yet they were not included in our counting. The reason is that their degeneracy is only of order $\mathcal{O}(1)$. These long-range-ordered configurations correspond to the band-touching points at the $\Gamma$ point in Fig.~\ref{fig:nem-bands}, which do not contribute (sub-)extensively. To further support the construction of the ground-state manifold and the distribution of the total magnetization of the nematic phase, we perform a statistical study of the ground-state manifold. Figure~\ref{fig:nematic_distribution} illustrates the distribution of the total magnetization obtained for over 180,000 states in a $L=10$ lattice, where all states possess the ground-state energy ${E}_0=-3J_{zz}$ as confirmed in the inset. The distribution agrees with our analytical construction, cf.~Fig.~\ref{fig:nem-GS-and-degeneracy}(b). We therefore conclude that our estimate captures those states that are entropically relevant, yielding a subextensive degeneracy that governs the physics in the thermodynamic limit.

%
%

\subsection{Classical Monte Carlo simulations for \texorpdfstring{$J_{z\pm}=\pm1/\sqrt{2}$}{Jzpm=1/sqrt(2)} }
\label{subsec:cMC-at-nematic}
\begin{figure}
    \centering
    \begin{overpic}[width=\columnwidth]{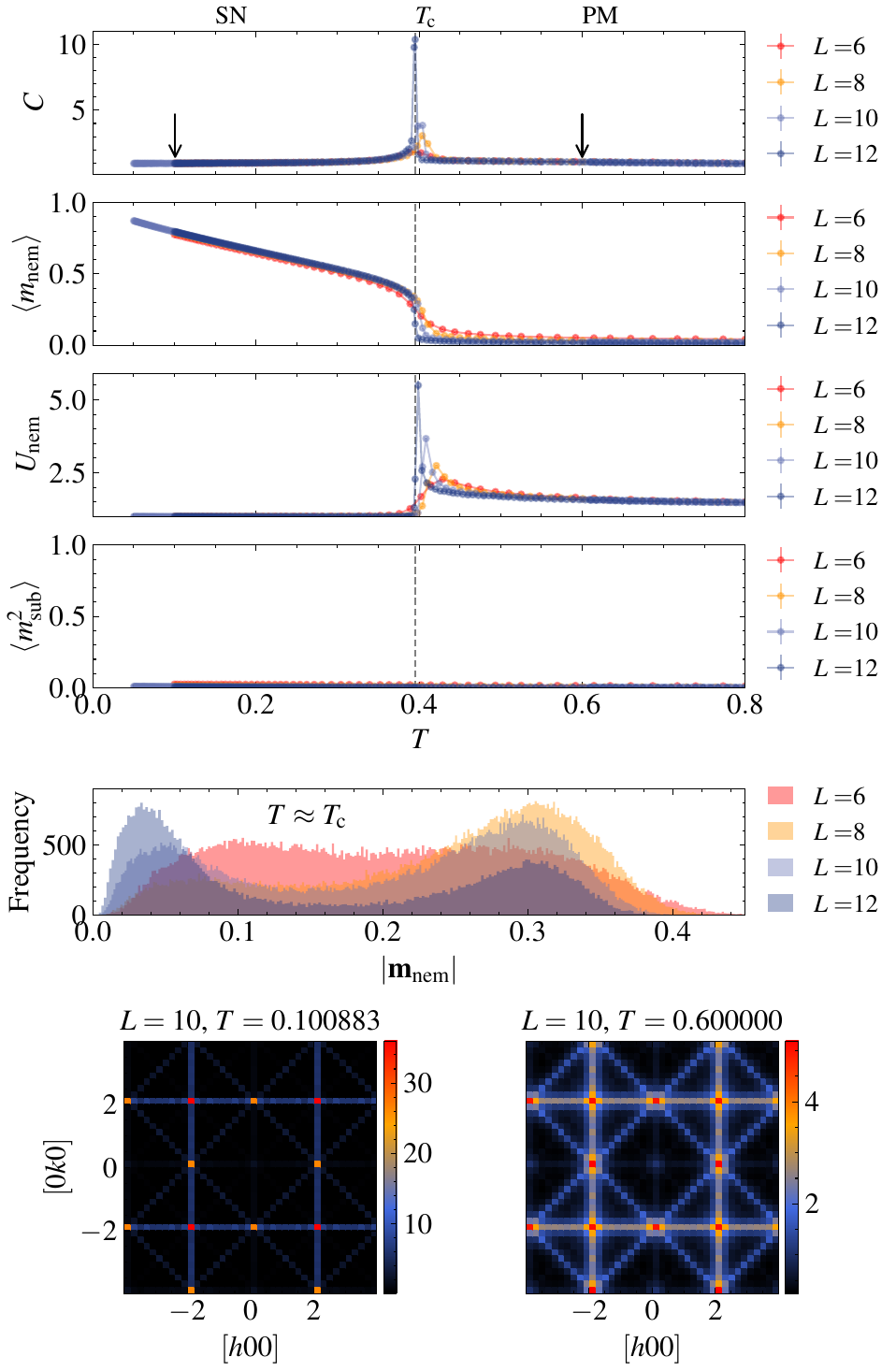}
    \put(0,98){(a)}
    \put(0,85){(b)}
    \put(0,72){(c)}
    \put(0,60){(d)}
    \put(0,45){(e)}
    \put(0,25){(f)}
    \put(33,25){(g)}
    \end{overpic}
\caption{Monte Carlo results at the exact nematic point $J_{z\pm} = 1/\sqrt{2}$. (a)~Specific heat, (b)~nematic order parameter, (c)~its Binder cumulant, and (d)~sublattice magnetic order parameter as functions of the temperature $T$ for different lattice sizes $L$.
The dashed line at $T_{\mathrm{c}}$ indicates the position of the first-order transition that separates the paramagnetic phase at high temperatures from the spin nematic phase at low temperatures. The sublattice magnetic order parameter remains zero in the spin-nematic phase, confirming the absence of any magnetic long-range order associated with the single-tetrahedron irreps.
(e)~Histogram of the nematic-order-parameter magnitude at the critical temperature. The double-peak structure indicates a first-order phase transition, consistent with the divergence observed in the Binder cumulant as the system size increases.
(f)~Equal-time spin structure factor for $L=10$ and $T=0.100883$ in the spin nematic phase.
(g)~Same as (f), but for $T=0.6$ in the paramagnetic phase.
}
\label{fig:jzpm-0.7-cMC} 
\end{figure}

In this subsection, we present the finite-temperature behavior of the system at the nematic point  $J_{z\pm}=1/\sqrt{2}$ obtained via classical Monte Carlo. Figure~\ref{fig:jzpm-0.7-cMC} illustrates the specific heat, the nematic and magnetic order parameters, the nematic Binder cumulant, the histogram of the nematic order parameter, and the spin structure factor obtained at two distinct temperatures. 
The finite-temperature behavior of the specific heat shows a strong first-order transition from the paramagnetic phase to the spin nematic phase, as visible in Fig.~\ref{fig:jzpm-0.7-cMC}(a), where the dashed line signals the transition temperature $T_{\rm c}$. Below the transition temperature, the nematic order parameter in Eq.~\eqref{eq:long_range_spin_nematic_OP} becomes nonzero, keeps on increasing with decreasing temperature, and approaches saturation in the limit  $T\to 0$, see Fig.~\ref{fig:jzpm-0.7-cMC}(b). At the transition, the nematic Binder cumulant has a peak that increases with system size, shown in Fig.~\ref{fig:jzpm-0.7-cMC}(c), typical of a first-order transition. 
As discussed in the previous subsections, since no dipole ordering is present, all magnetic order parameters vanish, implying the absence of magnetic long-range order at low temperature, as indicated by the sublattice magnetic order parameter shown in Fig.~\ref{fig:jzpm-0.7-cMC}(d). 
The histogram of the order-parameter magnitude has a two-peak structure, typical of a first-order transition, as visible in Fig.~\ref{fig:jzpm-0.7-cMC}(e). As expected, the region within the peaks is depopulated as the size increases, further supporting the first-order-transition interpretation.
The lack of magnetic long-range order is further supported by the absence of magnetic Bragg peaks in the equal-time spin structure factor, as illustrated in Figs.~\ref{fig:jzpm-0.7-cMC}(f)-(g). Instead of sharp peaks, the correlation functions develop continuous lines of scattering, connecting points of high symmetry, similar to the ones observed in the nematic phase at finite temperatures above the $A_2 \oplus T_{1-} \oplus T_2$ line~\cite{francini25}.  The observation of such continuous features in the equal-time spin structure factor is consistent with the subextensive degeneracy present in the nematic manifold. At temperatures above the critical temperature, where no symmetry breaking has taken place, these continuous lines of scattering are broader, and the system has access to all symmetry sectors. The broadness of these features is associated with thermal fluctuations, signaling the nonvanishing thermal population of high-energy modes. As the temperature is decreased below the critical temperature, the system selects one of the three cubic $C_3$ symmetry sectors while heavily suppressing high-energy modes, resulting in sharper features~\footnote{As discussed in Ref.~\cite{francini25}, the structure factor for a single $C_3$ symmetry sector would only display continuous lines along one cartesian axis.}.

\begin{figure}
    \centering
    \begin{overpic}[width=\columnwidth]{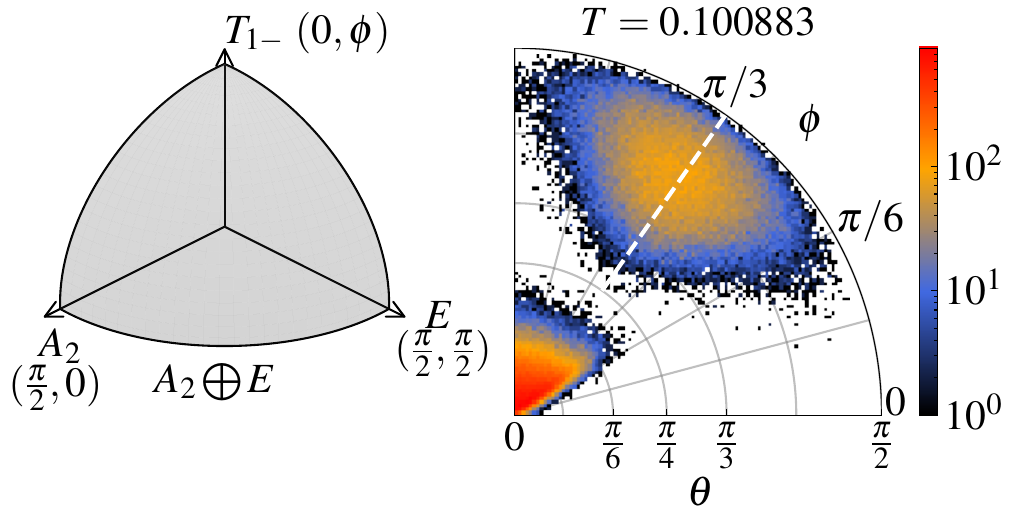}
    \put(0,45){(a)}
    \put(45,45){(b)}
    
    \end{overpic}
\caption{(a) Sketch of the $A_2\oplus E\oplus T_{1-}$ first octant, generated by the mixing angles $\theta$ and $\phi$ in Eq.~\eqref{eq:irreps mixing}. (b)~Irrep-magnitude distribution on the $A_2\oplus E\oplus T_{1-}$ first octant, stereographically projected, measured on many spin configurations obtained from Monte Carlo simulations at a low temperature for $L=10$ and $J_{z\pm} = 1/\sqrt{2}$. 
Two ``domains'' are visible, one close to the equator $\theta=\pi/2$ ($A_2\oplus E$ manifold) and one close to the north pole $\theta=0$ ($T_{1-}$ irrep). The angle $\phi$ in the $A_2\oplus E$ domain is centered around the value predicted by classical low-temperature expansion (white dashed line). 
}
    \label{fig:jzpm-0.7-nem-dom}
\end{figure}

In Sec.~\ref{subsec:GS-nematic-characterization}, the ground-state manifold of the nematic phase is described in terms of irreps, where it was found that the high-entropy  $A_2\oplus\psi_2$ single-tetrahedron configurations transform into single-tetrahedron $T_{1-}$ configurations by means of the additional subsystem symmetries. Tiling the lattice with these single-tetrahedron configuration leads to a multitude of full-lattice ground-state spin configurations, which do not exhibit magnetic long-range order. We now test the robustness of this construction at finite temperatures by inspecting the irrep-magnitude distributions of the $A_2$, $E$, and $T_{1-}$ in the $T<T_\mathrm{c}$ temperature regime obtained from Monte Carlo simulations. 
Using the mixing angles $\theta$ and $\phi$ defined in Eq.~\eqref{eq:A2ET1_mixing_angles}, we identify two ``domains'' of spin configurations on the $A_2\oplus E \oplus T_{1-}$ sphere at low temperatures, see Fig.~\ref{fig:jzpm-0.7-nem-dom}. 
The region close to the equator ($\theta>\pi/4$) in the distribution shown in Fig.~\ref{fig:jzpm-0.7-nem-dom}(b) corresponds to single tetrahedra whose spin configurations are mainly described by the $A_2\oplus E$ configurations. In the irrep-magnitude distribution, the white dashed line marks the mixing angle $\phi$ between the $A_2$ and the $E$ irreps predicted by a classical low-temperature expansion, see Sec.~\ref{sec:CLTE}. We note that the domain with a high $A_2\oplus E$ irrep contribution is centered around the value predicted by the classical low-temperature expansion at $J_{z\pm}\lessgtr1/\sqrt{2}$, supporting the assumption that the construction of the ground-state manifold presented in the previous subsection is still valid at finite temperatures in the nematic phase. The second region where most single-tetrahedron configuration cluster is observed near the $\theta=0$ point. Spin configurations near this point are mainly described by a
$T_{1-}$ irreps, as underlined by the intense area around $\theta=0$ ($\theta<\pi/4$) in Fig.~\ref{fig:jzpm-0.7-nem-dom}(b). 
The $T_{1-}$ mode consists of six single-tetrahedron states, such as the one in Eqs.~\eqref{eq:nematic-Tminus-gs-1} and \eqref{eq:nematic-Tminus-gs-2}, related by $C_3$ and time reversal symmetry, and corresponding to the $ x$,$y$, and $ z$ components of the irrep. Depending on the symmetry-breaking pattern of the cubic symmetry, all the tetrahedra in the $T_{1-}$ mode select one out of the three $x$, $ y$, or $ z$ components of the irrep. Due to the additional subsystem symmetries at the nematic point, the $T_{1-}$ selected component matches the $A_2\oplus \psi_2$ mode, as described in Sec.~\ref{subsec:GS-nematic-characterization}. Hence, it is possible to tile a full lattice configuration consisting of tetrahedra in the $T_{1-}$ or $A_2\oplus\psi_2$ modes. For additional details on the $T_{1-}$ mode in the nematic phase, we refer to Appendix~\ref{appendix:nematic-t1minus-details}.

This irrep-magnitude distribution obtained for the nematic phase strongly differs from the one shown in Fig.~\ref{fig:jzpm-0.4-cMC}(g) for the $A_2\oplus \psi_2$ low-temperature phase, where all the irreps cluster close to the equator. In the case of the nematic phase, the presence of the $T_{1-}$ irrep in the distribution is characteristic for the nematic phase. It is directly related to the subsystem symmetries discussed in Sec.~\ref{subsec:exact-symmetry-nematic}. Due to these additional symmetries, the energetically degenerate $T_{1-}$ mode also becomes entropically equivalent to the $A_2\oplus \psi_2$ irrep, see Sec.~\ref{subsec:GS-nematic-characterization}. As a consequence, the $T_{1-}$ mode appears in the ground-state manifold, as observed in Fig.~\ref{fig:jzpm-0.7-nem-dom}(b).
%
%
%
%
%
\subsection{Classical Monte Carlo simulations slightly away from \texorpdfstring{$J_{z\pm}=\pm1/\sqrt{2}$}{Jzpm=1/sqrt(2)}}
\begin{figure}[ht!]
    \centering
    \begin{overpic}[width=\columnwidth]{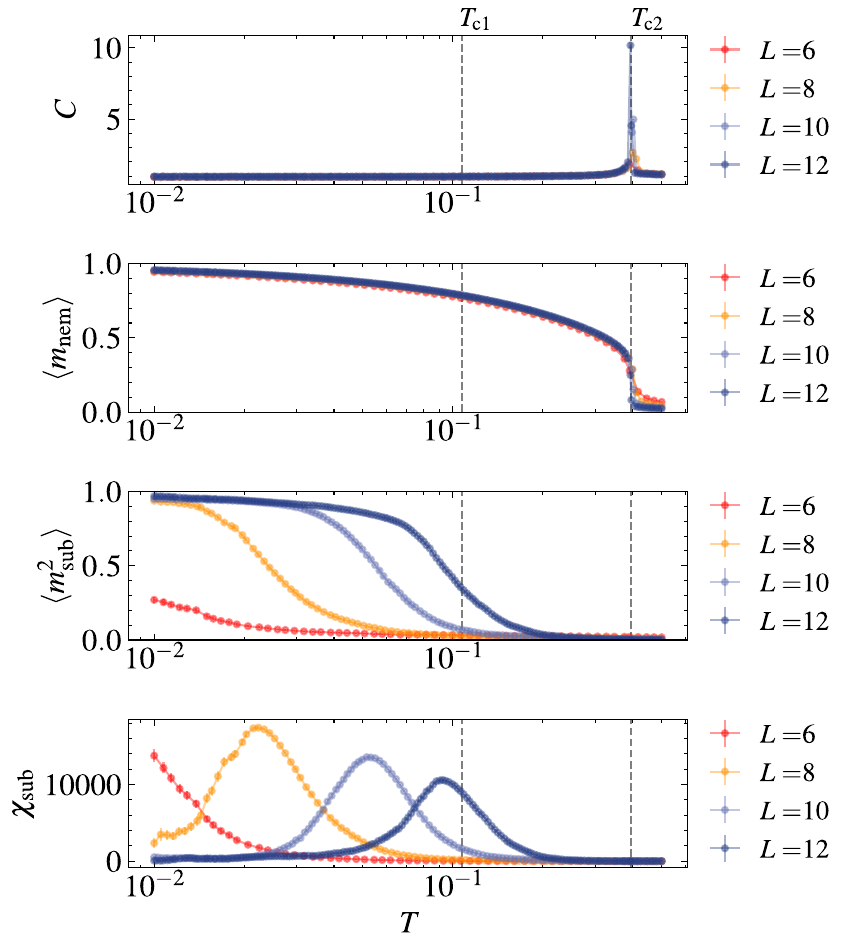}
    \put(0,95){(a)}
    \put(0,71){(b)}
    \put(0,47){(c)}
    \put(0,23){(d)}
    \put(20,97.5){$A_2\oplus\psi_2$}
    \put(57,97.5){SN}
    \put(72,97.5){PM}
    \end{overpic}
\caption{Monte Carlo results slightly away from the nematic point, at $J_{z\pm} = 0.74$.
(a)~Specific heat,
(b)~nematic order parameter,
(c)~sublattice magnetic order parameter, and
(d)~sublattice magnetic order parameter susceptibility as functions of the temperature $T$ for different lattice sizes $L$.
The dashed line at $T_{\rm c}$ indicates the position of the first-order transition into the nematic phase. The dashed line marked by $T_{\mathrm{c1}}$ highlights the crossing temperature into the $\mathbf{q}=0$ $A_2\oplus\psi_2$ phase. Strong finite-size effects modify the position of this temperature. Here we show our finite-size extrapolation using system sizes up to $L=12$.
}
\label{fig:jzpm-0.74}
\end{figure}

We conclude this section by presenting Monte Carlo results slightly away from the nematic point $J_{z\pm}=1/\sqrt{2}$. 
As a representative example, we show the data for $J_{z\pm}=0.74$, but the same qualitative behavior is observed in a range of values $0.5 \lesssim J_{z\pm} \lesssim 0.9$.
The behavior of the system is summarized in Fig.~\ref{fig:jzpm-0.74}. The specific heat [Fig.~\ref{fig:jzpm-0.74}(a)] shows a sharp peak that can be associated with a first-order transition. At the transition temperature $T_{\rm c2}$, marked by a dashed line, the magnitude of the nematic order parameter exhibits a jump, indicative of a first-order transition, as shown in Fig.~\ref{fig:jzpm-0.74}(b). Furthermore, the corresponding Binder cumulant and order-parameter histograms (not shown) near $T_{\rm c2}$ behave similarly to the case $J_{z\pm} = 1/\sqrt{2}$, corroborating the first-order nature of the transition.
Below $T_{\rm c2}$, both the specific heat and the nematic order parameter appear to evolve continuously toward a unit value as the temperature decreases.
However, in contrast to the situation at the exact nematic point [cf.~Fig.~\ref{fig:jzpm-0.7-cMC}(d)], the sublattice magnetic order parameter for $J_{z\pm} \neq 1/\sqrt{2}$, shown in Fig.~\ref{fig:jzpm-0.74}(c), starts to increase below a certain temperature, marking the onset of a $\mathbf{q}=0$ magnetic long-range-ordered phase. This low-temperature ordered phase corresponds to the mixed $A_2 \oplus \psi_2$ phase predicted by classical low-temperature expansion and is included in the antiferromagnetic sector of the ground-state manifold of the nematic phase.
We note, however, that the transition temperature $T_{\mathrm c1}$ at which the system crosses from the nematic phase to the long-range order  $A_2 \oplus \psi_2$ is strongly renormalized by size effects: as the system size increases, the temperature at which the sublattice magnetic order parameter becomes nonzero also increases significantly.
This pronounced finite-size effect may originate from the abundance of states that are nearly degenerate in both energy and entropy.
Here, it complicates the determination of the nature of the transition between the $A_2 \oplus \psi_2$ order at low temperatures and the spin-nematic order at intermediate temperatures from the numerical data.
However, the analytical solution at $J_{z\pm} = 1/\sqrt{2}$ allows us to infer that at least one phase transition must occur between the nematic phase and the $A_2 \oplus \psi_2$ magnetic phase somewhere in the phase diagram.
This follows from the fact that time-reversal symmetry remains unbroken above the exact nematic point at all temperatures, whereas the numerical data clearly indicate a time-reversal-symmetry-broken state for $J_{z\pm} \gtrsim 0.9$ at sufficiently low temperatures.
We therefore interpret the features observed in the sublattice order parameter at low temperatures as evidence of a true phase transition, even though, based on the numerics alone, they cannot be unambiguously distinguished from a crossover.

To estimate the transition temperature, we identify the local maximum of the sublattice-order-parameter susceptibility for different system sizes, shown in Fig.~\ref{fig:jzpm-0.74}(d). Extrapolation to the thermodynamic limit then provides an estimate for $T_{\mathrm{c1}}$, indicated by the dashed line in Fig.~\ref{fig:jzpm-0.74}.
This estimate suggests the existence of a finite temperature window in which the nematic phase is realized. A more precise determination of $T_{\mathrm{c1}}$ would require a more extensive study, including simulations of larger system sizes.
%
%
%
\section{Beyond the triple line}
\label{sec:beyond_triple}
So far, we have discussed the different ordered phases realized along the $A_2 \oplus E \oplus T_{1-}$ triple line. This analysis has revealed two novel phases: the $A_2 \oplus \psi_2$ mixed phase and the nematic phase. To assess the stability of these phases beyond fine-tuned parameter sets, we now perform classical Monte Carlo simulations in the vicinity of the $A_2 \oplus E \oplus T_{1-}$ line.

First, we study the $A_2\oplus E$ phase boundary. As discussed in Sec.~\ref{sec:Landau-theory}, the Landau theory in Eq.~\eqref{eq:landau} also describes a spin Hamiltonian whose interaction parameters fall on the $A_2\oplus E$ phase boundary, therefore identifying a more general set of spin models where the mixed $A_2\oplus \psi_2$ phase may be realized.
To study the $A_2\oplus E$ phase boundary numerically, we use the following parametrization
\begin{equation}
    J_{\pm}=-J_{zz}/2\,, \qquad J_{\pm\pm}=\frac{J_{zz}^2-J_{z\pm}^2}{J_{zz}}+\delta|J_{zz}|\,, \qquad J_{zz}<0\,, \label{eq:A2_E_plane} 
\end{equation}
where $\delta$ quantifies the deviation from the $A_2 \oplus E \oplus T_{1-}$ line toward the $A_2 \oplus E$ phase boundary.
We note that the value of $\delta$ is capped as a function of $J_{z\pm}$ to a maximum value where the resulting parametrization would describe a model where the $A_2$, $E$, and $T_2$ irreps are degenerate.
We have carried out Monte Carlo simulations for $J_{z\pm}=0.4$ and $J_{z\pm}=1.0$, both at $\delta=0.1$, finding behavior very similar to that shown in Figs.~\ref{fig:jzpm-0.4-cMC} and \ref{fig:jzpm-1.2-cMC} for the case $\delta=0$. This numerical analysis, together with the mean-field results presented in Sec.~\ref{sec:Landau-theory}, identifies the $A_2 \oplus E$ phase boundary as another region in parameter space where the $A_2 \oplus \psi_2$ phase is realized at low temperatures.

\begin{figure}
    \centering
    \begin{overpic}[width=\columnwidth]{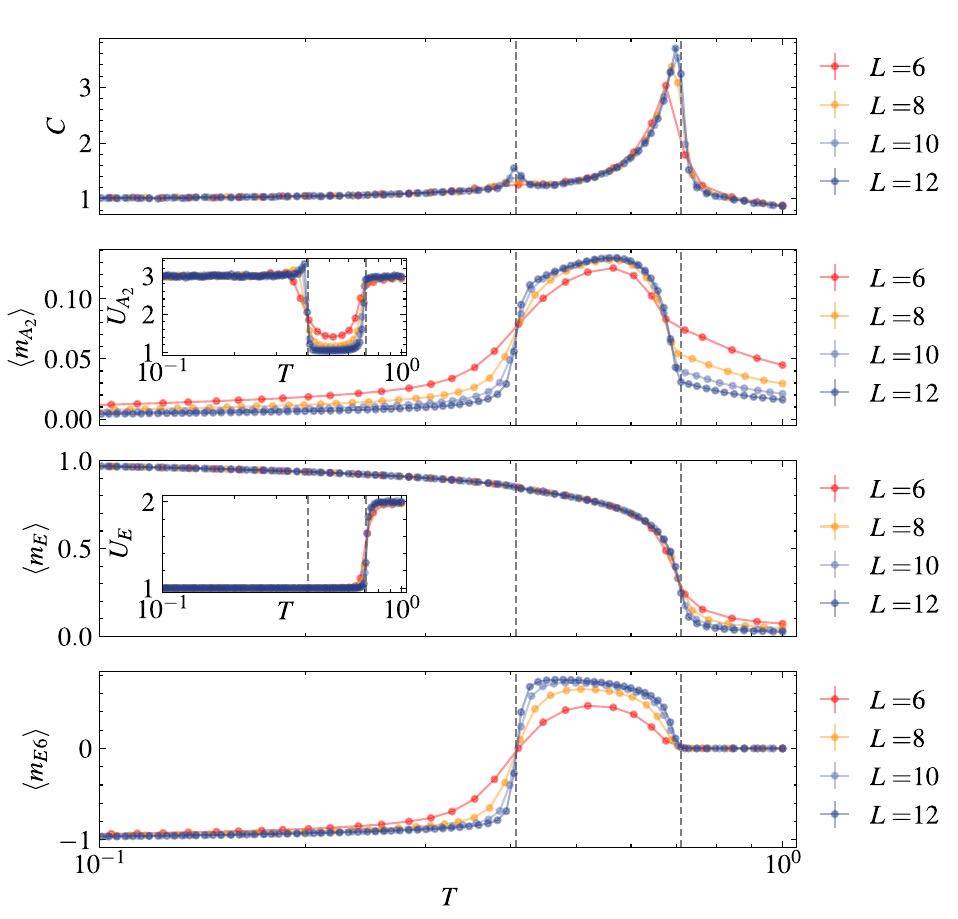}
    \put(0,95){(a)}
    \put(0,71){(b)}
    \put(0,47){(c)}
    \put(0,23){(d)}
    \put(20,93){$\psi_3$}
    \put(56,93){$A_2\oplus \psi_2$}
    \put(75,93){PM}
    \end{overpic}
\caption{%
Monte Carlo results for generic values of the couplings in the vicinity of the triple line, at $(J_{zz},J_\pm , J_{\pm\pm},J_{z\pm})=(-1,0.55,-0.84,0.4)$.
(a)~Specific heat,
(b)~$A_2$,
(c)~$E$, and
(d) $\langle \mathrm{m}_{E6}\rangle$ order parameters as functions of the temperature $T$ for different lattice sizes $L$.
Insets of panels (b) and (c) show the Binder cumulants of the corresponding order parameter.
The dashed lines separate the three distinct phases, namely the paramagnetic (PM), $A_2\oplus\psi_2$, and $\psi_3$ phases.
}
\label{fig:away_from_triple_line}
\end{figure} 

We now consider interaction parameters in the vicinity of the triple line, but in regimes where only one irrep has the lowest energy.
Figure~\ref{fig:away_from_triple_line} shows Monte Carlo results for a representative parameter set $(J_{zz}, J_\pm, J_{\pm\pm}, J_{z\pm}) = (-1, 0.55, -0.84, 0.4)$, for which the $E$ irrep has the lowest energy. These results indicate that the $A_2 \oplus \psi_2$ phase is stabilized in a finite temperature window above the low-temperature $E$ phase.
Specifically, the simulations identify three phases as signaled by the specific heat [see Fig.~\ref{fig:away_from_triple_line}(a)]~\footnote{Interestingly, and to the best of our knowledge, this is the first instance where a classical Monte Carlo simulation captures a peak in the specific heat for a transition between a phase with $\psi_2$ order ($A_2\oplus\psi_2$ in this case) and a phase with $\psi_3$ order.},
a paramagnetic phase at high temperatures followed by two magnetically ordered phases at intermediate and low temperatures. 
The symmetry-broken phase at intermediate temperatures corresponds to an $A_2\oplus \psi_2$ mixed phase, as seen from
(i)~the non-vanishing $\langle m_{A_2}\rangle$ and $\langle m_{E}\rangle$ parameters in the intermediate-temperature window [see Figs.~\ref{fig:away_from_triple_line}(b) and \ref{fig:away_from_triple_line}(c)], and 
(ii)~the evolution of the $\langle m_{E6} \rangle$ parameter, which acquires positive values in the intermediate phase, signaling $\psi_2$ order [see Fig.~\ref{fig:away_from_triple_line}(d)].
As expected, $\psi_3$ order is stabilized at low temperatures, as indicated by $\langle m_E \rangle$ approaching its saturation value and $\langle m_{E6} \rangle$ becoming negative~\cite{Noculak_HDM_2023}.
We note that the transition between the $A_2\oplus \psi_2$ and $\psi_3$ phases appears to be weakly first order, based on the slowly-developing peaks in the $A_2$ Binder cumulant, typical of a first-order transition, and in line with the expectations from a Landau-theory perspective~\cite{Javanparast2015}.
The histogram of the $A_2$-order-parameter magnitude also develops a double-peak structure very slowly with the system sizes.

For completeness, we have also investigated whether the nematic phase discussed in Sec.~\ref{sec:SN-phase} can be realized for more general interaction parameters. To this end, we have simulated spin models in the immediate vicinity of the exact nematic point, but away from the $A_2 \oplus E \oplus T_{1-}$ triple line. However, the system sizes currently accessible are insufficient to draw conclusive statements about the finite-temperature phase diagram in the thermodynamic limit. 
It is worth noting that the stability region of the nematic phase may be enhanced by quantum fluctuations~\cite{Gresista_QPLSL}. We leave this possibility for future work.
%
%
%
\section{Discussion and conclusion}
\label{sec:discussion-conclusion}

In this work, we consider a specific realization of the most general bilinear nearest-neighbor spin model on the pyrochlore lattice. The model is parametrized by the interaction coupling $J_{z\pm}$, which tunes along the line in parameter space where a ground-state energy degeneracy among the $A_2$, $E$, and $T_{1-}$ states occurs. Using large-scale classical Monte Carlo simulations, we show that this model exhibits a rich phase diagram, featuring two distinct $\mathbf{q}=0$ magnetic long-range-ordered phases and a novel nematic phase emerging near $|J_{z\pm}| = 1/\sqrt{2}$. 
For $|J_{z\pm}|\lesssim 0.5$, classical Monte Carlo simulations reveal two distinct $\mathbf{q}=0$ phases; a intermediate-temperature  $A_2$ phase and a low-temperatures \emph{mixed} $A_2\oplus \psi_2$ phase. In the mixed phase, the order parameters of both the $A_2$ and $E$ are non-vanishing and the precise single-tetrahedron spin configuration is parametrized by a mixing angle $\phi(J_{z\pm})$ between the two orders. In contrast, for $|J_{z\pm}|\gtrsim 0.9$, a $A_2\oplus \psi_2$  mixed $\mathbf{q}=0$ phase is realized already right below the high-temperature paramagnetic phase and remains stable down to the lowest temperatures considered. The coexistence of the $A_2$ and $E$ ordered phases in the mixed phase is possible only because of their ground-state energy degeneracy, which arises from the specific model considered, and the compatibility of the order parameters describing these phases~\cite{Javanparast2015}.
Using a classical-low temperature expansion, we demonstrate that the selection of the \emph{mixed} $A_2\oplus\psi_2$ phase is a consequence of a thermal order-by-disorder phenomenon~\cite{villain1980,Hickey_2024arxiv,NoculakPRB2024,zhitomirsky2012}, where the mixing angle $\phi(J_{z\pm})$ parametrizes the configurations in the 2-sphere describing the $A_2\oplus E$ set of states with the highest entropy contribution. Interestingly, we show that the evolution of the order-by-disorder-selected state as a function of the coupling $J_{z\pm}$ presents a discontinuity at $J_{z\pm}\sim 1.262$, where the spin orientation of the state with the highest entropy abruptly changes. 
We further show that the rich behavior of this model, including  the $\mathbf{q}=0$ phases realized, the qualitative behavior of the order parameters, and the order of the symmetry-breaking transition between these phases, is qualitatively captured by a Landau theory constructed using the order parameters of the $A_2$ and $E$ phases. 
Although the realization of the $A_2\oplus \psi_2$ mixed $\mathbf{q}=0$ phase falls within the traditional irrep framework used to describe the low-temperature phases of the general bilinear nearest-neighbor pyrochlore Hamiltonian~\cite{Wong2013,Yan2017_theory_multiphase}, it extends the conventional single-irrep classification by demonstrating that a stable ordered phase can exhibit coexisting $A_2$ and $\psi_2$ order parameters down to the lowest temperatures.
We note that a similar experimental proposal for a mixed state was provided in Ref.~\cite{Scheie_YbTiO} for $\rm Yb_2Ti_2O_7$, a compound which lies in close proximity to the $E$-$T_{1-}$ boundary. In this case however, classical Monte Carlo simulations of the classical model reveal either an $E$ phase or a $T_{1-}$ phase~\cite{Scheie_dynamics_YbTiO}.

For $J_{z\pm}$ in the vicinity of $J_{z\pm}= 1/\sqrt{2}$, our classical Monte Carlo simulations detect a nematic phase separated by a first-order phase transition from the paramagnetic phase. At $J_{z\pm} = 1/\sqrt{2}$, the model  acquires additional $\mathbb{Z}_2$ subsystem symmetries, where the sign of a cartesian component of the spins in the global basis can be flipped along spin chains which wrap around the system. In the nematic phase, the single-tetrahedron spin configurations are described by either a mixed $A_2\oplus \psi_2$, or a $T_{1-}$ spin configuration.
We demonstrate that these subsystem symmetries result in a subextensive degeneracy in the ground-state manifold and prevent the formation of any type of dipolar long-range order phase down to the lowest temperatures. We show that the stability of this nematic phase is not limited to the $J_{z\pm}=\pm1/\sqrt{2}$ points, but presents a cone of stability extending both in temperature and parameter space. Furthermore, we have identified two additional models which are dual to the nematic Hamiltonians, and therefore present equivalent thermodynamic signatures where a low-temperature nematic phase is realized. We note that the ground-state manifold of the nematic phase identified in this work can be mapped into the so-called densely-packed monopole manifold, realized in a four-body-term Ising Hamiltonian~\cite{Szabo_hidden,Slobinsky_monopole_liquid}. Indeed, the four-body-term Hamiltonian studied in Refs.~\cite{Szabo_hidden,Slobinsky_monopole_liquid} presents a similar subsystem symmetry as the one realized for the nematic model.
We have also shown that the novel mixed $A_2\oplus\psi_2$ phase is realized at low temperatures for interaction parameter along the A2+E+T1- line and in the  $A_2\oplus E$ boundary. Additionally, we demonstrated that this mixed phase is also observed at intermediate temperature windows in regions in parameter space where the E phase is the ground-state phase. The enlarged region of stability of this magnetic order makes its characterization particularly relevant for compounds lying in the vicinity of this model, such as $\mathrm{FeF}_3$~\cite{sadeghiSpinHamiltonianOrder2015}. Although, to our knowledge, no material has been identified that realizes the phases discussed here exactly, recent studies have shown that the application of chemical pressure~\cite{Pressure_Zhang} or hydrostatic pressure~\cite{Kermarrec2017_pressure} can induce a drift in the effective interaction parameters. In this context, such external control parameters may provide a route to tune pyrochlore compounds toward the regime described by our model.

Altogether, our work elucidates the rich behavior resulting from strong competition between distinct magnetic phases, where phenomena like thermal order-by-disorder and nematic phases can be realized in the pyrochlore lattice. It would be of great interest to study how quantum fluctuation may modify the observed phenomena. The introduction of these effects may result in the stabilization of distinct phases at higher temperatures~\cite{Hickey_2024arxiv}, and in the more extreme cases even prevent the formation of long-range order~\cite{Gresista_QPLSL}. It is possible that a similar selection phenomenon associated with the connectivity of the ground-state configurations yields a mixed $A_2\oplus \psi_2$ phase. Whether the selection of such phase stems from a quantum order-by-disorder~\cite{Canals_HDM_2008,Noculak_HDM_2023} or an order-by-singularity~\cite{ObS_1,ObS_2} mechanism would need to be clarified. Lastly, we expect that no conventional magnetic order is observed for the nematic points, where the additional subsystem symmetries remain valid symmetries in the quantum model.
%

\begin{acknowledgments}
We thank Ludovic Jaubert, Matthias Vojta, Subhankar Khatua, Kristian Tyn Kai Chung, Michel J. P. Gingras, and Jeff Rau for insightful discussions.
This work has been supported by the Deutsche Forschungsgemeinschaft (DFG) through 
Project No.\ 247310070 (SFB 1143, A01 \& A07), 
Project No.\ 390858490 (W\"urzburg-Dresden Cluster of Excellence \textit{ctd.qmat}, EXC 2147), and 
Project No.\ 411750675 (Emmy Noether program, JA2306/4-1).
D.L.-G. acknowledge financial support from the DFG through the W\"urzburg-Dresden Cluster of Excellence on Complexity,
Topology and Dynamics in Quantum Matter -- \textit{ctd.qmat} (EXC 2147, project-id 390858490) and
through SFB 1143 (project-id 247310070). D.L.-G. is supported by the Hallwachs-R\"ontgen Postdoc Program of \textit{ctd.qmat}.
The authors gratefully acknowledge the computing time made available to them on the high-performance computer Barnard at the NHR Center of TU Dresden. This center is jointly supported by the Federal Ministry of Education and Research and the state governments participating in the National High-Performance Computing (NHR) joint funding program~\cite{nhr-alliance}.

\end{acknowledgments}

\section*{Data availability}
%
The data that support the findings of this article are openly
available~\cite{data-availability}.
%

\appendix

\section{Coordinate frame conventions and irreducible representations}
\label{appendix:irreps-basis}
In this appendix, we provide details on the basis conventions employed in the main text. Moreover, we explicitly report the single-tetrahedron irrep in terms of spin components.
Each single-tetrahedron vertex is associated with a right-handed local cubic coordinate system defined as
\begin{equation}
\begin{split}
       \mathbf{x}_0 &=\frac{-\mathbf{e}_x-\mathbf{e}_y+2\mathbf{e}_z}{\sqrt{6}}\,,\quad \mathbf{y}_0=\frac{\mathbf{e}_x-\mathbf{e}_y}{\sqrt{2}}\,,\quad \mathbf{z}_0=\frac{\mathbf{e}_x+\mathbf{e}_y+\mathbf{e}_z}{\sqrt{3}}\,, \\
       \mathbf{x}_1 &=\frac{\mathbf{e}_x+\mathbf{e}_y+2\mathbf{e}_z}{\sqrt{6}}\,,\quad \mathbf{y}_1=\frac{-\mathbf{e}_x+\mathbf{e}_y}{\sqrt{2}}\,,\quad \mathbf{z}_1=\frac{-\mathbf{e}_x-\mathbf{e}_y+\mathbf{e}_z}{\sqrt{3}}\,, \\
       \mathbf{x}_2 &=\frac{\mathbf{e}_x-\mathbf{e}_y-2\mathbf{e}_z}{\sqrt{6}}\,,\quad \mathbf{y}_2=\frac{-\mathbf{e}_x-\mathbf{e}_y}{\sqrt{2}}\,,\quad \mathbf{z}_2=\frac{-\mathbf{e}_x+\mathbf{e}_y-\mathbf{e}_z}{\sqrt{3}}\,, \\
       \mathbf{x}_3 &=\frac{-\mathbf{e}_x+\mathbf{e}_y-2\mathbf{e}_z}{\sqrt{6}}\,,\quad \mathbf{y}_3=\frac{\mathbf{e}_x+\mathbf{e}_y}{\sqrt{2}}\,,\quad \mathbf{z}_3=\frac{\mathbf{e}_x-\mathbf{e}_y-\mathbf{e}_z}{\sqrt{3}}\,, \\
\end{split}
    \label{eq:local-coordinate-frame}
\end{equation}
where $\mathbf{e}_x$, $\mathbf{e}_y$ and $\mathbf{e}_z$ define a global cubic coordinate frame
\begin{equation}
\label{eq:global-coordinate-frame}
    \mathbf{e}_x=(1,0,0)\,\quad \mathbf{e}_y = (0,1,0)\,,\quad \mathbf{e}_z=(0,0,1)\,.
\end{equation}
Each spin on the $\mu$ sublattice of the pyrochlore lattice can be expressed as
\begin{equation}
\label{eq:global-coordinate-components}
    \mathbf{S}_\mu = \sum_{\alpha=x,y,z} s_\mu^\alpha \mathbf{e}_\alpha \quad \mathrm{with} \quad s_\mu^\alpha = \mathbf{S}_\mu\cdot \mathbf{e}_\alpha\,,
\end{equation}
where $s^\alpha_\mu$ denotes the spin components in the global coordinate frame. Equivalently, in the local basis, spins are decomposed as
\begin{equation}
    \label{eq:local-coordinate-components}
    \begin{split}
        \mathbf{S}_\mu &= S_\mu^x \mathbf{x}_\mu + S_\mu^y\mathbf{y}_{\mu} + S_\mu^z\mathbf{z}_\mu \\
        &=(\mathbf{S_\mu \cdot \mathbf{x}_\mu}) \mathbf{x}_\mu + (\mathbf{S_\mu \cdot \mathbf{y}_\mu})\mathbf{y}_{\mu} + (\mathbf{S_\mu \cdot \mathbf{z}_\mu})\mathbf{z}_\mu\,,
    \end{split}
\end{equation}
where $\mathbf{S}_\mu$ font is related to the spin components in the local basis notation, since it is mostly used throughout the paper. For example, the Hamiltonian in Eq.~\eqref{eq:general_bilinear_hamiltonian} contains spin components expressed in the local coordinate frame. However, the same Hamiltonian can be expressed in the global basis, with interaction parameters $\{J_1,J_2,J_3,J_4\}$, or in a coordinate-free notation with couplings $\{J_{\mathrm{H}},J_{\mathrm{Is}},J_{\mathrm{DM}},J_{\mathrm{PD}}\}$~\cite{KTC_2024_phase}. 

The Hamiltonian in Eq.~\eqref{eq:general_bilinear_hamiltonian} is reformulated as a sum of single-tetrahedron Hamiltionians $\mathcal{H}_\boxtimes$
\begin{equation*}
    \mathcal{H}=\sum_{\boxtimes}\mathcal{H}_\boxtimes\,,
\end{equation*}
where the sum runs on every tetrahedron on the pyrochlore lattice. The single-tetrahedron Hamiltonian can be formulated in terms of single-tetrahedron irreps $\mathbf{m}_I$ and their associated energy $a_I$,
\begin{equation}
\begin{split}
    \mathcal{H}_\boxtimes &=\frac{1}{2} \left(a_{A_2}m_{A_2}^2+ a_{E}\mathbf{m}_E^2+ a_{T_2}\mathbf{m}_{T_2}^2 + a_{T_{1,\mathrm{ice}}}\mathbf{m}_{T_{1,\mathrm{ice}}}^2 \right. \\ 
    &+\left. a_{T_{1,\mathrm{planar}}}\mathbf{m}_{T_{1,\mathrm{planar}}}^2 + a_{T_{1,\mathrm{mix}}} \mathbf{m}_{T_{1,\mathrm{ice}}}\cdot\mathbf{m}_{T_{1,\mathrm{planar}}} \right)\,,
\end{split}
\end{equation}
where the $a_{T_{1,\mathrm{mix}}}=-4J_{z\pm}$ is the energy associated to the mixing term between the $T_{1,\mathrm{ice}}$ and $T_{1,\mathrm{planar}}$ irreps. The irreps are linear combinations of the single-tetrahedron spin components. In Table~\ref {tab:irreps-def}, each irrep is reported together with its energy and its expression in terms of spin components in the local basis. The mixed $T_1$ irreps can be rotated into two new irreps $T_{1\pm}$ to suppress any mixing terms between the resulting $T_1$ irreps, namely
\begin{equation}
\label{eq:T1-rotation}
    \begin{pmatrix}
        \mathbf{m}_{T_{1-}} \\ \mathbf{m}_{T_{1+}}
    \end{pmatrix}
    =
    \begin{pmatrix}
        \cos{\alpha_{\mathrm{mix}}} & \sin{\alpha_{\mathrm{mix}}} \\
        -\sin{\alpha_{\mathrm{mix}}} & \cos{\alpha_{\mathrm{mix}}} \\
    \end{pmatrix}
    \begin{pmatrix}
        \mathbf{m}_{T_{1,\mathrm{ice}}} \\ \mathbf{m}_{T_{1,\mathrm{planar}}}
    \end{pmatrix},
\end{equation}
where the rotation angle is obtained by
\begin{equation}
    \label{eq:T1plusminus-mixing-angle}
    \tan2\alpha_{\mathrm{mix}}=\frac{2a_{\mathrm{mix}}}{a_{T_{1,\mathrm{ice}}}-a_{T_{1,\mathrm{planar}}}}=\frac{8J_{z\pm}}{J_{zz}+2J_{\pm}+4J_{\pm\pm}}\,.
\end{equation}
The rotated irreps are associated with energies 
\begin{equation}
    \label{eq:T1PM-energies}
    a_{T_{1\pm}} = \frac{1}{2}\left( a_{T_{1,\mathrm{ice}}}+a_{T_{1,\mathrm{planar}}} \pm \sqrt{(a_{T_{1,\mathrm{ice}}}-a_{T_{1,\mathrm{planar}}})^2 + a_{\mathrm{mix}}^2} \right)\,,
\end{equation}
where the $T_{1-}$ irrep is defined as the one always having the lower energy $a_{T_{1-}}<a_{T_{1+}}$.
In the non-Kramers case $J_{z\pm}=0$, the local $z$ component of the spins decouples from the local transverse components, reducing the $T_{1\pm}$ irreps to the spin-ice and planar, $T_{1,\mathrm{planar}}$ irreps. The $T_{1,\mathrm{ice}}$ spin-ice irrep consists of spins lying along the local $z$ direction and pointing either inward or outward the tetrahedron following the two-in-two-out rule, or ice rule. On the other hand, the $T_{1,planar}$ irrep is a splayed ferromagnetic configuration where the spins in every sublattice lie on their corresponding local $xy$ plane.~\cite{KTC_2024_phase}.
\begin{table*}[t]
\caption{Single-tetrahedron irrep and their typical naming in terms of spin components in the local coordinate frame.
}
    \centering
    \setlength{\extrarowheight}{2pt}
    \begin{tabularx}{\linewidth}{>{\hsize=.16\hsize\centering\arraybackslash}X >{\hsize=.16\hsize\centering\arraybackslash}X >{\hsize=.18\hsize\centering\arraybackslash}X >{\hsize=.50\hsize\centering\arraybackslash}X }
    \hline 
    \hline
        Irreps & Order & Irreps energy $a_I$ & Definition in terms of local spin components \\[1pt]
        \hline
        $m_{A_2}$ & All-in-all-out & $3J_{zz}$ &  $\frac{1}{2}(S_0^z + S_1^z +S_2^z + S_3^z)$ \\ \\
        $\mathbf{m}_E$ & $\Gamma_5$ & $-6J_{\pm}$ & $\frac{1}{2}\begin{pmatrix}
            S_0^x + S_1^x +S_2^x + S_3^x \\ 
            S_0^y + S_1^y +S_2^y + S_3^y
        \end{pmatrix}$ \\ \\
        $\mathbf{m}_{T_{2}}$ & Palmer-Chalker & $2J_{\pm}-4J_{\pm\pm}$ & $\frac{1}{4}\begin{pmatrix}
            S_0^y + S_1^y - S_2^y - S_3^y \\ 
            -\sqrt{3}S_0^x - S_0^y + \sqrt{3}S_1^x + S_0^y -\sqrt{3}S_2^x  -S_1^y +\sqrt{3}S_3^x + S_3^y \\
            \sqrt{3}S_0^x - S_0^y - \sqrt{3}S_1^x + S_0^y -\sqrt{3}S_2^x + S_1^y +\sqrt{3}S_3^x - S_3^y 
        \end{pmatrix}$ \\ \\
        $\mathbf{m}_{T_{1,\mathrm{ice}}}$ & Spin Ice & $-J_{zz}$ & $\frac{1}{2}\begin{pmatrix}
            S_0^z + S_1^z -S_2^z - S_3^z \\
            S_0^z - S_1^z +S_2^z - S_3^z \\
            S_0^z - S_1^z -S_2^z + S_3^z
        \end{pmatrix}$ \\ \\
        $\mathbf{m}_{T_{1,\mathrm{planar}}}$ & Splayed Ferromagnet & $2J_{\pm}+4J_{\pm\pm}$ & $\frac{1}{4}\begin{pmatrix}
            2S_0^x + 2S_1^x -2S_2^x - 2S_3^x \\
            -S_0^x + \sqrt{3}S_0^y + S_1^x - \sqrt{3} S_0^y -S_2^x  +\sqrt{3}S_1^y +S_3^x - \sqrt{3}S_3^y \\
            -S_0^x - \sqrt{3}S_0^y + S_1^x + \sqrt{3} S_0^y +S_2^x  +\sqrt{3}S_1^y -S_3^x - \sqrt{3}S_3^y
        \end{pmatrix}$ \\[2.25\baselineskip]
        \hline
        \hline
    \end{tabularx}
    \label{tab:irreps-def}
\end{table*}
%

\section{Identification of accidental symmetry manifolds}
\label{appendix:accidental_symm}
In this appendix, we provide details on the identification of the accidental symmetry manifolds of Sec.~\ref{subsec:mixing-angles}. Each spin in a tetrahedron can be expressed as a function of irreps
\begin{equation*}
    \mathbf{S}_i =\mathbf{S}_i(m_{A_2},\mathbf{m}_E,\mathbf{m}_{T_{1-}})
\end{equation*}
or, equivalently, as a function of the five angles in Eq.~\eqref{eq:angle-representation},
\begin{equation}
    \mathbf{S}_i =\mathbf{S}_i(\theta,\phi,\eta,\beta,\gamma)=\mathbf{S}_i(\boldsymbol{\pi})
\end{equation}
where the vector $\boldsymbol{\pi}=(\theta,\phi,\eta,\beta,\gamma)$ stands as a shorthand notation for the $S^5$ angles. Each spin in the tetrahedron must also respect the strong spin constraint
\begin{equation}
\label{eq:strong-spin-constraint}
    \mathbf{S}_i\cdot \mathbf{S}_i = S^2, \quad \forall i \in \boxtimes\,,
\end{equation}
leading to four different equations. However, only three of these equations are independent. In principle, these equations allow us to obtain the physically relevant configuration $\boldsymbol{\pi}_0$ respecting both the strong constraint in Eq.~\eqref{eq:strong-spin-constraint} and the  $A_2\oplus E \oplus T_{1-}$ constraint in Eq.~\eqref{eq:irreps-constraint}.

In case where the constraints are satisfied on entire submanifolds, i.e the $S^2$ submanifold described by the $A_2\oplus E$ subspace described in Sec.~\ref{subsec:mixing-angles}, the analysis can be extended by taking into account small fluctuations around some configuration $\boldsymbol{\pi}_0$, keeping also in mind that all the constraints still need to be satisfied. Expanding the tetrahedron spins as 
\begin{equation*}
    S_i^{\alpha}(\boldsymbol{\pi}_0+\delta\boldsymbol{\pi})\approx S_i^{\alpha}(\boldsymbol{\pi}_0)+\delta\boldsymbol{\pi}\cdot\nabla \left. S_i^{\alpha} \right|_{\boldsymbol{\pi}_0}+\mathcal{O}(\delta\boldsymbol{\pi}^2)\,,
\end{equation*}

with a gradient operator $\nabla=(\partial_\theta,\partial_\phi,\partial_\eta,\partial_\beta,\partial_\gamma)$, and imposing the strong spin constraint on $\mathbf{S}_i(\boldsymbol{\pi}_0+\delta\boldsymbol{\pi})$, we get four equations of the form 
\begin{equation}
    \delta\boldsymbol{\pi}\cdot \left.\nabla(\mathbf{S}_i\cdot \mathbf{S}_i)\right|_{\boldsymbol{\pi}_0}=0,
\end{equation}
where only three of them are independent~\cite{Benton2014_thesis}. For nonzero fluctuations, the system of equations in Eq.~\eqref{eq:fluctuation-equation} could have one or more solutions in $(\theta,\phi,\eta,\beta,\gamma)$, which correspond to a single-tetrahedron spin configuration.

\section{Stereographic projection}
\label{appendix:stereo-proj}
In this appendix, we present details on the stereographic projection used in the main text. A point on the unitary sphere $P=(x,y,z)$ is mapped to a point on the plane $P'=(X,Y)$ via stereographic projection
\begin{equation}
    \label{eq:stereo-proj}
    (X,Y)=\left(\frac{x}{1+z},\frac{y}{1+z}\right)\,,
\end{equation}
where the north pole $(0,0,1)$ of the sphere is mapped to the origin of the plane, the equator of the sphere $(x,y,0)$ to the unitary circle on the plane, and the south pole to infinity. The stereographic projection is a conformal mapping, which preserves angles from the sphere to the plane. However, it introduces some distortion in the radial direction on the plane, enlarging the objects close to the equator. 

The stereographic projection can be used to visualize the mixing between the degenerate irreps along the $A_2\oplus E\oplus T_{1-}$ line. Within the angle definitions in Eq.~\eqref{eq:angle-representation}, we can define an $S^2$ sphere parametrized by the angles $(\theta,\phi)$ via the absolute values of the three irreps. In this case, the sphere is reduced to the first octagon. Its north pole represents the $T_{1-}$ irreps, while the equator corresponds to the $A_2\oplus E$ mixed state, with special points of only $A_2$ and $E$ irreps, see Fig.~\ref{fig:jzpm-0.4-cMC} for an example of this mapping for an $A_2\oplus E$ phase.

The stereographic projection also helps the visualization of the degenerate $A_2\oplus E$ sphere and its U(1) orbits of soft modes. In this case, the sphere is spanned by the angles $(\phi,\eta)$ of Eq.~\eqref{eq:angle-representation}, and we restricted the representation of the sphere to the northern hemisphere to avoid infinities due to the southern hemisphere representation, see Fig.~\ref{fig:CLTE_map}(a), \ref{fig:CLTE_map}(c), \ref{fig:CLTE_map}(e), and \ref{fig:CLTE_map}(g). Nevertheless, time-reversal symmetry links states on the northern hemisphere to ones on the southern hemisphere, ensuring a complete equivalence between the two regions.

\section{Further details on the classical low-temperature expansion}
\label{appendix:CLTE}
\begin{figure}
    \centering
    \begin{overpic}[width=\columnwidth]{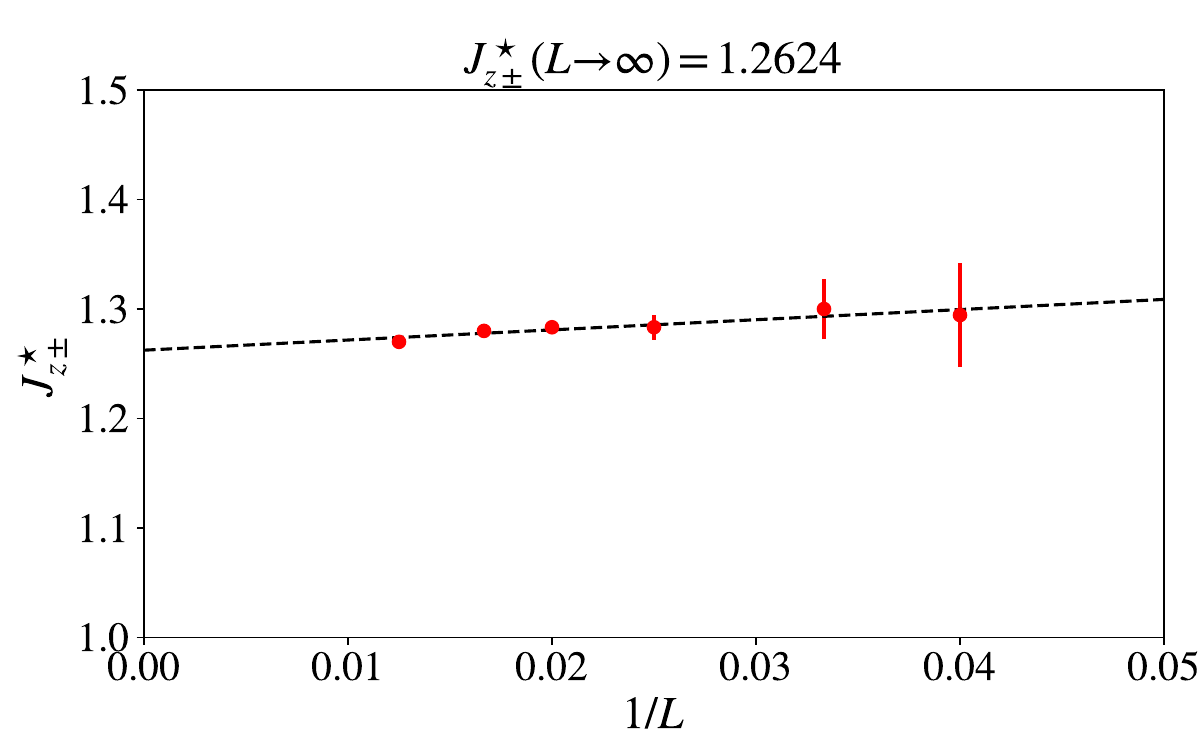}
    \end{overpic}
    \caption{ Evolution of the critical crossover $J_{z\pm}^\star$ value signaling the change in the maximal entropy state. In this figure, we show data for $L\in\{25,30,40,50,60,80\}$. The title of this figure provides the numerical value of the $J_{z\pm}^\star$ in the thermodynamic limit. }
    \label{fig:CLTE_Jzpm_star}
\end{figure}

In the main text, we discussed how the low-temperature selection of the $\mathbf{q}=0$ state evolves as a function of the coupling parameter $J_{z\pm}$. In particular, we identified two high-entropy branches in the $A_2\oplus E$ manifold, where the highest entropy branch changes at a value $J_{z\pm}^\star$. In this appendix, we numerically determine the value of $J_{z\pm}^\star$. In Fig.~\ref{fig:CLTE_Jzpm_star}, we illustrate the value of $J_{z\pm}^\star$ at which the two high-entropy branches have an equal entropy as a function of system size $L$ used to perform the calculation. A naive linear $1/L$ fit of the different computed system sizes predicts that the $J_{z\pm}^\star\simeq 1.2624$ in the thermodynamic limit.

\begin{figure}
    \centering
    \begin{overpic}[width=\columnwidth]{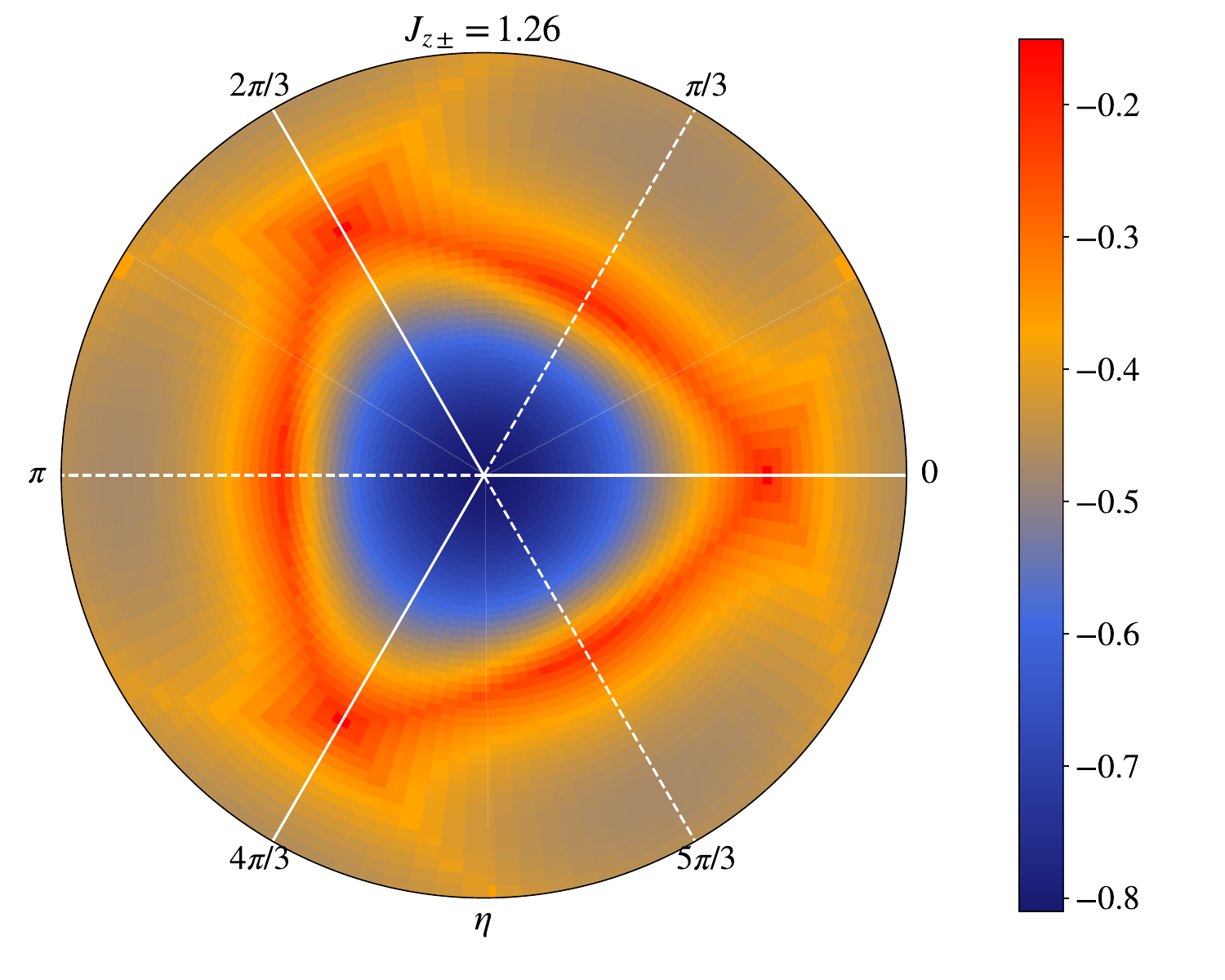}
    \end{overpic}
    \caption{Entropy on the $A_2\oplus E$ sphere measured at an interaction parameter $J_{z\pm}$ at which the two high-entropy points in the U(1) manifold are approximately degenerate in the free-energy. }
    \label{fig:CLTE_Jzpm_star_continuous}
\end{figure}

The error bars in Fig.~\ref{fig:CLTE_Jzpm_star} are obtained as an upper limit of the Riemann-sum integration error by the product of the maximal value obtained by the entropy by the integrated function in Eq.~\eqref{eq:CLTE-free-energy}, times the volume of the integration range divided by the number of $\mathbf{q}$ points used in a Riemann integration.  We note that, for interaction parameters near to $J_{z\pm}^\star$ an approximately uniform entropy is measured along a continuous path in the $A_2\oplus E$ sphere, see Fig.~\ref{fig:CLTE_Jzpm_star_continuous}.

Interestingly, at $J_{z\pm}^\star$, the angular difference between the two branches, i.e. $\phi_1-\phi_2$, is approximately $\pi/3$. Using the equations defining the angles $\phi_1$ and $\phi_2$, namely Eq.~\eqref{eq:A2E-two-soft-modes} and Eq.~\eqref{eq:A2E-second states}, and assuming that $\phi_1-\phi_2=\pi/3$ at $J_{z\pm}^\star$ we obtain the value
\begin{equation}
    \label{eq:A2E-critical-jzpm}
    J_{z\pm}^\star = \frac{1+\sqrt{1+\kappa^2}}{\kappa}\quad\mathrm{with}\quad \kappa=\tan\left(2\arctan\frac{\sqrt{3}+\sqrt{11}}{2}\right),
\end{equation}
yielding an approximate value of $J_{z\pm}^\star\simeq 1.26216$, which approximates the value obtained from the finite size analysis in Fig.~\ref{fig:CLTE_Jzpm_star}. Although the angle difference of $\pi/3$ suggests that certain symmetry would be the driving mechanism of the entropy degeneracy between these branches, we note that the symmetries of the pyrochlore include rotations by $2\pi/3$, not $\pi/3$. We leave the study of the mechanism behind this degeneracy as future work.

\section{Further details on the Landau theory}
\label{appendix:landau}
%
In this appendix, we provide further details on the Landau theory presented in Sec.~\ref{sec:cMC}. First, we identify all invariant terms and derive the free energy in Eq.~\eqref{eq:landau}. Next, we discuss the minimization of the free energy and visualize the selection of the global minimum at the phase transition from the $A_2$ to the $A_2 \oplus \psi_2$ phase. Finally, an additional analysis of the transition from the paramagnetic phase to the $A_2 \oplus \psi_2$ phase and phase diagrams, similar to Fig.~\ref{fig:LG}, for distinct sets of parameters is presented. 

To identify the symmetry-allowed terms of the free energy in Eq.~\eqref{eq:landau}, we first study the action of the symmetry transformations of the tetrahedral group $T_d$ on the irreps $m_{A_2}$ and $\mathbf{m}_{E}$, as shown in Table~\ref {tab:irreps-def}. The single-tetrahedron group $T_d$ exhibits four types of symmetries: a $C_3$ symmetry around the local $\langle 111 \rangle$ direction, $C_2$ rotations around the global $x$, $y$, and $z$ axes, reflections $\sigma$, and an $S_4$ symmetry consisting of reflections followed by a $\pi/2$ rotation \cite{Rau2018FrustratedQR}. In addition, the time reversal symmetry $\tau$ needs to be considered. 
\begin{table}[t]
\caption{Assumed factors of $m_{A_2}$ and $\mathbf{m}_E = (m_{E}^x,m_{E}^y)^\text{T}$ under  symmetry operations of $T_d$. $\mathbf{m}_E$ transforms according to Eq.~\eqref{eq:c3_transformation} under $C_3$.}
    \centering
    \setlength{\extrarowheight}{3pt}
    \begin{tabularx}{\columnwidth}{>{\hsize=.20\hsize\centering\arraybackslash}X >{\hsize=.13\hsize\centering\arraybackslash}X >{\hsize=.13\hsize\centering\arraybackslash}X >{\hsize=.13\hsize\centering\arraybackslash}X >{\hsize=.13\hsize\centering\arraybackslash}X
    >{\hsize=.13\hsize\centering\arraybackslash}X
    >{\hsize=.13\hsize\centering\arraybackslash}X}
    \hline 
    \hline
        Irrep & $I$ & $C_3$ & $C_2$ & $\sigma$ & $S_4$ & $\tau$ \\
        \hline
        $m_{A_2}$ & 1 & 1 & 1 & -1 & -1 & -1 \\
        $m_{E}^x$ & 1 & * & 1 & -1 & -1 & -1 \\ 
        $m_{E}^y$ & 1 & * & 1 & 1 & 1 & -1 \\[3pt]
        \hline
        \hline
    \end{tabularx}
    \label{tab:irrep_transformation}
\end{table}
While the irreps stay invariant or assume a minus sign under most symmetry operations in Table~\ref{tab:irrep_transformation}, the $\mathbf{m}_E$ irrep transforms according to
\begin{equation}
    \label{eq:c3_transformation}
    \begin{pmatrix}
        m_{E}^x \\[0.5ex]
        m_{E}^y
    \end{pmatrix}
    \xrightarrow{C_3}
    \begin{pmatrix}
        \cos \frac{2\pi n}{3} & -\sin \frac{2\pi n}{3} \\[0.5ex]
        \sin \frac{2\pi n}{3} & \cos \frac{2\pi n}{3} 
    \end{pmatrix}
    \begin{pmatrix}
        m_{E}^x \\[0.5ex]
        m_{E}^y
    \end{pmatrix}
\end{equation}
under the $C_3$ symmetry, corresponding to a $2\pi n/3$ rotation ($n\in\{0,1,2\}$) in the local $xy$ plane. Following reference \cite{Javanparast2015}, we introduce the complex variable
\begin{equation}
    m_{xy} = \sqrt{(m_{E}^x)^2+(m_{E}^y)^2}\  e^{i\eta}=|\mathbf{m}_E| e^{i\eta} \text{,}
\end{equation}
where $|\mathbf{m}_E|$ corresponds to the magnitude and $\eta$ to the direction of the magnetic moment in the local $xy$ plane. Applying a $C_3$ symmetry operation shifts the angle $\eta$ by $2\pi n/3$ ($n\in\{0,1,2\}$). This shift imply that only terms of the form  $m_{xy}^k$  are invariant under $C_3$ if $k\in3\mathbb{Z}$. In Table~\ref{tab:c3_invariant_transfomation} the transformation under the tetrahedral group $T_d$ of the real and imaginary parts of $m_{xy}^k$ is displayed.
\begin{table}[t]
\caption{Assumed factors of $C_3$ invariant terms under symmetry operations of $T_d$.}
    \centering
    \setlength{\extrarowheight}{3pt}
    \begin{tabularx}{\columnwidth}{>{\hsize=.20\hsize\centering\arraybackslash}X >{\hsize=.13\hsize\centering\arraybackslash}X >{\hsize=.13\hsize\centering\arraybackslash}X >{\hsize=.13\hsize\centering\arraybackslash}X >{\hsize=.13\hsize\centering\arraybackslash}X
    >{\hsize=.13\hsize\centering\arraybackslash}X
    >{\hsize=.13\hsize\centering\arraybackslash}X}
    \hline 
    \hline
        Term & $I$ & $C_3$ & $C_2$ & $\sigma$ & $S_4$ & $\tau$ \\
        \hline
        $|\mathbf{m}_E|^3\cos{3\eta}$ & 1 & 1 & 1 & -1 & -1 & -1 \\
        $|\mathbf{m}_E|^3\sin{3\eta}$ & 1 & 1 & 1 & 1 & 1 & -1 \\ 
        $|\mathbf{m}_E|^6\cos{6\eta}$ & 1 & 1 & 1 & 1 & 1 & 1 \\
        $|\mathbf{m}_E|^6\sin{6\eta}$ & 1 & 1 & 1 & -1 & -1 & 1 \\[2pt]
        \hline
        \hline
    \end{tabularx}
    \label{tab:c3_invariant_transfomation}
\end{table}

To identify all symmetry allowed terms, we first note that the squares of the order parameters, that is $m_{A_2}^2$ and $|\mathbf{m}_E|^2 = (m_{E}^x)^2+(m_{E}^y)^2$ are invariant under $T_d$. Consequently, higher-order terms obtained from the product of these 2nd-order terms result in invariant 4th-order terms. We consider these terms to be ``isotropic'' in the $\Gamma_5$ manifold as they obtain the same value for all the states lying on the $\Gamma_5$ manifold characterized by the angle $\eta$. To construct anisotropic terms, we multiply $m_{A_2}$ by $|\mathbf{m}_E|^3\cos{3\eta}$, where the anisotropy is reflected in the  $\cos{3\eta}$. In particular, we note that the term $m_{A_2}|\mathbf{m}_E|^3\cos{3\eta}$ is invariant since the minus signs in Table~\ref{tab:irrep_transformation} and Table~\ref{tab:c3_invariant_transfomation} under the $\sigma$, $S_4$ and $\tau$ transformations cancel out.

Following the same procedure as for the 4th order terms, we consider products of invariant $2$nd and $4$th order terms to obtain the allowed $6$th order terms, which include both isotropic and anisotropic terms. 
All $T_d$ invariant terms are summarized in Table~\ref{tab:invariant_terms}. 
\begin{table}
    \centering
    \caption{Invariant terms in $m_{A_2}$ and $\mathbf{m}_E$ up to sixth order.}
    \begin{tabular}{>{\centering\arraybackslash}m{0.19\columnwidth}>{\centering\arraybackslash}m{0.79\columnwidth}}
    \hline \hline \\[-2ex]
        2nd order & $m_{A_2}^2$, $|\mathbf{m}_E|^2$ \\[0.5ex]
        \\[-2ex]
        4th order & $m_{A_2}^4$, $|\mathbf{m}_E|^4$, $m_{A_2}^2|\mathbf{m}_E|^2$, $m_{A_2}|\mathbf{m}_E|^3\cos3\eta$\\[0.5ex]
        \\[-2ex]
        6th order & $m_{A_2}^6$, $|\mathbf{m}_E|^6$, $m_{A_2}^4|\mathbf{m}_E|^2$, $m_{A_2}^2|\mathbf{m}_E|^4$,\\[0.2ex]
        & $m_{A_2}^3|\mathbf{m}_E|^3\cos3\eta$, $m_{A_2}|\mathbf{m}_E|^5\cos3\eta$, $|\mathbf{m}_E|^6\cos6\eta$\\ \\[-2ex]
    \hline \hline
    \end{tabular}
    \label{tab:invariant_terms}
\end{table}

By including all terms up to fourth order, in addition to the sixth order symmetry anisotropic and isotropic terms, $|\mathbf{m}_E|^6\cos{6\eta}$ \cite{Rau2018FrustratedQR} and $|\mathbf{m}_E|^6$, respectively~\footnote{We introduce both of these terms to ensure that the free energy does not diverge to negative infinity, yielding an unphysical behavior.},
we obtain the free energy of Eq.~\eqref{eq:landau}.

\begin{figure}
    \centering
    \begin{overpic}[width=\columnwidth]{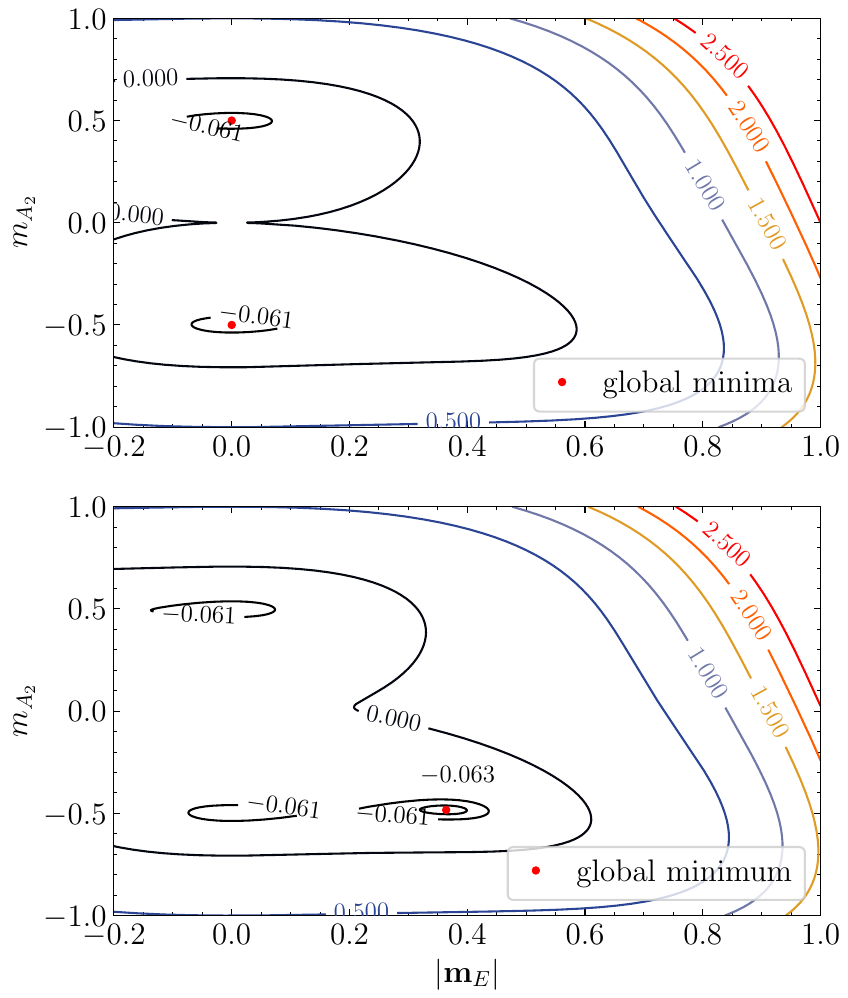}
    \put(2,100){(a)}
    \put(2,51){(b)}
    \put(45.5,21.2){\rotatebox{250}{\linethickness{0.3pt}\line(1,0){2.1}}}
    \end{overpic}
    \caption{Contour plots of the free energy in Eq.~\eqref{eq:landau} for the fixed parameters along the dashed line between point 2 and 4 in Fig.~\ref{fig:LG}, i.e. $\{r_1,r_2,r_{xyz},r_3,\omega,r_4,f_6\}=\{-0.5,1,1,1,2,1,0.5\}$. The parameter $r_0$ is set to (a) $0$, displaying an $A_2$ phase and (b) $-0.05$, where the system transitioned into the $A_2\oplus\psi_2$ phase.}
    \label{fig:contour}
\end{figure}

We find the value of the parameters $|\mathbf{m}_E|$, $m_{A_2}$ and $\eta$ as for the minimal free-energy configurations  minimize Eq.~\eqref{eq:landau} as a function of $r_0$ and $r_1$, while leaving all other variables constant, using the Nelder-Mead (downhill simplex) method of the python scipy library \cite{Gao2012}. We found that, for the sets of parameters considered in this work, $\eta=n\pi/3$ where $n\in\{0,\dots,5\}$, corresponding to a $\psi_2$ state. The minimization process identifies free-energy landscapes as the ones shown in Fig.~\ref{fig:contour}, for the sets of parameters $(r_0,r_1)=(0,-0.5)$, and $(r_0,r_1)=(-0.05,-0.5)$. As discussed in the main text, these parameters are located on the path $2\rightarrow4$ in Fig.~\ref{fig:LG}(a), where a first-order transition from the $A_2$ to the $A_2\oplus\psi_2$ phase is observed. In Fig.~\ref{fig:contour}(a) the $|\mathbf{m}_E|$ component of the global minima vanishes, meaning that the system is an $A_2$ phase. In the immediate vicinity of the phase transition, displayed in Fig.~\ref{fig:contour}(b), a separate minimum with a finite value of $|\mathbf{m}_E|$ becomes the new global minimum, which implies that the system transitioned into an $A_2\oplus\psi_2$ phase. The discontinuous evolution of the global minima exposes the  first-order phase transition between the two phases. Due to the high competition between the different minima, we use multiple starting points for the Nelder-Mead method, and select the solution that provides the lowest free energy.

\begin{figure}
    \centering
    \begin{overpic}[width=\columnwidth]{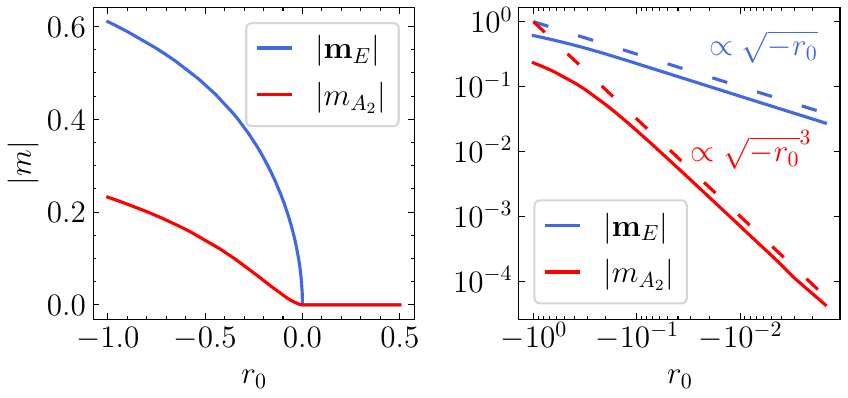}
    \put(2,48){(a)}
    \put(52,48){(b)}
    \end{overpic}
    \caption{Magnitude of the order parameters $|\mathbf{m}_E|$ and $|m_{A_2}|$ as a function of the parameter $r_0$ for the fixed set of parameters $\{r_1,r_2,r_{xyz},r_3,r_4,\omega,f_6\}=(0.5, 1, 1, 1, 2, 0.5)$ displayed with (a) normal scaled axes and (b) log scaled axes for negative values of $r_0$, where the dashed lines are proportional to powers of $-r_0$.}
    \label{fig:log_plot}
\end{figure}

In order to obtain the phase diagrams in Fig.~\ref{fig:LG}, we identify the global minimum of the free energy in Eq.~\eqref{eq:landau} for each coordinate $(r_1,r_0)$ and the fixed set of parameters $\{r_2,r_{xyz},r_3,\omega,r_4,f_6\}$. In Fig.~\ref{fig:log_plot} we expand Fig.~\ref{fig:LG}(d), which displays the second-order phase transition from the paramagnetic phase to the $A_2 \oplus \psi_2$ phase, by a logarithmic scaled plot of the order parameters in Fig.~\ref{fig:log_plot}(b). Using the logarithmic scale we determine the dependence of the order parameters for small values of $-r_0$ as $|\mathbf{m}_E|\propto \sqrt{-r_0}$ and $|m_{A_2}| \propto \sqrt{-r_0}^3$ yielding $|m_{A_2}| \propto |\mathbf{m}_E|^3$. We note that a similar dependency of the order parameter in the $A_2 \oplus \psi_2$ phase was already found in reference \cite{Javanparast2015}. 

The analysis presented so far is mainly focused in the set of parameters $\{r_2,r_{xyz},r_3,\omega,r_4,f_6\}$ considered for Fig.~\ref{fig:LG} in the main text. We note however that 
the phases and the qualitative behavior at the phase transitions remain unchanged under different sets of parameters. In the following we analyze how the phase diagram is modified under varying the parameters $r_{xyz}$ and $\omega$, and show that the qualitative behavior is kept. In particular, we consider the effect of varying the two coupling terms, namely $r_{xyz}m_{A_2}^2|\mathbf{m}_E|^2$ and $\omega m_{A_2} |\mathbf{m}_E|^3 \cos{3\eta}$, that include both order parameters $|\mathbf{m}_E|$ and $|m_{A_2}|$. The effect of these anisotropic two terms in the Landau theory can be understood in the following way; the $r_{xyz}m_{A_2}^2|\mathbf{m}_E|^2$ is nonnegative and consequently increases the free energy, while $\omega m_{A_2} |\mathbf{m}_E|^3 \cos{3\eta}$ always lowers the minimum of the free energy since it is the only odd term in $m_{A_2}$ and therefore adapts the sign to be negative. As a result we interpret $r_{xyz}m_{A_2}^2|\mathbf{m}_E|^2$ as a term that suppresses the coupling between the order parameters and $\omega m_{A_2} |\mathbf{m}_E|^3 \cos{3\eta}$ as coupling-enhancing.

\begin{figure}
    \centering
    \begin{overpic}[width=\columnwidth]{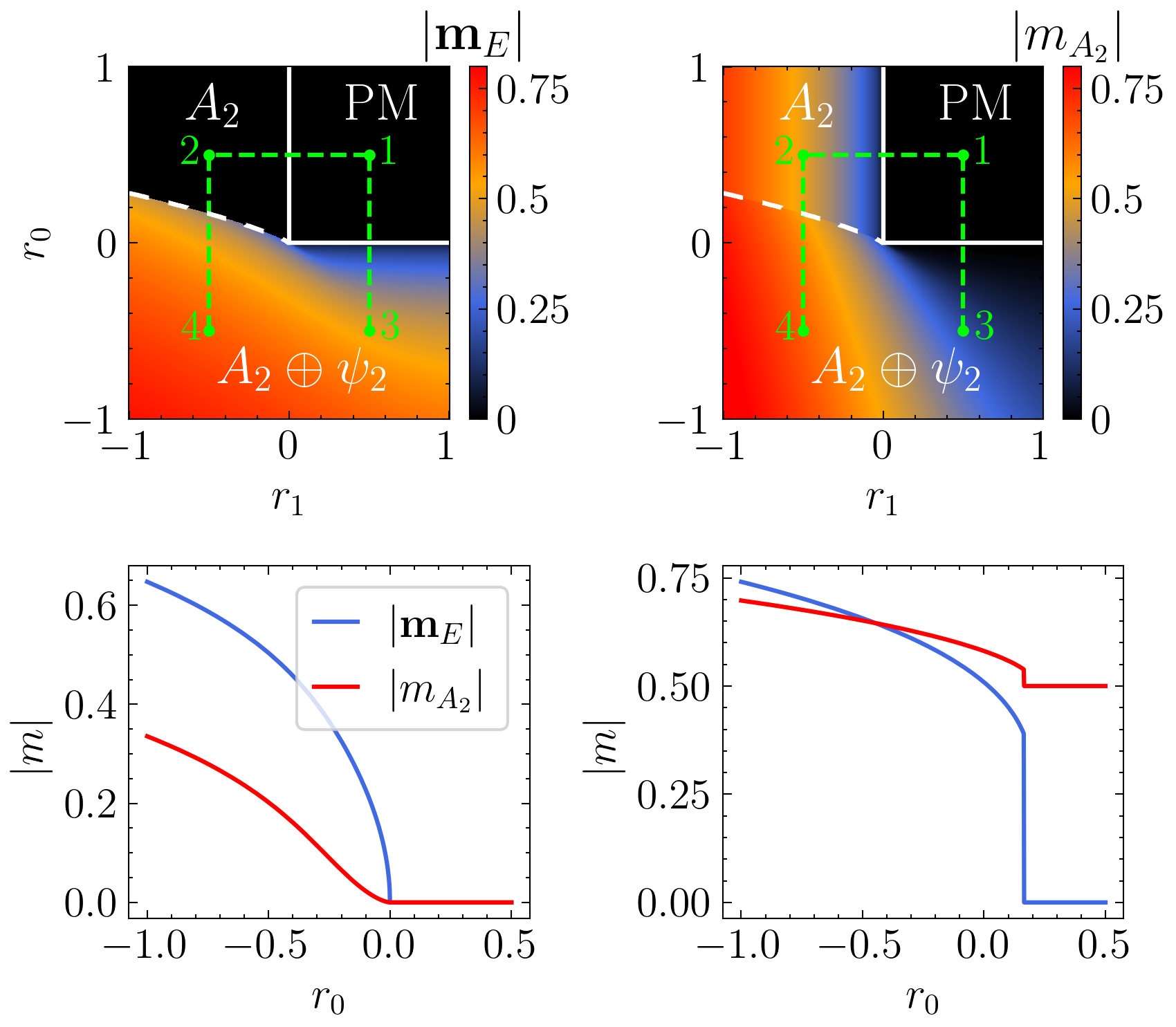}
    \put(2,83){(a)}
    \put(52,83){(b)}
    \put(2,42){(c)}
    \put(52,42){(d)}
    \end{overpic}
    \caption{Magnitude of the order parameters (a) $|\mathbf{m}_E|$ and (b) $|m_{A_2}|$ as a function of the parameters $r_0$ and $r_1$ for the fixed set of parameters $\{r_2,r_{xyz},r_3,r_4,\omega,f_6\}=(1, 0.2, 1, 1, 2, 0.5)$. Panel (c) displays the order parameters along the path $1\rightarrow3$ and panel (d) along the path $2\rightarrow4$, where the labels are the same as in panel (c).}
    \label{fig:coupling_enhancing}
\end{figure}

\begin{figure}
    \centering
    \begin{overpic}[width=\columnwidth]{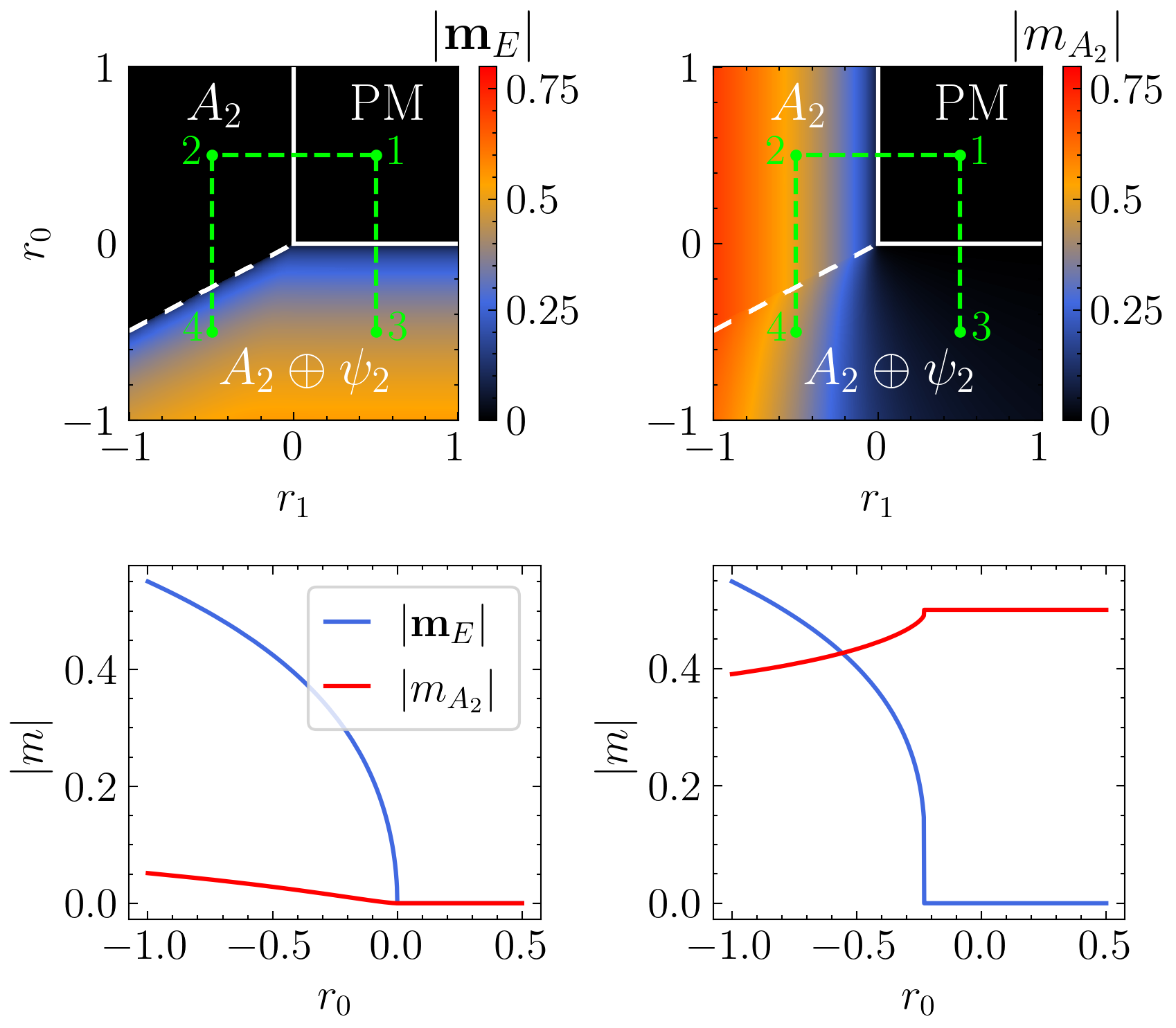}
    \put(2,83){(a)}
    \put(52,83){(b)}
    \put(2,42){(c)}
    \put(52,42){(d)}
    \end{overpic}
    \caption{Magnitude of the order parameters (a) $|\mathbf{m}_E|$ and (b) $|m_{A_2}|$ as a function of the parameters $r_0$ and $r_1$ for the fixed set of parameters $\{r_2,r_{xyz},r_3,r_4,\omega,f_6\}=(1, 1, 1, 1, 0.5, 0.5)$. Panel (c) displays the order parameters along the path $1\rightarrow3$ and panel (d) along the path $2\rightarrow4$, where the labels are the same as in panel (c).}
    \label{fig:coupling_suppressed}
\end{figure}

In Fig.~\ref{fig:coupling_enhancing}(a)-(b) we illustrate the magnitudes of $|\mathbf{m}_E|$ and $|m_{A_2}|$ for the case $\omega \gg r_{xyz}$. Consequently, the term $r_{xyz}m_{A_2}^2|\mathbf{m}_E|^2$ is suppressed relative to $\omega m_{A_2} |\mathbf{m}_E|^3 \cos{3\eta}$ leading to a strong coupling between the order parameters, which results in a larger magnitude of the order parameters in the $A_2\oplus\psi_2$ phase compared to Fig.~\ref{fig:LG}. Additionally, the phase transition from the $A_2$ to the $A_2\oplus\psi_2$ phase in Fig.~\ref{fig:coupling_enhancing}(d) takes place for larger values of $r_0$ and displays a discontinuous upward jump in $|m_{A_2}|$ compared to Fig.~\ref{fig:LG}(e). 
In the case $\omega \ll r_{xyz}$, shown in Fig.~\ref{fig:coupling_suppressed}, the magnitudes of order parameters are smaller compared to Fig.~\ref{fig:LG} with $|m_{A_2}|$ almost vanishing, while the phase transition from the $A_2$ to the $A_2\oplus\psi_2$ phase in Fig.~\ref{fig:coupling_suppressed}(d) exhibits a small downward jump in $m_{A_2}$.
As previously mentioned, the general characteristics of the phase transitions discussed in Sec.~\ref{sec:cMC} remain unchanged under the different sets of parameters: the transitions of the paramagnet to the $A_2$ phase and the paramagnet to the $A_2\oplus\psi_2$ phase are both of second order, and the transitions from $A_2$ to $A_2\oplus\psi_2$ display a second-order phase transition with the direction of the jump in $|m_{A_2}|$ depending on the values of the parameters. This observation implies that the general features discussed in the main text are qualitatively preserved for a variety of interaction parameters $\{r_2,r_{xyz},r_3,r_4,\omega,f_6\}$, which appear to be relevant for the model describing the  $A_2\oplus E\oplus T_{1-}$ line.

\section{Nematic order parameter in global and local frame}
\label{appendix:nem-OP-transformation}
In this appendix, we provide details on the nematic order parameter, its definition in the local and global frame, and the transformation rules under the group $T_d$.
First, consider the five symmetric traceless $3\times3$ matrices $\Lambda^{(\alpha)}$ introduced in Eq.~\eqref{eq:symmetric_traceless_matrices},
\begin{equation}
\begin{split}
    \Lambda^{(\alpha)}= \Biggl\{ \frac{1}{\sqrt{3}}
    &\begin{pmatrix}
        -1 & 0 & 0 \\
        0 & -1 & 0 \\
        0 & 0 & 2
    \end{pmatrix},
    \begin{pmatrix}
        -1 & 0 & 0 \\
        0 & 1 & 0 \\
        0 & 0 & 0
    \end{pmatrix},
    \begin{pmatrix}
        0 & 0 & 0 \\
        0 & 0 & 1 \\
        0 & 1 & 0
    \end{pmatrix}, \\
    &\begin{pmatrix}
        0 & 0 & 1 \\
        0 & 0 & 0 \\
        1 & 0 & 0
    \end{pmatrix},
    \begin{pmatrix}
        0 & 1 & 0 \\
        1 & 0 & 0 \\
        0 & 0 & 0
    \end{pmatrix}
    \Biggl\}.
\end{split}
\label{eq:gellmann-matrices}
\end{equation}
These matrices form a basis for the real symmetric traceless matrices. Specifically, the single-site traceless symmetric tensor $\mathsf{q}_i^{\mu\nu}=s_i^{\mu}s_i^\nu-\frac{1}{3}\delta^{\mu\nu}$, can be decomposed as~\cite{hecker_24,Nematic_order_Shanon2006PRL,Nematic_order_Shanon2015PRB,Taillefumier2017_xxz}
\begin{equation}
    \label{eq:nematic-OP-components-global}
    d_{\mathrm{glob},i}^{(\alpha)}=\frac{\sqrt{3}}{2} \operatorname{Tr}[\mathsf{q}_i \Lambda^{(\alpha)}]\,
\end{equation}
where the spins $\mathbf{s}_i$ of length $S$ are defined in the \textit{global} coordinate frame. Hence, the nematic tensor $\mathsf{q}_i$ is expressed in the global frame, as well as the components $d_{\mathrm{glob},i}^{\alpha}$ of the single-site nematic order parameter.
The normalization is chosen such that the magnitude becomes $|\mathbf{d}_{\mathrm{glob},i}|=S^2$ in the nematic phase at low temperatures. Note that the magnitude remains $S^2$ also in the presence of a $\mathbf{q}=0$ order.
Before moving on to the $T_d$ symmetry group, it is convenient to show how these quantities transform under a global orthogonal transformation in the spin space:
\begin{align}
    \label{eq:global-transf-rules}
        \mathbf{s}_i &\rightarrow \mathbf{s}_i'=\mathsf{O}\mathsf{s}_i\,\quad \mathrm{with} \quad \mathsf{O}\mathsf{O}^\top = \mathbbm{1}_{3\times 3}\,, \nonumber\\
        \mathsf{q}_i &\rightarrow \mathsf{q}_i'= \mathsf{O}\mathsf{q}_i\mathsf{O}^\top\,,\nonumber\\
        \mathbf{d}_{\mathrm{glob},i} &\rightarrow \mathbf{d}_{\mathrm{glob},i}'= \mathsf{M}\mathbf{d}_{\mathrm{glob},i}\quad \mathrm{with} \quad \mathsf{O}^\top \Lambda^{(\alpha)}\mathsf{O}=\mathsf{M}^{\alpha\beta}\Lambda^{(\beta)}\,,
\end{align}
where the matrix $\mathsf{O}$ is an orthogonal $3\times3$ matrix, while the matrix $\mathsf{M}$ is a $5\times5$ matrix acting on the single-site nematic order parameter. The matrix $\mathbbm{1}_{p\times p}$ represents a $p\times p$ identity matrix.
From the relation $\mathsf{O}^\top \Lambda^{(\alpha)}\mathsf{O}=\mathsf{M}^{\alpha\beta}\Lambda^{(\beta)}$, and from the property $\operatorname{Tr}[\Lambda^{(\alpha)}\Lambda^{(\beta)}]=2\delta^{\alpha\beta}$, it is possible to show that the $5\times 5$ $\mathsf{M}$ matrix is also orthogonal, i.e. $\mathsf{M}\mathsf{M}^\top=\mathbbm{1}_{5\times5}$. Hence, under global rotations of the spin moments, the nematic vector rotates accordingly, and its magnitude is preserved.

The Hamiltonian in Eq.~\eqref{eq:general_bilinear_hamiltonian} is symmetric under the $T_d$ group, involving discrete transformation both in spin and lattice space. Its action in the global frame involves  a permutation acting on the lattice sites and a rotation acting on the spin component. As a consequence, the single-site nematic order parameter transforms as
\begin{equation}
    \label{eq:nem-OP-Td-global}
    \mathbf{d}_{\mathrm{glob},i}\rightarrow \mathbf{d}_{\mathrm{glob},i}'= \pi_{ij}\mathsf{T}\mathbf{d}_{\mathrm{glob},j}
\end{equation}
where $\pi_{ij}$ represents the site permutation while the matrix $\mathsf{T}$ refers to the action of a discrete rotation of the $T_d$ group on the single-site nematic order parameter. It is known that in the presence of the $T_d$ symmetry group, the nematic order parameter splits into a doublet component, called $\eg$, and a triplet component $\tg$~\cite{hecker_24}:
\begin{equation}
    \label{eq:nematic-op-split}
    \mathbf{d}_{\mathrm{glob},i}=\left( \mathbf{d}_{i}^{\eg}, \mathbf{d}_i^{\tg} \right)\,,
\end{equation}
where the $\eg$ ($\tg$) doublet (triplet) is constructed from the diagonal (off-diagonal) matrices in Eq.~\eqref{eq:gellmann-matrices}. Under the transformation in Eq.~\eqref{eq:nem-OP-Td-global}, the doublet and triplet components do not mix, meaning that in the global coordinate frame either the doublet or the triplet is realized, but they cannot be simultaneously nonzero~\cite{hecker_24}.
The nematic phase with doublet components is characterized by spins aligned with the $\pm\hat{\mathbf{x}}$, $\pm\hat{\mathbf{y}}$, or $\pm\hat{\mathbf{z}}$ global Cartesian directions, i.e. towards the faces of a cube. Conversely, the triplet nematic phase has magnetic moments aligned along cubic diagonals. 

In our calculations, we consider the single-site nematic order parameter defined in a \textit{local} coordinate frame, see Eq.~\eqref{eq:single_site_spin_nematic_vector}. 
This definition introduces additional single-site rotation from the global to the local frame, allowing a mixing of the doublet and triplet components. 
%

\section{Finite-size scaling of continuous phase transitions}
\label{appendix:FSS-continuous-transition}
In this appendix, we present further details on the analysis of the continuous phase transitions and on the method we used to extract the critical temperature and the critical exponents $\nu$ and $\gamma/\nu$. Continuous phase transitions, or critical phenomena, are characterized by power-law scaling of physical observables due to emergent scale invariance at criticality. 
However, on finite-size systems, the correlation length can grow only up to the largest linear size of the system. As a consequence, the scaling of observables in finite-size systems is modified. Considering a dimensionless renormalization-group invariant quantity $R$ (e.g., a Binder cumulant), the finite-size scaling around criticality is asymptotically given by 
\begin{equation}
    \label{eq:FSS-RG-inv}
    R=f_R((T-T_{\mathrm{c}})L^{1/\nu} )+\mathcal{O}(L^{-\omega})\,,
\end{equation}
where $T_{\mathrm{c}}$ is the critical temperature, $L$ is the linear size of the system, $\nu$ is the correlation-length critical exponents, $\omega$ is the exponent related to scaling corrections, and $f_R$ is a (up to a rescaling of its argument) universal function that depends only on the observable and the universality class of the transition~\cite{pelissetto02}. 
Neglecting scaling corrections, quantities that scale as in Eq.~\eqref{eq:FSS-RG-inv} are expected to show a common crossing point at the critical temperature. Moreover, data points for different system sizes collapse on a common scaling curve once they are plotted as a function of the scaling argument $(T-T_{\mathrm{c}})L^{1/\nu}$.

Dimensionful observables have a slightly different scaling. Around the transition, the scaling for the susceptibility reads 
\begin{equation}
    \label{eq:FSS-susc-scaling}
    \chi_I=L^{\gamma/\nu}f_{\chi_I}((T-T_{\mathrm{c}})L^{1/\nu} )+\mathcal{O}(L^{-\omega})\,,
\end{equation}
where $\chi_I$ is the susceptibility for the $I$-th irreps, the critical exponent $\gamma/\nu$ quantifies the divergence at criticality, and $f_{\chi_I}$ represents the scaling function of $\chi_I$. Hence, the susceptibility curve for different system sizes collapses on a common curve once they are rescaled with $L^{-\gamma/\nu}$ and plotted as a function of $(T-T_{\mathrm{c}})L^{1/\nu}$.
\begin{table}[t]
\caption{Critical quantities extracted from finite-size-scaling collapse for the three representative cases of continuous transition discussed in the main text.}
    \centering
    \setlength{\extrarowheight}{3pt}
    \begin{tabularx}{\columnwidth}{>{\hsize=.25\hsize\centering\arraybackslash}X >{\hsize=.25\hsize\centering\arraybackslash}X >{\hsize=.25\hsize\centering\arraybackslash}X >{\hsize=.25\hsize\centering\arraybackslash}X }
    \hline 
    \hline
        $J_{z\pm}$ & $T_{\mathrm{c}}$ & $\nu$ & $\gamma/\nu$  \\
        \hline
        $0$ & $0.957(1)$ & $0.635(8)$ & $1.205(15)$ \\
        $0.4$ & $0.752(2)$ & $0.62(3)$ & $1.66(4)$  \\ 
        $1.2$ & $0.440(2)$ & $0.65(2)$ & $1.60(2)$ \\[2pt]
        \hline
        \hline
    \end{tabularx}
    \label{tab:fss-scaling}
\end{table}

\begin{figure*}[t]
    \centering
    \begin{overpic}[width=0.8\textwidth]{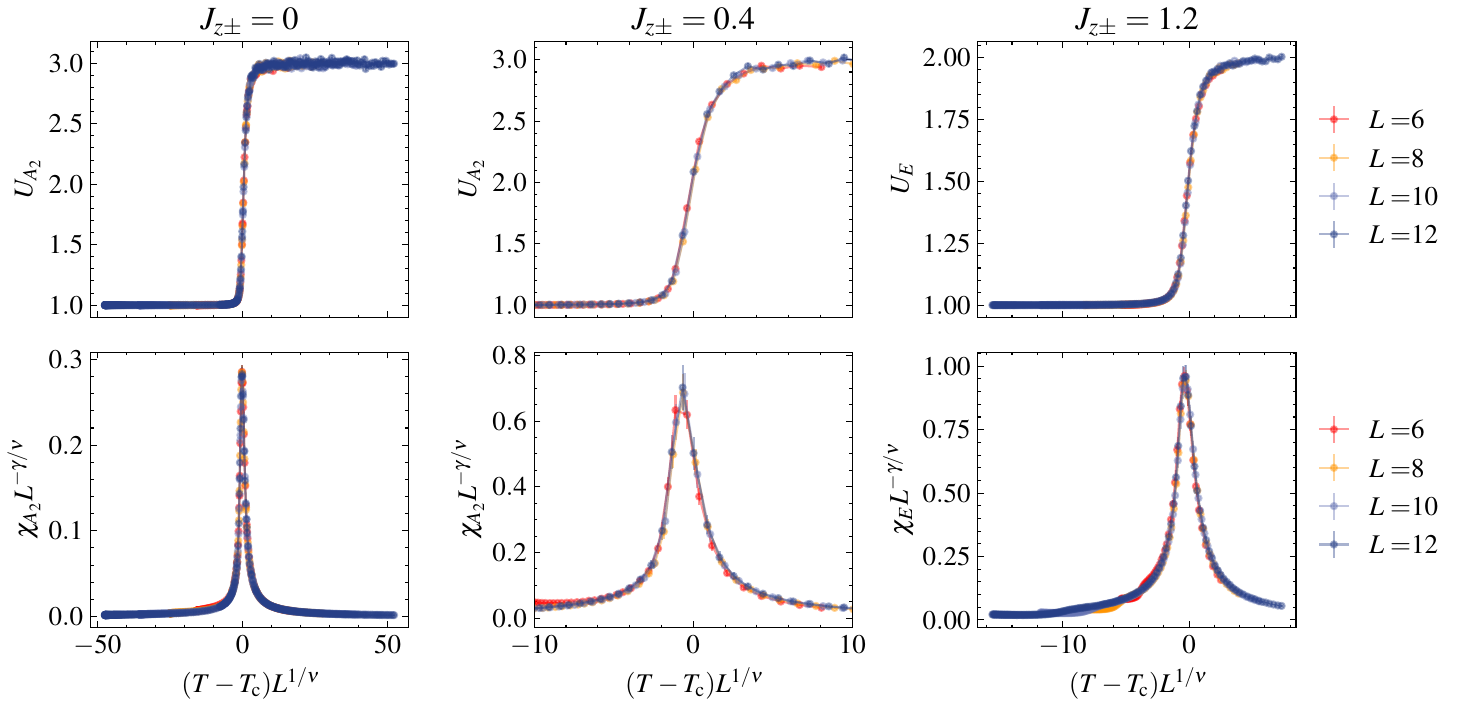}
        \put(0,45){(a)}
        \put(30,45){(b)}
        \put(60,45){(c)}
        \put(0,23){(d)}
        \put(30,23){(e)}
        \put(60,23){(f)}
    \end{overpic}
    \caption{Finite-size scaling collapses for the Binder cumulant (top) and susceptibilities (bottom) for different values of $J_{z\pm}$. The resulting critical temperatures and critical exponents are reported in Tab.~\ref{tab:fss-scaling}.}
    \label{fig:fss-scaling-continuous}
\end{figure*}

The finite-size scaling collapse represents a powerful tool for determining the critical exponents and the critical temperature of a continuous phase transition. In our analysis, we employed a Bayesian inference method to find the optimal scaling curve and the associated critical temperatures and exponents~\cite{harada11}. The finite-size-scaling collapses for Binder cumulant and susceptibilities are reported in Fig.~\ref{fig:fss-scaling-continuous} for the three representative continuous transitions discussed in the main text. The respective quantities extracted from each scaling collapse are reported in Tab.~\ref{tab:fss-scaling}. The quality of the scaling of the Binder cumulant and susceptibility is good for each one of the transition considered: from the paramagnetic phase to the $A_2$ phase in the non-Kramers case [Fig.~\ref{fig:fss-scaling-continuous}(a)-(d)], from the paramagnetic phase to the $A_2$ phase in the Kramers case for $J_{z\pm}=0.4$ [Fig.~\ref{fig:fss-scaling-continuous}(b)-(e)], as well as from the paramagnetic phase to the $A_2\oplus \psi_2$ phase in the Kramers case for $J_{z\pm}=1.2$ [see Fig.~\ref{fig:fss-scaling-continuous}(c)-(f)].

\section{Further details on \texorpdfstring{$A_2\oplus\psi_2$}{A2+ψ2} detection from Monte Carlo data}
\label{appendix:E6-op-and-details}
In this appendix, we provide details on the detection of the $A_2\oplus \psi_2$ phase, in particular, how to determine whether the components of the $E$ phase are $\psi_2$ or $\psi_3$. 
We used the $\langle m_{E6} \rangle$ parameter defined in Eq.~\eqref{eq:e6} to determine the nature of the $E$ components in the $\mathbf{q}=0$ $A_2\oplus E$ phase. Our simulations show that, throughout the $J_{z\pm}$ range, whenever the $A_2 \oplus E$ phase is realized, the $\langle m_{E6} \rangle$ parameter becomes positive, exposing the nature of the mixed state as the $A_2 \oplus \psi_2$ phase.

\begin{figure}[tb!]
    \centering
    \begin{overpic}[width=\columnwidth]{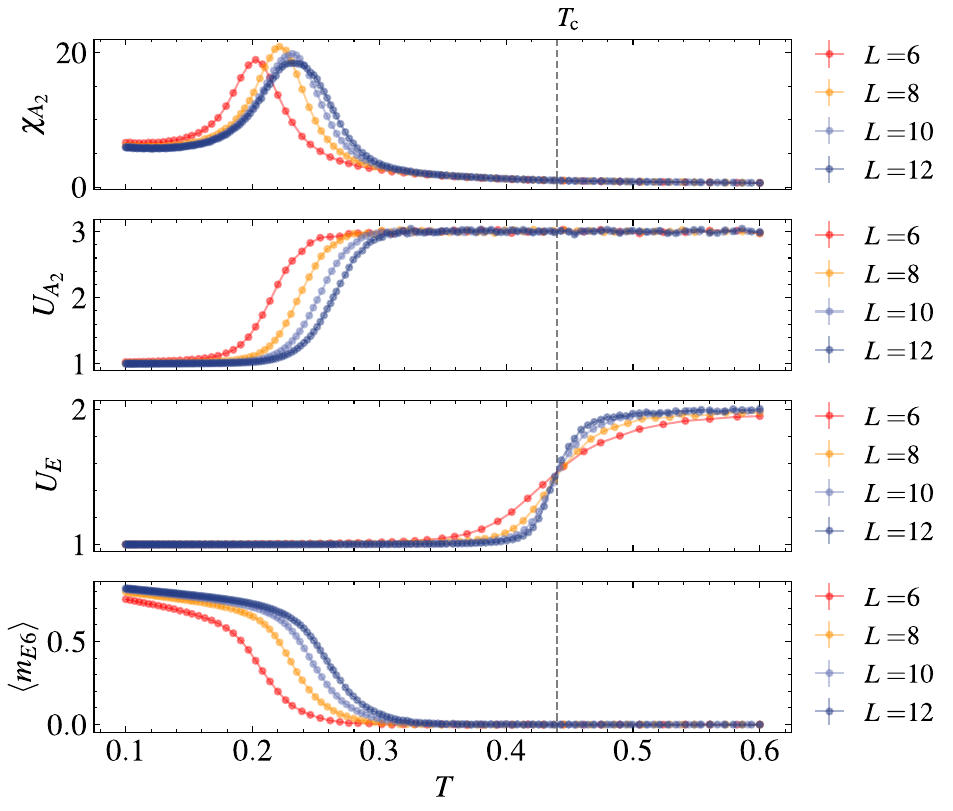}
    \put(0,78){(a)}
    \put(0,60){(b)}
    \put(0,42){(c)}
    \put(0,24){(d)}
    \put(25,82.5){$A_2\oplus \psi_2$}
    \put(70,82.5){PM}
    \end{overpic}
    \caption{Monte Carlo results at $J_{z\pm} = 1.2$. (a)~$A_2$ susceptibility, (b)~$A_2$ Binder parameter, (c)~$E$ Binder parameter, and (d)~$\langle m_{E6} \rangle$ parameter as a function of temperature $T$. The observables in panel (a-c) are related to the long-range order parameter displayed in Fig.~\ref{fig:jzpm-1.2-cMC} in the main text.  The vertical dashed line indicates the position of the phase transition. The continuous transition at $T_{\mathrm{c}}$ drives the system from the paramagnetic phase into the $A_2\oplus \psi_2$ long-range order phase. The $A_2$ related observables in panel (a)-(b) do not show the typical features of symmetry-breaking phase transitions, such as diverging susceptibility maxima with increasing sizes or a common crossing point of the Binder parameter. Then, the crossing point of the $E$ Binder parameter can be taken as the transition temperature where the $\mathbb{Z}_2 \times\mathbb{Z}_3$ symmetry breaks. The $\langle m_{E6} \rangle$ parameter in panel (d) acquires positive values at low-$T$, pointing to a $\psi_2$ realization of the $E$ irrep.
}
\label{fig:jzpm-1.2-appendix}
\end{figure} 

In the following, we show an example of the temperature evolution of the $\langle m_{E6} \rangle$ parameter along with the susceptibility associated with the $A_2$ order parameter, and the Binder parameters of the $A_2$ and $E$ phases  for the case $J_{z\pm}=1.2$. 
As shown in Fig.~\ref{fig:jzpm-1.2-cMC} in the main text, the $A_2$ order parameter starts to increase at a lower temperature than the temperature at which the $E$ order parameter becomes nonzero. As discussed in the main text, the appearance of a nonvanishing $m_{A_2}$ parameter, however, is not associated with a phase transition. Indeed, the $A_2$ susceptibility in Fig.~\ref{fig:jzpm-1.2-appendix}(a) lacks any divergencies of its maxima as a function of increasing sizes. On the contrary, it seems to approach a common crossover curve.
Furthermore, a common crossing point at the transition temperature is typically present in the Binder parameter, whereas the $A_2$ Binder parameter in Fig.~\ref{fig:jzpm-1.2-appendix}(b) displays no evidence for such a crossing point, strengthening the crossover hypothesis. 
The position of the symmetry-breaking transition can be determined from the intersection of the $E$ Binder parameter, as displayed in Fig.~\ref{fig:jzpm-1.2-appendix}(c).
Altogether, the data shown in Fig.~\ref{fig:jzpm-1.2-cMC} of the main text and in Fig.~\ref{fig:jzpm-1.2-appendix} suggest a $\mathbf{q}=0$ phase consisting of a mixed $A_2 \oplus E$ phase. Here, the $\langle m_{E6} \rangle$ parameter proves to be the tool to determine the $\psi_2$ nature of the $E$ irrep. As reported in Fig.~\ref{fig:jzpm-1.2-appendix}(d), the $\langle m_{E6} \rangle$ parameter becomes positive at low temperature, signaling the presence of the $\psi_2$ component and ruling out any possibility of $\psi_3$ in the state.
To summarize, the phase reduces to the mixed $A_2 \oplus \psi_2$ phase as marked in the phase diagram of Fig.~\ref{fig:phase-diagram}.
We note that similar arguments also hold for the other low temperature regions of mixed $A_2\oplus E$ phase, implying that, throughout the phase diagram, the phase is labeled as $A_2 \oplus \psi_2$.

\section{Further details on \texorpdfstring{$T_{1-}$}{T1-} irreps in the nematic phase}
\label{appendix:nematic-t1minus-details}

\begin{figure}[tb!]
    \centering
    \begin{overpic}[width=\columnwidth]{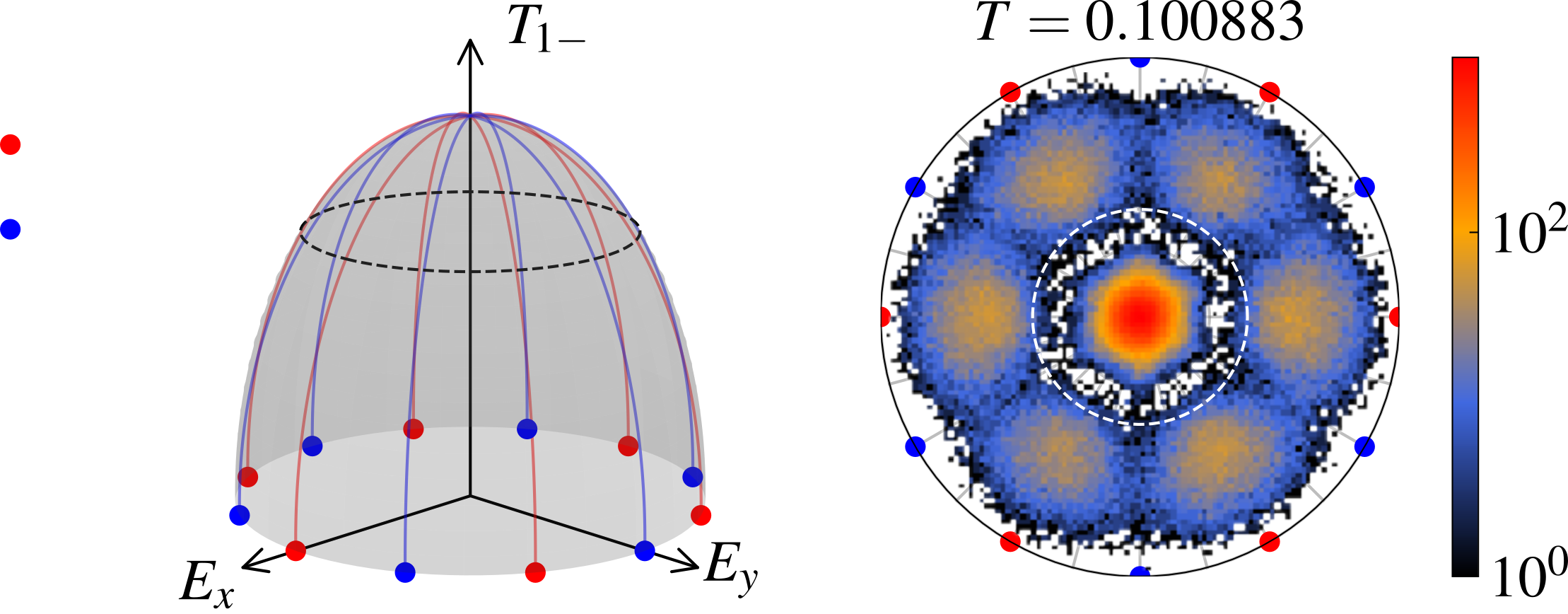}
    \put(0,35){(a)}
    \put(45,35){(b)}
    \put(4,29){$\psi_2$}
    \put(4,23.5){$\psi_3$}
    \end{overpic}
    \caption{(a) Sketch of the $T_{1-}\oplus E$ hemisphere. The red points and lines highlight the position of the $\psi_2$ angle, while the blue lines relate to the $\psi_3$ position. (b) Irrep distributions on the $T_{1-}\oplus E$ hemisphere for $L=10$ at $J_{z\pm} = 1/\sqrt{2}$. The temperature is $T=0.100883$ with a total magnitude $|\mathbf{m}_{\boxtimes}|\approx 0.99$. Two domains are visible, one close to $\theta=\pi/2$, $A_2\oplus E$ configurations, and one close to $\theta=0$, $T_{1-}$ configurations. In the $A_2\oplus E$ domain, the distribution of $\eta$ points to a $\psi_2$ realization. However, in the $T_{1-}$ domain it is not clear if $\psi_2$ or $\psi_3$ are realized.
    }
    \label{fig:jzpm-0.7-eta-dom}
\end{figure} 
\begin{figure}[tb!]
    \centering
    \begin{overpic}[width=\columnwidth]{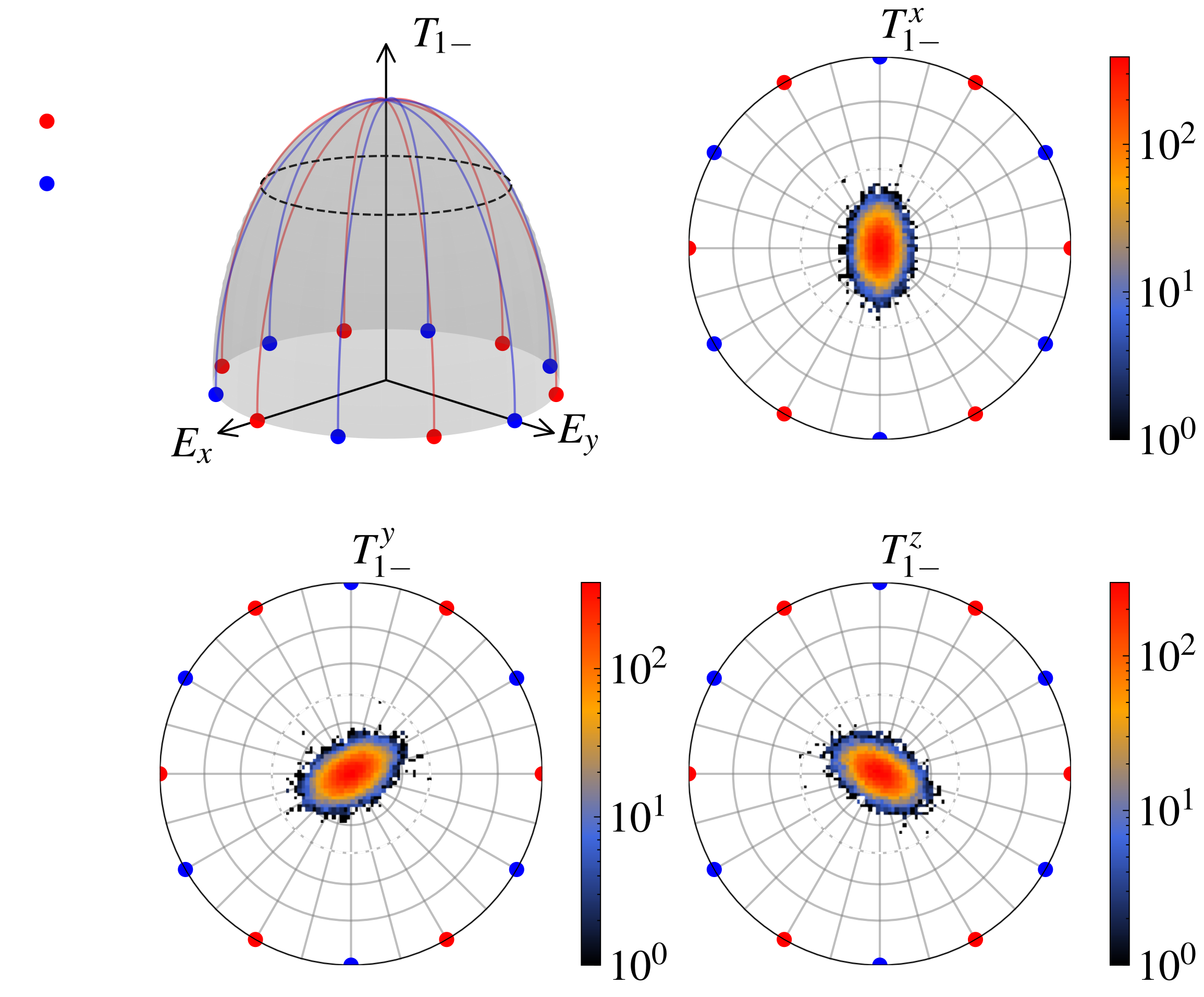}
    \put(15,80){(a)}
    \put(55,80){(b)}
    \put(15,35){(c)}
    \put(55,35){(d)}
    \put(7,72){$\psi_2$}
    \put(7,66.5){$\psi_3$}
    \end{overpic}
    \caption{Irreps distributions in three different sectors of the $T_{1-}$ domain at $T=0.100883$ at the nematic point $J_{z\pm} = 1/\sqrt{2}$ on the $T_{1-}\oplus E$ hemisphere sketched in panel (a). In this domain, the single-tetrahedron $T_{1-}$ irreps points along the $x,y,z$ directions, so we can decouple the three sectors. (b) Along $x$ the $T_{1-}^x$ couples with $\psi_3$ component of the $E$ irreps via $\eta=\pm\pi/2$. (c) Along $y$ the $T_{1-}^y$ couples with $\psi_3$ component of the $E$ irreps via $\eta=\pi/6$ or $\eta=-5\pi/6$. (d) Along $z$ the $T_{1-}^z$ couples with $\psi_3$ component of the $E$ irreps via $\eta=-\pi/6$ or $\eta=5\pi/6$.}
    \label{fig:jzpm-0.7-eta-T1-dom}
\end{figure} 

In this appendix, we provide further details on the $T_{1-}$ mode in the nematic phase. As discussed in the main text, the nematic phase features tetrahedra in the $A_{2}\oplus\psi_2$ irrep and in the $T_{1-}$ irrep, see Sec.~\ref{subsec:cMC-at-nematic}. The cause of the $T_{1-}$ appearance is tracked back to the exact additional subsystem symmetries at the nematic point, rendering the energetically degenerate $T_{1-}$ irrep also entropically equivalent to the $A_2\oplus\psi_2$ mode, as described in Sec.~\ref{subsec:exact-symmetry-nematic} and Sec.~\ref{subsec:GS-nematic-characterization}.

Figure~\ref{fig:jzpm-0.7-nem-dom} in the main text shows a finite number of tetrahedra in the $T_{1-}$ mode at the nematic point at low temperature. Small fluctuations away from the pure $T_{1-}$ mode tend to avoid the $A_2$ mode while favoring the $E$ mode. Hence, the irrep-magnitude distributions are represented on the $T_{1-}\oplus E$ manifold, a hemisphere with pole corresponding to the $T_{1-}$ irreps and the equator being the $E$ manifold, as sketched in Fig.~\ref{fig:jzpm-0.7-eta-dom}(a).
Figure~\ref{fig:jzpm-0.7-eta-dom}(b) shows two regions, one close to the equator for small $T_{1-}$ and one around the pole for dominating $T_{1-}$. As expected, the irreps at the equator are distributed around $\psi_2$ components, since it belongs to the $A_2\oplus E$ sector of the nematic phase. However, around the pole, a hexagonal shape is visible, hinting at the mixing of $T_{1-}$ states with $\psi_3$ fluctuations.
This hypothesis is corroborated once the $A_2\oplus E$ sector is masked out and single components of the $T_{1-}$ irreps are observed. In Fig.~\ref{fig:jzpm-0.7-eta-T1-dom}, the three components of $T_{1-}$ are singled out: in Fig.~\ref{fig:jzpm-0.7-eta-T1-dom}(b), $T_{1-}^x$ components couple only with $\psi_3$ states with $\eta=\{\pi/2,3\pi/2\}$, while in Fig.~\ref{fig:jzpm-0.7-eta-T1-dom}(c), $T_{1-}^y$ components couple only with $\psi_3$ states with $\eta=\{\pi/6,7\pi/6\}$, and in Fig.~\ref{fig:jzpm-0.7-eta-T1-dom}(d), $T_{1-}^z$ components couple only with $\psi_3$ states with $\eta=\{5\pi/6,11\pi/6\}$.
%
%
%

\bibliographystyle{longapsrev4-2}
\bibliography{pyrochlore-nematic}

\end{document}